\documentstyle[12pt]{article}

\newcommand{\be}{\begin{equation}}
\newcommand{\ee}{\end{equation}}

\def\psnormal{\textwidth=16cm\textheight=22cm
          \oddsidemargin=0.5cm\evensidemargin=0cm
          \topmargin=0cm\parindent=1cm}
\psnormal

\begin{document}
\pagestyle{empty}
\rightline{{\bf CERN--TH.6194/91}}
\rightline{{\bf IEM--FT--38/91}}
\rightline{{\bf  US--FT--7/91}}

\vspace{0.2cm}
\begin{center}
{\bf COMPLETE STRUCTURE OF $Z_n$ YUKAWA COUPLINGS}
\vspace{0.7cm}

J.A. CASAS${}^*$, F. GOMEZ${}^{\dagger}\;$ and C. MU\~NOZ${}^{**}$
\vspace{0.5cm}

${}^{*}$ Instituto de Estructura de la Materia (CSIC),\\
Serrano 119, 28006--Madrid, Spain
\vspace{0.3cm}

${}^{\dagger}$ Departamento de F\'{\i}sica de Part\'{\i}culas,\\
Universidad de Santiago, 15706--Santiago, Spain
\vspace{0.3cm}

${}^{**}$ CERN, CH--1211 Geneva 23, Switzerland
\vspace{0.3cm}

\end{center}

\centerline{\bf Abstract}
\vspace{0.2cm}

\noindent We give the complete twisted Yukawa couplings for all
the $Z_n$ orbifold constructions in the most general case, $i.e.$
when orbifold deformations are considered. This includes a
certain number of tasks. Namely, determination of the allowed
couplings, calculation of the explicit dependence of the Yukawa
couplings values on the moduli expectation values ($i.e.$ the
parameters determining the size and shape of the compactified
space), etc. The final expressions are completely explicit,
which allows a counting of the {\em different} Yukawa couplings
for each orbifold (with and without deformations). This
knowledge is crucial to determine the phenomenological
viability of the different schemes, since it is directly related
to the fermion mass hierarchy. Other facts concerning
the phenomenological profile of $Z_n$ orbifolds are also
discussed, e.g. the existence of non--diagonal entries in the
fermion mass matrices, which is related to a non--trivial structure of the
Kobayashi--Maskawa matrix. Finally some theoretical results are given,
e.g. the no--participation of (1,2) moduli in twisted Yukawa couplings.
Likewise, (1,1) moduli associated with fixed tori which are involved in the
Yukawa coupling, do not affect the value of the coupling.

\noindent
\vspace{0.2cm}
\begin{flushleft}
{\bf CERN--TH.6194/91} \\
{\bf IEM--FT--38/91} \\
{\bf US--FT--7/91}\\
{July 1991}
\end{flushleft}
\psnormal
\psnormal
\newpage
\pagestyle{plain}
\pagenumbering{arabic}
\section{Introduction and brief review}

In the last few years an enormous effort has been made in order to establish
the
 connection between
string theories \cite{green} (especially $E_8 \times E_8$ heterotic string
\cite{gross})
and low energy
physics. Different schemes for constructing classical string vacua  have
arisen during this time. Using these schemes it has been possible to build up
 four--dimensional
strings that resemble the Standard Model in many aspects, e.g. $SU(3) \times
SU(2) \times U(1)_Y $ gauge group and three generations of particles with the
 correct
representations [3--6]. In spite of these achievements
there remain many pending
 questions. In
particular there is a large number of classical vacuum states, which reduces
the
 predictive
power of the theory. At present there are no dynamical criteria to prefer a
 particular
vacuum, so the best we can do is to study their phenomenological
characteristics
 in order to
select the viable vacua. In this respect, orbifold compactifications
\cite{dixon} have proved to be
very interesting four--dimensional string constructions since they can pass
 succesfully a
certain number of low energy tests \cite{jcasas}. However, not all the
 experimental constraints
 have been
used in order to probe the phenomenological potential of orbifolds. The best
 example of this
is the observed structure of fermion masses and mixing angles \cite{jacasas}.

Concerning the last point, a crucial ingredient in order to relate theory and
 observation is
the complete knowledge of the theoretical Yukawa couplings. This knowledge
 includes a certain
number of aspects:

\begin{enumerate}
\item[i)] Physical states that enter the couplings.
\item[ii)] Allowed couplings.
\item[iii)] Numerical values of the Yukawa couplings and dependence of these
 values on the
physical parameters  that define the string vacuum (e.g. the size of the
 compactified space).
\item[iv)] Number of different Yukawa couplings and phenomenological viability
 of the scheme
($i.e.$ fitting of the observed pattern of fermion masses and mixing angles by
the theoretical Yukawa couplings).
\end{enumerate}

In this sense only for prime Abelian orbifolds, $i.e.$ $Z_3$ and $Z_7$ are the
Yukawa couplings completely known \cite{hamidi,jacasas,gomez}. For the
other orbifolds points i) and ii) have
recently been studied in ref.\cite{kobayashi}. General expressions of
$Z_n$ Yukawa couplings have been
determined in ref.\cite{burwick}. However, although very useful, they are not
explicit enough to
lucidate points iii) and iv), specially when deformations of the compactified
 space are
considered. Undoubtedly, a better knowledge of the Yukawa couplings is of
 utmost importance
to select or discard explicit string constructions with a highly non--trivial
 test. This is
the main purpose of this paper, $i.e.$ to answer points i), ii), iii), iv)
for {\em all} the $Z_n$
orbifolds. Besides the phenomenological motivation, there are strong
theoretical
 reasons to
completely determine the Yukawa couplings. In particular it is the only way to
 know the moduli
dependence of the matter Lagrangian and, in consequence, the superpotential.
 This allows the
examination of the properties of the action under target-space modular
 transformations (e.g. $R
\rightarrow 1/R$ ) \cite{ferrara}. It is also necessary in order to discuss
supersymmetry breaking dynamics \cite{lalak}
and cosmological implications (note that moduli play the role of Brans-Dicke
 fields in four
dimensions) of these theories.

Let us review briefly $Z_n$ orbifold constructions. A $Z_n$ orbifold is
 constructed by
dividing $R^6$ by a six--dimensional lattice $\Lambda$ modded by some $Z_n$
symmetry, called the point group P. The space group S is defined
as $S=\Lambda\times P$, $i.e.$ $S=\{(\gamma,u);\ \gamma\in P,\
u\in\Lambda\}$. The requirement of having $N=1$ supersymmetry in
four dimensions and the absence of tachyons restrict the number
of possible point groups \cite{dixon}. The complete list is given in the
first two columns of Table 1, where the so--called twist
$\theta$ ($i.e.$ the generator of P) is represented in an
orthogonal complex basis of the six--dimensional space.
$\Lambda$ must be chosen so that $\theta$ acts
crystallographically on it. If the realization of $\theta$ on
the lattice coincides with the Coxeter element of a rank--six
Lie algebra root lattice, the orbifold is of the Coxeter
type. A list of Coxeter orbifolds, taken from ref. \cite{markushevich}, is
given in
the third column of Table 1. Some additional examples of Coxeter
orbifolds can be found in ref. \cite{kobayashi}. The lattice of the
$Z_8$--II orbifold, $SO(5)\times SO(8)$, corresponds in fact to
a generalized Coxeter orbifold where the Coxeter element has
been multiplied by an outer automorphism. Non--Coxeter orbifolds
can also be constructed. An example of a non--Coxeter orbifold
(the $Z_4$ one with $[SO(4)]^3$ lattice) is studied in Section 3.
 The total number of possible lattices associated with each $Z_n$
orbifold can be found in ref. \cite{ono}. As will become clear in the text,
 some
properties of the Yukawa couplings for a particular $Z_n$ orbifold depend
on the lattice chosen, while others do not.

It is important to  point out that in a string orbifold
construction the lattice $\Lambda$ can get deformations
compatible with the point group \cite{jacasas,gomez}. These degrees of freedom
correspond to the untwisted moduli surviving compactification.
Deformations play an important role on the value of the Yukawa couplings.

We are interested in the couplings between twisted fields (the
untwisted sector has already been studied \cite{hamidi} and is physically
less interesting \cite{jacasas}). As we will see, these couplings present a
very rich range, which is extremely attractive as the
geometrical origin of the observed variety of fermion masses
\cite{hamidi,ibanez,jacasas}. A
twisted string satisfies $x(\sigma=2\pi)=gx(\sigma=0)$ as the
boundary condition, where $g$ is an element of the space group
whose point group component is non--trivial. Owing to the boundary
condition a twisted string is attached to a fixed point of $g$.
Physical twisted fields are associated with conjugation classes
of the space group rather than with particular elements \cite{dixon}. For
example $\{hgh^{-1},\ h\in S \}$ is the conjugation class of $g$. For
prime orbifolds conjugation classes are in one--to--one
correspondence with the fixed points of $P$. For non--prime
orbifolds the situation is a bit more involved since two
different fixed points under $\theta^n$ may be connected by
$\theta^m,\ m<n$. Then both of them correspond to the same conjugation
class.

The paper is organized as follows. In Section 2 we expound the various steps
necessary to obtain the final spectrum of Yukawa couplings for each
 orbifold,
taking the $Z_6$--I as a guide example. These steps include: determination of
 the
geometrical structure of the orbifold and deformation parameters, physical
 states,
calculation of explicit Yukawa couplings, and counting of different couplings.
 Section 3
is devoted to a comparative study of the $[SO(4)]^3$  and $[SU(4)]^2$ $Z_4$
 orbifolds. This
shows which properties of the couplings depend on the lattice chosen and which
 do not.
Furthermore, the $[SO(4)]^3$ case provides an example of a non--Coxeter
 orbifold. Besides
this, the $Z_4$ orbifold allows one to see the physical meaning of a (1,2)
modulus (absent in
the $Z_6$--I case) and its effect in the Yukawa coupling values. The complete
 results for the
rest of $Z_n$ orbifolds are given in Appendix 1 and summarized  in Table 1.

\section{The method}

Several steps are necessary in order to obtain the final spectrum of
Yukawa couplings for each orbifold. We explain these steps in
the present section, taking the $Z_6$--I orbifold as an illustrative example.
 The
reason for this choice is that prime orbifolds ($Z_3$ and $Z_7$) have already
been studied in depth in references \cite{hamidi,jacasas,gomez}. A complete
exposition of the method followed here can be found in ref. \cite{gomez}. It
is,
however, convenient to discuss the present example in some detail, since
 non--prime
orbifolds exhibit certain features which are absent in the prime ones.

\subsection{Geometrical structure}

The twist $\theta$ of the $Z_6$--I
orbifold has the form (see Table 1)
\be
\theta={\rm diag}(e^{i\alpha},e^{i\alpha},e^{-2i\alpha}),\;\;\;\;
\alpha=\frac{2\pi}{6}
\label{diag61}
\ee
in the complex orthogonal basis $\{(\tilde e_1,\tilde e_2),\
(\tilde e_3,\tilde e_4),\ (\tilde e_5,\tilde e_6)\ \}$. Very
often it is more suitable to work in the lattice basis
$\{e_1,...,e_6\}$, which in this case is simply a set of simple
roots of $G_2^2\times SU(3)$. Then $\theta$ acts as the Coxeter
element
\begin{eqnarray}
\begin{array}{lcl}
\theta e_1 = -e_1- e_2 ,& & \theta e_2 = 3e_1+2e_2 ,\\
\theta e_3 = -e_3- e_4 ,& & \theta e_4 = 3e_3+2e_4 ,\\
\theta e_5 =  e_6 ,& & \theta e_6 = -e_5- e_6 .
\end{array}
\label{teta61}
\end{eqnarray}
Note that we have labelled as $e_5,e_6$ the simple roots of
$SU(3)$. As mentioned above $\Lambda$ can get
deformations compatible with the point group. These degrees of
freedom correspond to the Hermitian part of the five untwisted
(1,1) moduli surviving compactification $N_{1\bar 1},N_{2\bar
2},N_{3\bar 3},N_{1\bar 2},N_{2\bar 1},$  where $N_{i\bar j}=
|i>_R \otimes \alpha_{\bar j L}^{-1}|0>_L;\;\; |0>_L$ ($|0>_R$) being
the left (right) vacuum, $\alpha_L$ is an oscillator operator
and $i$ ($\bar j$) is a holomorphic (antiholomorphic) index.
Note that under a deformation the actuation of $\theta$ on the
lattice basis, eq. (\ref{teta61}), remains the same. Then $P$
invariance imposes the following relations:
\begin{eqnarray}
\begin{array}{ccc}
|e_2|\;=\; \sqrt{3}\;|e_1|, & |e_4|\;=\;\sqrt{3}\;|e_3|, & |e_5|\;=\;|e_6|, \\
\alpha_{12}\;=\; -\sqrt{3}/2, & \alpha_{34}\;=\;-\sqrt{3}/2, &
 \alpha_{56}\;=\;-1/2, \\
\alpha_{24}\;=\;\alpha_{13},& \alpha_{23}\;=\;-\sqrt{3}\alpha_{13}-\alpha_{14},
 &
\alpha_{ij}=0 \;\;\; i=1,2,3,4 \;\; j=5,6
\end{array}
\label{reldef61}
\end{eqnarray}
where $\alpha_{ij}=\cos\theta_{ij}$ and
$e_ie_j=|e_i||e_j|\cos\theta_{ij}$. Therefore we can take the
5 {\em deformation degrees of freedom}
\begin{eqnarray}
\begin{array}{c}
R_i=|e_i|;\;\;\;i=1,3,5 \\
\alpha_{13},\;\alpha_{14}
\end{array}
\label{pardef61}
\end{eqnarray}
$R_i$ are global scales of the three sublattices
($G_2,G_2,SU(3)$), and for $\alpha_{13}=\alpha_{14}=0$ we
recover the rigid $G_2\times G_2\times SU(3)$ lattice. The
connection between the lattice basis $(e_1,...,e_6)$ and the
orthogonal basis $(\tilde e_1,...,\tilde e_6)$ in which $\theta$
takes the form (\ref{diag61}) is given by
\begin{eqnarray}
e_i&=& A_i\;[\cos(\varphi_1^i) \tilde e_1 + \sin(\varphi_1^i) \tilde e_2] +
B_i\;[\cos(\varphi_2^i) \tilde e_3 + \sin(\varphi_2^i) \tilde e_4 ] \nonumber\\
e_{i+1}&=& -\sqrt{3}[A_i\;(\sin(\frac{\pi}{3}-\varphi_1^i) \tilde e_1 +
\cos(\frac{\pi}{3}-\varphi_1^i) \tilde e_2) + B_i\;(
 \sin(\frac{\pi}{3}-\varphi_2^i) \tilde e_3 +
 \cos(\frac{\pi}{3}-\varphi_2^i) \tilde e_4)] \nonumber\\
e_j&=& R_5\;[ \cos((j-5)\frac{2\pi}{3}+\phi)
\tilde e_5 + \sin((j-5) \frac{2\pi}{3}+\phi) \tilde e_6 ]\;;\;\;i=1,3\;\;j=5,6
\label{latort61}
\end{eqnarray}
where $\varphi_1^1,\;\varphi_1^3,\;\varphi_2^1,\;\varphi_2^3,\;\phi$ are
 arbitrary angles that are
irrelevant for our results and
\[
\begin{array}{ll}
A_1 = R_1 \frac{1}{\sqrt{2}}[\Delta_1 \pm \Delta_3]^{1/2}&
B_1 = R_1 \frac{1}{\sqrt{2}}[\Delta_2 \mp \Delta_3]^{1/2}\\
A_3 = k_2 R_3 \sqrt{2}[\Delta_1 \pm \Delta_3]^{-1/2}&
B_3 = k_1 R_3 \sqrt{2}[\Delta_2 \mp \Delta_3]^{-1/2}
\end{array}
\]
with
\[
\begin{array}{c}
k_i=2 R_1R_3  (-1)^i [\alpha_{13}
 \sin(\frac{\pi}{3}+\varphi_i^1-\varphi_i^3)+\alpha_{14}
\cos(\varphi_i^1-\varphi_i^3)]
[\sin(\varphi_2^1-\varphi_2^3-\varphi_1^1+\varphi_1^3)]^{-1}\;\; i=1,2 \\
\Delta_1=1+k_2^2-k_1^2\;\;\;\;\;\Delta_2=1-k_2^2+k_1^2\;\;\;\;\;\Delta_3=[(1-k_1
 ^2-k_2^2
 )^2-4k_1^2k_2^2]
^{1/2} .
\end{array}
\]

Let us consider now the fixed points under the action of the
point group. $f_n$ is a fixed point under $\theta^n$ if it
satisfies $f_n=\theta^n f_n+u,\ u\in\Lambda$. As $Z_6$--I is a
non--prime orbifold, a point fixed by $\theta^n\;(n \neq 1)$ may be
not fixed by $\theta^m\;(m \neq n)$. Consequently, the fixed
points under $\theta$, $\theta^2$ and $\theta^3$ must be
considered separately ($\theta^4$, $\theta^5$ are simply the
antitwists of $\theta^2$ and $\theta$). It is easy to check from
(\ref{teta61}) that there are three different fixed points under
$\theta$. Working in the lattice basis their coordinates (up to
lattice translations) are

\begin{eqnarray}
f_1^{(1)}&=& g_1^{(0)} \otimes g_1^{(0)} \otimes \hat g_1^{(0)}, \nonumber \\
f_1^{(2)}&=& g_1^{(0)} \otimes g_1^{(0)} \otimes \hat g_1^{(1)}, \nonumber \\
f_1^{(3)}&=& g_1^{(0)} \otimes g_1^{(0)} \otimes \hat g_1^{(2)}
\label{fix161}
\end{eqnarray}
with
\begin{eqnarray}
\begin{array}{llll}
g_1^{(0)} = (0,0), & \hat g_1^{(0)} = (0,0), &
\hat g_1^{(1)} = (\frac{1}{3},\frac{2}{3}), & \hat g_1^{(2)} =
 (\frac{2}{3},\frac{1}{3}) .
\end {array}
\end {eqnarray}
Similarly under $\theta^2$ there are 27 fixed points. 12 of them
are connected to the others by a $\theta$ rotation
\begin{eqnarray}
\begin{array}{ll}
f_2^{(1)}= g_2^{(0)} \otimes g_2^{(0)} \otimes \hat g_2^{(0)}, &  \\
f_2^{(2)}= g_2^{(0)} \otimes g_2^{(0)} \otimes \hat g_2^{(1)}, &  \\
f_2^{(3)}= g_2^{(0)} \otimes g_2^{(0)} \otimes \hat g_2^{(2)}, &  \\
f_2^{(4)}= g_2^{(0)} \otimes g_2^{(1)} \otimes \hat g_2^{(0)}\sim
g_2^{(0)} \otimes g_2^{(2)} \otimes \hat g_2^{(0)},&
f_2^{(10)}= g_2^{(1)} \otimes g_2^{(1)} \otimes \hat g_2^{(0)}\sim
g_2^{(2)} \otimes g_2^{(2)} \otimes \hat g_2^{(0)}, \\
f_2^{(5)}= g_2^{(0)} \otimes g_2^{(1)} \otimes \hat g_2^{(1)}\sim
g_2^{(0)} \otimes g_2^{(2)}\otimes \hat g_2^{(1)} ,&
f_2^{(11)}= g_2^{(1)} \otimes g_2^{(1)} \otimes \hat g_2^{(1)}\sim
g_2^{(2)} \otimes g_2^{(2)} \otimes \hat g_2^{(1)} , \\
f_2^{(6)}= g_2^{(0)} \otimes g_2^{(1)} \otimes \hat g_2^{(2)}\sim
g_2^{(0)} \otimes g_2^{(2)} \otimes \hat g_2^{(2)} ,&
f_2^{(12)}= g_2^{(1)} \otimes g_2^{(1)} \otimes \hat g_2^{(2)}\sim
g_2^{(2)} \otimes g_2^{(2)} \otimes \hat g_2^{(2)} , \\
f_2^{(7)}= g_2^{(1)} \otimes g_2^{(0)} \otimes \hat g_2^{(0)}\sim
g_2^{(2)} \otimes g_2^{(0)} \otimes \hat g_2^{(0)} , &
f_2^{(13)}= g_2^{(1)} \otimes g_2^{(2)} \otimes \hat g_2^{(0)}\sim
g_2^{(2)} \otimes g_2^{(1)} \otimes \hat g_2^{(0)} , \\
f_2^{(8)}= g_2^{(1)} \otimes g_2^{(0)} \otimes \hat g_2^{(1)}\sim
g_2^{(2)} \otimes g_2^{(0)} \otimes \hat g_2^{(1)} ,&
f_2^{(14)}= g_2^{(1)} \otimes g_2^{(2)} \otimes \hat g_2^{(1)}\sim
g_2^{(2)} \otimes g_2^{(1)} \otimes \hat g_2^{(1)} , \\
f_2^{(9)}= g_2^{(1)} \otimes g_2^{(0)} \otimes \hat g_2^{(2)}\sim
g_2^{(2)} \otimes g_2^{(0)} \otimes \hat g_2^{(2)} , &
f_2^{(15)}= g_2^{(1)} \otimes g_2^{(2)} \otimes \hat g_2^{(2)}\sim
g_2^{(2)} \otimes g_2^{(1)} \otimes \hat g_2^{(2)}
\end{array}
\label{fix261}
\end{eqnarray}
with
\begin {eqnarray}
\begin {array}{lll}
g_2^{(0)} = (0,0), & g_2^{(1)} = (0, \frac{1}{3}), & g_2^{(2)}
 =(0,\frac{2}{3}),\\
\hat g_2^{(0)} = (0,0), & \hat g_2^{(1)} = (\frac{1}{3}, \frac{2}{3}), &
\hat g_2^{(2)} = (\frac{2}{3},\frac{1}{3}) .
\end {array}
\end {eqnarray}
Consequently, there are 15 conjugation classes under $\theta^2$.
Finally, under $\theta^3$, there are 16 fixed tori that are the
direct product of 16 fixed points in the sublattice
$(e_1,...,e_4)$ by the 2--torus defined by the sublattice
$(e_5,e_6)$. (Notice that $\theta^3$ is trivial in the $SU(3)$
root lattice.) 15 of these fixed points are connected between themselves
by $\theta$ rotations
\begin{eqnarray}
\begin {array}{lclcl}
f_3^{(1)}= g_3^{(0)} \otimes g_3^{(0)}, & & & & \\
f_3^{(2)}= g_3^{(1)} \otimes g_3^{(0)}&\sim& g_3^{(2)} \otimes g_3^{(0)}&\sim&
g_3^{(3)} \otimes g_3^{(0)},\\
f_3^{(3)}= g_3^{(0)} \otimes g_3^{(1)}&\sim& g_3^{(0)} \otimes g_3^{(2)}&\sim&
g_3^{(0)} \otimes g_3^{(3)},\\
f_3^{(4)}= g_3^{(1)} \otimes g_3^{(1)}&\sim& g_3^{(2)} \otimes g_3^{(2)}&\sim&
g_3^{(3)} \otimes g_3^{(3)},\\
f_3^{(5)}= g_3^{(2)} \otimes g_3^{(1)}&\sim& g_3^{(3)} \otimes g_3^{(2)}&\sim&
g_3^{(1)} \otimes g_3^{(3)},\\
f_3^{(6)}= g_3^{(3)} \otimes g_3^{(1)}&\sim& g_3^{(1)} \otimes g_3^{(2)}&\sim&
g_3^{(2)} \otimes g_3^{(3)}
\end{array}
\label{fix361}
\end{eqnarray}
with
\begin{eqnarray}
\begin{array}{llll}
g_3^{(0)} = (0,0), & g_3^{(1)}= (0,\frac{1}{2}), &
g_3^{(2)}= (\frac{1}{2},0), & g_3^{(3)}= (\frac{1}{2},\frac{1}{2}) .
\end{array}
\label{pfix361}
\end{eqnarray}
The direct product under the $(e_5,e_6)$ torus has been
understood. Consequently, there are 6 conjugation classes under $\theta^3$.
A similar analysis for other orbifolds can be found in Appendix 1.

\subsection{Physical states}

The next step is to determine which are the physical states.
These must be invariant under a total $Z_6$ transformation which,
besides the twist $\theta$ in the 6--dimensional space, includes
a $Z_6$ gauge transformation, usually represented by a shift
$V^I$ (the so--called embedding) on $\Lambda_{E_8 \times E_8}$ and a shift
$v^t$
on $\Lambda_{SO(10)}$. Accordingly one has to construct for each $\theta^k$
 sector
linear combinations of states, associated with $\theta^k$ fixed points, that
are
eigenstates of $\theta$ \cite{ohtsubo,kobayashi}. If $f_k$ is a fixed point
of $\theta^k$ such that
$l$ is the smallest number giving $\theta^lf_k=f_k+u,\;u\in\Lambda$, then the
eigenstates of $\theta$ have the form
\begin{eqnarray}
\begin{array}{c}
|f_k>+\;e^{-i\gamma}|\theta f_k>+... +\; e^{-i(l-1)\gamma}|\theta^{l-1}
 f_k>\\
\gamma=\; \frac{2\pi p}{l}\;,\;\;\; p=\;1,2,...,l
\end{array}
\label{fisstgen}
\end{eqnarray}
with eigenvalue $e^{-i\gamma}$ (obviously, if $k=1$, then $l=1$
and eq. (\ref{fisstgen}) is trivial). Under a $Z_6$ transformation
the complete state gets a phase \cite{ohtsubo,kobayashi}
\begin{eqnarray}
\Delta(k,e^{i\gamma})&=&\exp \left\{ 2\pi i [-\frac{1}{2}k(\sum_{I} (V^I)^2 -
\sum_{t} (v^t)^2 ) + \right. \nonumber\\
& & \left. +\sum_{I} (P^I+kV^I)V^I - \sum_{t} (p^t + k v^t)v^t]
\right\} \exp\{i\gamma\} ,
\label{DELTA}
\end{eqnarray}
where $p^t$ is the NSR part momentum put on the $SO(8)$ weight lattice and
$P^I$ is the transverse 8--dim. momentum ($E_8 \times E_8$ root momentum)
fulfilling the right--mover and left--mover massless conditions respectively.
Then $\Delta(k,e^{i\gamma})=1$ for physical states.

Let us apply this to the $E_8\times E_8$ heterotic string
compactified on the $Z_6$--I orbifold with
$V^I=\frac{1}{6}(1,1,-2,0,...,0)(0,...,0)$, $i.e.$ the standard embedding.
The unbroken gauge group is $(E_6\times SU(2)\times U(1))\times
E_8$. In the $\theta$ sector there are three physical states
transforming as (27)'s of $E_6$ corresponding to $|f_1^{(1)}>,
|f_1^{(2)}>, |f_1^{(3)}>$ respectively (see eq. (\ref{fix161})).
In the $\theta^2$ sector we can construct 27 eigenstates of
$\theta$ (see eq. (\ref{fix261}))
\begin{eqnarray}
|f_2^{(1)}>,\;\; |f_2^{(2)}>,\;\; |f_2^{(3)}>,\;\;
\left\{|f_2^{(j)}>+ e^{-i\gamma} |\theta
 f_2^{(j)}>\right\}_{j=4,...,15}\;,\;\;\;
\gamma=\pi,2\pi .
\label{fisst261}
\end{eqnarray}
After some algebra, only the symmetric
combinations survive ($i.e.$ $\Delta(k,e^{i\gamma})\;=\;1$ for them),
giving rise to 15 (27)'s under $E_6$. Similarly in the
$\theta^3$ sector we can construct 16 eigenstates of
$\theta$ (see eq. (\ref{fix361}))
\begin{eqnarray}
|f_3^{(1)}>,\;\;
\left\{|f_3^{(j)}>+ e^{-i\gamma}|\theta
f_3^{(j)}>+ e^{-i2\gamma}|\theta^2
f_3^{(j)}> \right\}_{j=2,...,6}\;,\;\;\;
\gamma=\frac{2\pi}{3},\;\frac{4\pi}{3},2\pi .
\label{fisst361}
\end{eqnarray}
In this case there survive 6 (27)'s, corresponding to the
symmetric combinations, and 5 ($\overline{27}$)'s, corresponding to
the $\gamma=2\pi/3$ combinations. We have performed a similar analysis
for each orbifold.

\subsection{Allowed Yukawa couplings}

Let us now analyse the allowed Yukawa couplings between physical
states. A twisted string associated with a fixed point $f$ and a
rotation $\theta^j$ is closed due to the action of
$g=(\theta^j,(I-\theta^j)f)$, so the corresponding conjugation
class is given by $\{
(\theta^k,u)(\theta^j,(I-\theta^j)f)(\theta^{-k},-\theta^{-k}u)
\}$, with $k\in Z,\; u\in \Lambda$. After some algebra, the general
expression of the conjugation class of $g$ is
\be
\left( \; \theta^j,\; (I-\theta^j) \left[ (f+\Lambda)\cup(\theta
f+\Lambda)\cup....\cup(\theta^{j-1} f+\Lambda) \right] \right) .
\label{conclgen}
\ee
The set of translations $(1-\theta^j)\;\{\bigcup
(\theta^kf+\Lambda), k=0,...,j-1\}$ is called the coset
associated with $\theta^j$ and $f$ (note that the cosets
associated with $f$ and $\theta^kf$ are obviously the same). For
a trilinear coupling of twisted fields $T_1T_2T_3$ to be allowed,
the product of the respective conjugation classes should contain
the identity. This implies, in particular, that the product of
the three point group elements
$\theta^{j_1}\theta^{j_2}\theta^{j_3}$ should be 1 (this is the
so--called point group selection rule). For the $Z_6$--I
orbifold this implies that only $\theta\theta^{2}\theta^{3}$,
$\theta^2\theta^{2}\theta^{2}$ and $\theta\theta\theta^{4}$
couplings have to be considered. A straightforward application of
the H--momentum conservation \cite{hamidi,cvetic,nilles} shows that the
$\theta\theta\theta^{4}$ couplings are also forbidden.
Furthermore for the $\theta\theta^{2}\theta^{3}$
couplings we must require
\begin{eqnarray}
(I,0) &\in& \left( \theta, \;(I-\theta)(f_1+\Lambda) \right) \;
\left( \; \theta^2,\; (I-\theta^2) \left[ (f_2 + \Lambda) \cup (\theta
 f_2+\Lambda)
\right] \right) \nonumber\\
& &\left( \theta^3,\;(I-\theta^3) \left[ (f_3 + \Lambda) \cup (\theta f_3 +
\Lambda ) \cup (\theta^2 f_3 + \Lambda) \right] \right) ,
\label{selruge61}
\end{eqnarray}
which leads to the so--called space group selection rule for the
coupling $\theta\theta^{2}\theta^{3}$ in the $Z_6$--I orbifold
\be
f_1\ +\ (I+\theta)f_2\ -\ (I+\theta+\theta^2)f_3\;\in\;\Lambda
\label{selru161} .
\ee
It should be noticed that if the space group selection rule
(\ref{selru161}) is satisfied for three fixed points $f_1,f_2,f_3$,
then it is also satisfied for $\theta^{k_1}f_1$,
$\theta^{k_2}f_2$, $\theta^{k_3}f_3$, and, consequently, for all
the physical combinations
$\sum_{k_1}e^{-ik_1\gamma_1}|\theta^{k_1}f_1>$,
$\sum_{k_2}e^{-ik_2\gamma_2}|\theta^{k_2}f_2>$,
$\sum_{k_3}e^{-ik_3\gamma_3}|\theta^{k_3}f_3>$, see
eq. (\ref{fisstgen}). For the case at hand, $i.e.$ the $Z_6$--I
orbifold, one can consider 270 kinds of couplings
$f_1^{(j_1)}f_2^{(j_2)}f_3^{(j_3)}$ of the
$\theta\theta^{2}\theta^{3}$ type, from which only 90 are
allowed: those for which the $e_5,e_6$ components ($i.e.$ the
$SU(3)$ sublattice projection) of $f_1-f_2$ are vanishing. So,
if we write the fixed points
\be
\left.
\begin {array}{rcl}
f_1 &=& g_1^{(0)} \otimes g_1^{(0)} \otimes \hat g_1^{(k_1)}\\
f_2 &=& g_2^{(i_2)}\otimes g_2^{(j_2)}\otimes \hat g_2^{(k_2)}\\
f_3 &=& g_3^{(i_3)}\otimes g_3^{(j_3)}\otimes [\alpha(e_5)+\beta(e_6)]
\end {array}
\right\}
\begin {array} {l}
k_1,k_2,i_2,j_2=0,1,2,\\
i_3,j_3=0,1,2,3,\\
\alpha,\beta \in R .
\end {array}
\ee
the selection rule is
\be
\begin {array}{lcr}
k_1 &=& k_2 .
\end {array}
\ee
At this point it is important to stress the following fact: for two given
 $f_1,\;f_2$,
the third fixed point $f_3$ (corresponding to $\theta^3$) is not determined
uniquely \footnote{This also happens for given $f_1,\;f_3$ but not for
$f_2,\;f_3$.}. We say
that the space group selection rule {\em is not diagonal}. From the physical
 point
of view this is extremely important, since it allows for non--diagonal fermion
 mass
matrices and, hence, a non--trivial Kobayashi--Maskawa matrix. This feature is
absent for prime orbifolds. On the other hand the space selection rule for the
$\theta^2\theta^2\theta^2$ couplings simply reads
\be
f_1\ +\ f_2\ +\ f_3\;\in\;\Lambda
\label{selru261}
\ee
which is {\em diagonal}. Note, however, that in this case the selection rule
can
be satisfied by some representatives of the conjugation classes and not by
 others.
In this case there are 3375 couplings to consider, from which only 369 are
 allowed.
These are

\be
\left.
\begin {array}{lcr}
i_1+i_2+i_3&=&0 \\
j_1+j_2+j_3&=&0 \\
k_1+k_2+k_3&=&0
\end {array}
\right\}
\;\;
mod.\;3 .
\ee

\noindent {denoting}
\be
\left.
\begin {array}{rcl}
f_1 &=& g_2^{(i_1)}\otimes g_2^{(j_1)} \otimes \hat g_2^{(k_1)}\\
f_2 &=& g_2^{(i_2)}\otimes g_2^{(j_2)} \otimes \hat g_2^{(k_2)}\\
f_3 &=& g_2^{(i_3)}\otimes g_2^{(j_3)} \otimes \hat g_2^{(k_3)}
\end {array}
\right\}\;\;\;
\begin {array} {l}
i_1,i_2,i_3=0,1,2\\
j_1,j_2,j_3=0,1,2\\
k_1,k_2,k_3=0,1,2 .
\end {array}
\ee
Let us finally note that the product of the $\theta$ eigenvalues
of the physical combinations of fixed points involved in the
coupling should be one, otherwise the coupling is vanishing. For
example, in the $Z_6$--I orbifold the following
$\theta\theta^{2}\theta^{3}$ coupling
\be
|f_1^{(j_1)}>\; (|f_2^{(j_2)}> + |\theta f_2^{(j_2)}>)\;
(|f_3^{(j_3)}> + e^{-\frac{2\pi i}{3}} |\theta f_3^{(j_3)}> + e^{-\frac{4 \pi
 i}{3}}
|\theta^2 f_3^{(j_3)}>)
\label{excou161}
\ee
is forbidden on these grounds, since, due to $\theta$
invariance, it is equal to
\be
|f_1^{(j_1)}>\; (|f_2^{(j_2)}> + |\theta f_2^{(j_2)}>)\;
|f_3^{(j_3)}> \;\;
(1+e^{-\frac{2\pi i}{3}}+e^{-\frac{4\pi i}{3}})
 = 0 .
\label{excou261}
\ee
This result can also be obtained for the standard embedding case from
gauge invariance since the state considered in the $\theta^3$
sector corresponds to a ($\overline{27}$), while the others are
(27)'s. In consequence all the couplings to be considered in the
$Z_6$--I involve symmetric combinations of fixed points ($i.e.$
$\theta$ eigenvalue $=1$) exclusively. We have performed the
previous analysis for all the $Z_n$ orbifolds. In all cases only couplings
between symmetric combinations of
fixed points survive. We do not really know what is the
fundamental principle behind this rule (if any), but it has
important consequences. For instance, in ref. \cite{ohtsubo} it was
suggested that the phases of the non--zero $\theta$ eigenvalue
states (see eq. (\ref{fisstgen})) could be the geometrical origin
of the phases of the Kobayashi--Maskawa matrix. Clearly, the
present rule excludes this possibility.

\subsection{Calculation of Yukawa couplings}

We are interested in couplings of the type $\psi\psi\phi$ (i.e.
fermion--fermion--boson). A trilinear string scattering
amplitude is given by the correlator
$<V_1(z_1)V_2(z_2)V_3(z_3)>$ of the vertex operators creating
the corresponding states. Complete expressions for the vertex
operators of the fields under consideration can be found in
refs. \cite{hamidi,cvetic}. As has been pointed out \cite{hamidi}
the non--vanishing
Yukawa couplings are essentially given by the bosonic twist
correlator $<\sigma_1(z_1)\sigma_2(z_2)\sigma_3(z_3)>$, where
$\sigma_i$ represents a twist field creating the appropriate
twisted ground state. According to subsection 2.2 $\sigma$ fields
for physical states are, in general, linear combinations of
$\sigma$ fields associated with specific rotations and fixed
points, say $\sigma_{\theta^j,f}$. For example, for a physical
state in the $\theta^j$ sector whose twist part is given by
$\sum_{k=0,...,l-1}e^{-ik\gamma}|\theta^{k}f>$ (the meaning of
$\gamma$ and $l$ is given in eq. (\ref{fisstgen})) the
corresponding twist field is simply
$\sum_{k=0,...,l-1}e^{-ik\gamma}\sigma_{\theta^j,\theta^{k}f}$.
According to the result of the previous subsection only
symmetric combinations ($\gamma=2\pi$) are relevant for trilinear
couplings, so the correlator
$<\sigma_1(z_1)\sigma_2(z_2)\sigma_3(z_3)>$ associated with a
$\theta^{j_1}\theta^{j_2}\theta^{j_3}$ will take the form
\begin {eqnarray}
\lefteqn{<\sigma_1(z_1)\;\sigma_2(z_2)\;\sigma_3(z_3)>=} \nonumber\\
& &\left( \sqrt{l_1l_2l_3} \right) ^{-1}\;
\sum_{k_1=0}^{l_1 -1} \;\sum_{k_2=0}^{l_2 -1} \;\sum_{k_3=0}^{l_3 -1}
<\sigma_{\theta^{j_1},\theta^{k_1}f^{(1)}}(z_1)\;
\sigma_{\theta^{j_2},\theta^{k_2}f^{(2)}}(z_2)\;
\sigma_{\theta^{j_3},\theta^{k_3}f^{(3)}}(z_3)> ,
\label{sigcorr1}
\end{eqnarray}
where the square root is a normalization factor. The correlation
functions on the right--hand side are evaluated following
standard lines \cite{hamidi}. They are defined by
\begin {eqnarray}
\lefteqn{<\sigma_{\theta^{j_1},\theta^{k_1}f^{(1)}}(z_1)
\sigma_{\theta^{j_2},\theta^{k_2}f^{(2)}}(z_2)
\sigma_{\theta^{j_3},\theta^{k_3}f^{(3)}}(z_3)>=}\nonumber\\
& & \int D X \; e^{-S} \;\sigma_{\theta^{j_1},\theta^{k_1}f^{(1)}}(z_1)
\sigma_{\theta^{j_2},\theta^{k_2}f^{(2)}}(z_2)
\sigma_{\theta^{j_3},\theta^{k_3}f^{(3)}}(z_3) .
\label{sigcorr2}
\end {eqnarray}
Owing to the Gaussian character of the action $S$
\be
S= \frac{1}{4 \pi} \int d^2 z ( \partial{X} \; \bar{\partial} \bar {X} + \bar
 {\partial}
X  \partial\bar{X} ) \;,
\label{action1}
\ee
where $X=X_1+iX_2$ and a sum over the three complex coordinates
is understood, the scattering amplitude can be separated into a
classical and a quantum part \cite{hamidi}
\be
Z=Z_{qu}\sum_{<X_{cl}>}\exp(-S_{cl}) .
\label{Zcorr}
\ee
The quantum contribution represents a global factor for all
the couplings with the same
$\theta^{j_1}\theta^{j_2}\theta^{j_3}$ pattern in a given
orbifold; so the physical information mostly resides in the
classical contribution. Eventually, a total normalization factor,
which depends on the size of the compactified space,
has to be determined with the help of the four--point
correlation function. The final task lies in writing the
couplings in terms of the physically significant parameters,
i.e. those that parametrize the size and shape of the orbifold.

Let us consider, for the sake of definiteness, a
$\theta\theta^2\theta^3$ coupling in our guide example, the
$Z_6$--I orbifold. The classical contribution, see
eq. (\ref{Zcorr}), to a
$<\sigma_{\theta}\sigma_{\theta^2}\sigma_{\theta^{-3}}>$
correlator has been determined in references \cite{gomez,burwick}, so we escape
here the details of the calculation. The result is that the
contribution of the classical (instantonic) solutions to the
classical action, eq. (\ref{action1}), for a three--point correlation
on the sphere, is
\begin{eqnarray}
S_{cl}^i&=& \frac {1}{4 \pi} \frac {|\sin (2\pi k_i/N)|\;|\sin (3\pi k_i/N)|}
{|\sin(\pi k_i/N)|}\; |v_i|^2
\nonumber \\
v&\in& (f_2-f_3+\Lambda)_{\perp}
\label{Scl1}
\end{eqnarray}
where $i=1,2,3$ denotes the corresponding complex coordinate
(and thus the projection over the associated $z$--plane), the
fields $X^i$ are twisted by $\exp(2\pi k_i/N)$ ($k_1=1, k_2=1,
k_3=2$ and $N=6$ for the $Z_6$--I orbifold) and
$(f_2-f_3+\Lambda)_{\perp}$ selects only $(f_2-f_3+\Lambda)$
shifts that are orthogonal to the invariant plane (this means
that we can choose $(f_2-f_3)_{i=3}=0$ and
$\Lambda=\Lambda_{G_2\times G_2}$). Several comments are in order
here. First it is clear that the $i=3$ plane ($i.e.$ the invariant
plane) does not contribute to the classical action, and the
coupling for $i=3$ behaves much as an untwisted one. In fact in
the invariant plane the three strings must be attached to the
same fixed point, $i.e.$ $(f_1)_{i=3}=(f_2)_{i=3}=(f_3)_{i=3}$.
These facts are general for all the couplings, in any orbifold,
when fixed tori are involved. Second, the $v$--coset in
(\ref{Scl1}) does not depend on $f_1$ since in the calculation
of $X_{cl}(z)$, $z_1$ has been sent to infinity by using
$SL(2,C)$ invariance. We call this the 2--3 picture (for more
details see ref. \cite{gomez}). Equation (\ref{Scl1}) can be expressed in the
1--2 and 1--3 pictures as well
\begin{eqnarray}
S_{cl(1-2)}^i&=& \frac {1}{16 \pi} \frac { |\sin(\pi k_i/N)| }{ |\sin(2 \pi
 k_i/N)| \;
|\sin(3 \pi k_i/N)| } \;
|v_i^{(12)}|^2 \nonumber \\
 v^{(12)}&\in& (I-\theta^2)
(f_1-f_2+\Lambda_{12})_{\perp}\;,
\label{Scl2}
\end{eqnarray}
\begin{eqnarray}
S_{cl(1-3)}^i&=& \frac {1}{16 \pi} \frac {|\sin(\pi k_i/N)|}
{|\sin(2 \pi k_i/N) |\; |\sin(3 \pi k_i/N)| }\;
|v_i^{(13)}|^2 \nonumber \\
v^{(13)}&\in& (I-\theta^3)(f_1-f_3+\Lambda_{13})_{\perp}\;,
\label{Scl3}
\end{eqnarray}
where
\begin{eqnarray}
\begin{array}{c}
\Lambda_{12}= (I+\theta+\theta^2)\Lambda + \omega\;,\;\;
\Lambda_{13}= (I+\theta)\Lambda+ \omega \;, \\
\omega = (I+\theta+\theta^2)f_3 - (\theta +\theta^2)f_2 -f_1 .
\end{array}
\label{L1213}
\end{eqnarray}
We can check by using the space group selection rule
(\ref{selru161}) that there is a one--to--one correspondence
between $S_{cl(1-2)},S_{cl(1-3)},S_{cl(2-3)}$. The 2--3 picture,
eq. (\ref{Scl1}) is the most convenient one since
$\Lambda_{12},\Lambda_{13}$ are subsets of the original lattice
$\Lambda$. Furthermore $S_{cl(1-2)}$ and $S_{cl(1-3)}$ depend on
the three fixed points considered $f_1,f_2,f_3$; while
$S_{cl(2-3)}$ depends only on $f_2,f_3$. \footnote{This difference
can be understood recalling that for $f_2,f_3$ given, the space
group selection rule (\ref{selru161}) determines $f_1$ uniquely,
which does not hold for the other two possibilities.}

We can now write the complete form of the correlator using
eqs. (\ref{Zcorr},\ref{Scl1})
\be
<\sigma_{\theta}\sigma_{\theta^2}\sigma_{\theta^{3}}>=\;N\; \sqrt{l_2\; l_3}\;
\sum_{u\in\Lambda_{\perp}} \exp \left\{ -\frac{1}{2 \pi} \sin (\frac{\pi}{3})
\left[ (f_{23}+u)_1^2 + (f_{23}+u)_2^2 \right] \right\} ,
\label{sigcorr3}
\ee
where $(f_{23}+u)_i$ is the $i$--plane projection of $(f_2-f_3
+u)$, $\Lambda_\perp = \Lambda_{G_2\times G_2}$, and $N$ is the
properly normalized quantum part \cite{burwick}
\be
N=\; \sqrt{V_{\perp}} \; \frac {1}{2 \pi} \;\frac {\Gamma(\frac{5}{6})
\Gamma(\frac{2}{3})}{\Gamma(\frac {1}{6}) \Gamma(\frac {1}{3}) } ,
\label{Nz6}
\ee
with $V_{\perp}$ the volume of the $G_2\times G_2$ unit cell.
General expressions for the couplings similar to
eq. (\ref{sigcorr3}) can be found in ref. \cite{burwick} for all the $Z_n$
orbifolds.
 We have
performed the calculation in all the  cases, checking that the results of
the mentioned reference are correct.

Expression (\ref{sigcorr3}) is not explicit enough for most purposes. For
example it does not allow
examination of the transformation properties of the Yukawa couplings under
 target--space
modular transformations (e.g. $R\; \rightarrow \; 1/R$). From a
phenomenological
 point of
view eq. (\ref{sigcorr3}) does not exhibit the dependence of the value of the
 coupling on
physical quantities, $i.e.$ those that parametrize the size and shape of the
 compactified
space. In fact eq. (\ref{sigcorr3}) is not even good enough to calculate the
 final value of
the coupling numerically, especially when deformations are considered. The key
point in order to do this is to write $(f_{23}+u)_i$ in terms of
 $(e_1,...,e_6)$,
$i.e.$ the lattice basis. This can be done with the help of the results of
subsection 2.1, see eq. (\ref{latort61}). Then the correlator (\ref{sigcorr3})
appears as an explicit function of the
deformation parameters of the compactified space. Substituting the resulting
 expression in
(\ref{sigcorr1}) we obtain the final Yukawa coupling, which can be writen in a
 quite compact way
\begin{eqnarray}
C_{\theta\theta^2\theta^3}&=& N \; \sqrt{l_2\; l_3}\;
\;\sum_{\vec{u} \in Z^4} \exp \left[-\frac{\sqrt{3}}{4 \pi}
(\vec{f_{23}}+\vec{u})^{\top} M (\vec{f_{23}}+\vec{u}) \right] \nonumber\\
&=& \sqrt{V_{\perp}}\;\; \sqrt{l_2\; l_3}\; \frac {1}{2 \pi}\; \frac
 {\Gamma(\frac{5}{6})
\Gamma(\frac{2}{3})}{\Gamma(\frac {1}{6}) \Gamma(\frac {1}{3}) }
\; \vartheta
\left[
\begin{array}{c}
\vec{f_{23}} \\ 0
\end {array}
\right]
[ 0,\; \Omega]
\label{C123}
\end{eqnarray}
where $\vec{f_{23}}$ represents the first four components of $(f_2-f_3)$
($i.e.$
 those
corresponding to the $G_2 \times G_2$ sublattice basis $(e_1,...,e_4)$ ), and
\be
\vartheta
\left[
\begin{array}{c}
\vec{f_{23}} \\ 0
\end {array}
\right]
[ 0,\; \Omega]
= \sum_{\vec{u} \in Z^4} \exp \left[ i \pi
(\vec{f_{23}}+\vec{u})^{\top} \Omega (\vec{f_{23}}+\vec{u}) \right]\;,\;\;\;\;
\Omega=\frac {i \sqrt{3}}{4 \pi^2} M
\label{thetajac}
\ee
with
\begin {eqnarray}
\Omega= \frac {i \sqrt{3}}{4 \pi^2}
\left(
\begin{array}{cccc}
R_1^2 & -\frac{3}{2} R_1^2 & R_1 R_3 \alpha_{13} & \sqrt{3} R_1 R_3 \alpha_{14}
 \\
-\frac{3}{2} R_1^2 & 3 R_1^2 & -R_1 R_3 (3\alpha_{13}+\sqrt{3}\alpha_{14}) &
3 R_1 R_3 \alpha_{13} \\
R_1 R_3 \alpha_{13} & -R_1 R_3 (3\alpha_{13}+\sqrt{3}\alpha_{14}) & R_3^2 &
-\frac{3}{2} R_3^2 \\
\sqrt{3} R_1 R_3 \alpha_{14} & 3 R_1 R_3 \alpha_{13} & -\frac{3}{2} R_3^2 &
 R_3^2
\end {array}
\right)
\label{Omega61}
\end {eqnarray}
where the deformation parameters $R_i^2,\; \alpha_{ij}$ have been
defined in eqs. (\ref{reldef61}, \ref{pardef61}).

It is worthwhile to have a look at eq. (\ref{C123}) to realize
which are the physical quantities on which the value of the
coupling depends. First $C_{\theta\theta^2\theta^3}$ depends on
the relative positions in the lattice of the relevant fixed
points to which the physical fields are attached. This
information is condensed in $\vec{f_{23}}$. Second
$C_{\theta\theta^2\theta^3}$ depends on the size and shape of
the compactified space, which is reflected in the orbifold
compactification parameters $(R_i^2, \alpha_{13}, \alpha_{14})$
appearing in $\Omega$ and (implicitely) in $V_{\perp}$. Note
that both pieces of information appear in a completely distinguishable way
from each other in eq. (\ref{C123}). Notice also that  the deformation
parameter $R_5^2$ does not appear in $\Omega$. This is due to
the fact that $R_5$ parametrizes the size of the $i=3$
sublattice, $i.e.$ the fixed torus, and we have learnt that for
$i=3$ the coupling is equivalent to an untwisted one. This is a
general fact for all the orbifold couplings in which fixed tori
are involved (e.g. it does not occur for the
$\theta^2\theta^2\theta^2$ coupling of the $Z_6$--I orbifold, see below).
We say that $R_5$ is not an {\em effective deformation parameter} for the
$\theta\theta^2\theta^3$ couplings. The number of effective deformation
parameters (4 in this case) is
physically relevant since it is strongly related to the number of different
Yukawa couplings
and their corresponding sizes.

For the other twisted coupling  $\theta^2\theta^2\theta^2$ in the $Z_6$--I
orbifold, the expression of the coupling can be
calculated in the same way as in the $\theta\theta^2\theta^3$ case, and is
given
 by
\begin {eqnarray}
C_{\theta^2\theta^2\theta^2} &=& F(l_1,l_2,l_3) \;
 N \sum_{v \in (f_3-f_2+\Lambda)} \exp [-\frac
{\sqrt{3}}{8\pi}  \; |v|^2]  \nonumber\\
&=& F(l_1,l_2,l_3) \; N \sum_{\vec{u} \in Z^6} \exp [-\frac
{\sqrt{3}}{8\pi}  \; (\vec{f_{23}}+\vec{u})^{\top} M (\vec{f_{23}}+\vec{u})]
 \nonumber\\
&=& F(l_1,l_2,l_3) \; N \vartheta
\left[
\begin{array}{c}
\vec{f_{23}} \\
0
\end {array}
\right]
[0, \Omega] ,
\end {eqnarray}
where $F=1$ for $l_1=1$ or $l_2=1$ or $l_3=1$ and $F=\frac{1}{\sqrt{2}}$ for
$l_1=l_2=l_3=2$. $l_i$ is the number of elements in the conjugation class
associated with $f_i$, see eq. (\ref{fisstgen}). $\vec{f_{23}}$ represents the
 components
of $(f_2-f_3)$ in the lattice basis $(e_1,...,e_6)$. The global normalization
 factor
and the $\Omega$ matrix are given by
\be
\begin{array}{c}
N= \sqrt{V_{\Lambda}}\; \frac{3^{3/4}}{8 \pi^3} \;
\frac{\Gamma^6(\frac{2}{3})}{\Gamma^3(\frac{1}{3})}\\
\Omega = i\frac{\sqrt{3}}{8\pi^2}\;
\left(
\begin {array}{cccccc}
a & -\frac{3}{2}a & b & c & 0 & 0\\
-\frac{3}{2}a & 3 a & -3b-c & 3 b & 0 & 0 \\
b & -3b-c & d & -\frac{3}{2}d & 0 & 0\\
c & 3 b  & -\frac{3}{2} d & 3 d & 0 & 0 \\
0 & 0 & 0 & 0 & e & -\frac{1}{2} e \\
0 & 0 & 0 & 0 & -\frac{1}{2} e & e
\end {array}
\right)\;\;\;
\begin{array}{l}
a= R_1^2 \\
b= R_1R_3 \alpha_{13} \\
c= \sqrt{3} R_1 R_3 \alpha_{14} \\
d= R_3^2 \\
e= R_5^2 .
\end{array}
\end{array}
\ee
Clearly the number of effective parameters is 5.

We have performed a similar analysis for all the trilinear twisted
couplings in all the $Z_n$ orbifolds, giving the number of effective
deformation parameters in each case. The results are expounded in
Appendix 1.

\subsection{Accidental symmetries and the number of different couplings}

We are now ready to count the {\em number of different couplings} that appear
in
 each
 orbifold.
{}From the physical point of view this is one of the most relevant questions
about
 a
string construction, since it is directly related to the possibility of
 reproducing the
observed pattern of fermion masses and mixing angles. Unfortunately, this
task is probably the
most tedious part of the work presented here. Again, we expound in some detail
the analysis for the two possible couplings in the $Z_6$--I orbifold. Let us
 begin
with the twisted coupling $\theta\theta^2\theta^3$.The corresponding results
for
other orbifolds can be found in Appendix 1.

The first point is that the $\Omega$ matrix  appearing in the Jacobi theta
 function of the
coupling, eqs. (\ref{C123}--\ref{Omega61}), is universal for all the
$\theta\theta^2\theta^3$ couplings. This means that the differences between the
 Yukawa
 couplings come
exclusively from the sum in $(f_{23}+u)$ in the classical part of the
 correlation. Hence, two
couplings
\begin {eqnarray}
C\sim\vartheta \left[ \begin{array}{c} \vec{f_{23}} \\ 0 \end{array} \right]
 \left[ \;
0,\Omega \right]\;\; \mbox{ and }\;\;
C'\sim\vartheta\;\left[ \begin{array}{c} \vec{f_{23}}'\\ 0 \end{array} \right]
 \; \left[
\; 0,\Omega \right]
\label{ccou61}
\end {eqnarray}
\vspace{.5 cm}
will have the same value if there exists an integer unimodular transformation
 $U$ $(i.e.
\;\;U\in GL(4,Z),\; \mid U \mid = \pm1 ) $ such that
\begin {eqnarray}
U^{\top}\;\Omega\;U &=& \Omega \\
\label{uo61}
U\;\vec{f_{23}} &=& \vec{f_{23}}'\;+\;\vec{v}\;\;,\;\;\;\;\vec{v}\in Z^4 .
\label{uf61}
\end {eqnarray}
Then, if the previous equations are true for some $U$, there is a one--to--one
 correspondence
between the terms of the series defining \(\vartheta \left[ \begin{array}{c}
 \vec{f_{23}} \\
 0 \end{array}
\right] \), see eq. (\ref{thetajac}), and those of \(\vartheta \left[
 \begin{array}{c}
\vec{f_{23}}' \\ 0 \end{array} \right] \). So we have to look for $U$--matrices
 satisfying
(\ref{uo61}). There is a set of  $U$--matrices that always fulfil (\ref{uo61}).
 These are \(
\{ I,-I \} \) and \( \{ {\theta^n, n\in Z} \} \). To check the latter, note
that
 when the sum
(\ref{thetajac}) is expressed in the complex orthogonal basis, the exponent
 takes a diagonal
form
\begin{eqnarray}
\sum_{u\in{\Lambda_\bot}}\; \exp{ \{ \;a_1(f_{23}+u)_1^2+a_2(f_{23}+u)_2^2\;\}
}
\label{ort61}
\end{eqnarray}
as can be seen from (\ref{sigcorr3}). Then the terms multiplying the
 coefficients
 $a_i$ are
unchanged under $\theta^n$ twists, since these correspond to make rotations in
 each $i$--plane.
This argument is always valid because the factorization (\ref{ort61}) is a
 consequence of the
fact that the classical contributions can be computed in each $i$--plane
 separately. In
addition to the group generated by $\{-I,\theta\}$ , there can be "accidental
$U$--symmetries" leaving $\Omega$ unchanged in eq. (\ref{uo61}). Some of these
 symmetries con
be spontaneously broken when deformations are taken into account. After
 inspection it turns
out that, for the case at hand, these accidental symmetries are generated by
\begin{eqnarray}
U_1=\left(
\begin{array}{r r}
e^{i\alpha} & \\
 & 1
\end{array}
\right)
\;\;\;
U_2=\left(
\begin{array}{r r}
1 & \\
 & e^{i\alpha}
\end{array}
\right)
\;\;\;
U_3=\left(
\begin{array}{r r}
0 &1 \\
1 & 0
\end{array}
\right)
\label{us61}
\end{eqnarray}
(when expressed in the complex orthogonal basis) plus products of these
matrices
 by $\{-I,\;
\theta \}$. $U_1,\;U_2,\;U_3$ are broken when deformations are
 considered.

Now, two Yukawa couplings $C,\;C'$ are equal (in the non--deformed case)
if $\vec{f_{23}}$ and $\vec{f_{23}}'$ are
connected as in (\ref{uf61}) by one of the $U$--matrices mentioned above. The
 analysis has to
be performed for the 90 $\theta\theta^2\theta^3$ allowed couplings,
see subsection 2.3.
The result is that for the rigid $G_2 \times G_2 \times SU(3)$ lattice
 $(\;i.e.\; R_1^2 =
R_3^2=R_5^2\;,\;\alpha_{13}=\alpha_{14}=0\;)$ there are 10 different couplings,
 corresponding
to the following set of $\vec{f_{23}}$ shifts (in $G_2 \times G_2$)

\be
\begin{array}{l}
l_3=1\;l_2=1\;:
\begin{array}{lll}
(0,0) \otimes (0,0), & &
\end{array}\\
l_3=3\;l_2=1\;:
\begin{array}{lll}
(0,0) \otimes (0,\frac{1}{2}), &(0,\frac{1}{2}) \otimes (0,\frac{1}{2}), &
\end{array}\\
l_3=1\;l_2=2\;:
\begin{array}{lll}
(0,0) \otimes (0,\frac{1}{3}), &(0,\frac{1}{3}) \otimes (0,\frac{1}{3}), &
\end{array}\\
l_3=3\;l_2=2\; \left\{
\begin{array}{lll}
(0,0) \otimes (0,\frac{1}{6}), &(0,\frac{1}{6}) \otimes (0,\frac{1}{6}), &
(0,\frac{1}{2}) \otimes (0,\frac{1}{3}), \\
(0,\frac{1}{3}) \otimes (0,\frac{1}{6}), & (0,\frac{1}{2}) \otimes
 (0,\frac{1}{6}) & .
\end{array}
\right.
\end{array}
\ee
The meaning of $l_i$ and its influence in the couplings are given
in eqs. (\ref{fisstgen}),(\ref{sigcorr1}).
With deformations the symmetry of the $\Omega$ matrix is smaller,
as explained above, and it turns out to be 30 different couplings
\be
\begin{array}{l}
l_3=1\;l_2=1\;:
\begin{array}{lll}
(0,0) \otimes (0,0), & &
\end{array}\\
l_3=1\;l_2=2\;:
\begin{array}{llll}
(0,0) \otimes (0,\frac{1}{3}), &(0,\frac{1}{3}) \otimes (0,0), &
(0,\frac{1}{3}) \otimes (0,\frac{1}{3}), &(0,\frac{1}{3}) \otimes
 (0,\frac{2}{3}),
\end{array}\\
l_3=3\;l_2=1\;
\left\{
\begin{array}{llll}
(0,0) \otimes (0,\frac{1}{2}), &(0,\frac{1}{2}) \otimes (0,0), &
(0,\frac{1}{2}) \otimes (0,\frac{1}{2}), & (0,\frac{1}{2}) \otimes
 (\frac{1}{2},0), \\
(0,\frac{1}{2}) \otimes (\frac{1}{2},\frac{1}{2}),& & &
\end{array}
\right.\\
l_3=3\;l_2=2\; \left\{
\begin{array}{llll}
(0,0) \otimes (0,\frac{1}{6}) &(0,\frac{1}{6}) \otimes (0,0), &
(0,\frac{1}{6}) \otimes (0,\frac{1}{6}), & (0,\frac{1}{6}) \otimes
 (0,\frac{5}{6}), \\
(0,\frac{1}{6}) \otimes (\frac{1}{2},\frac{1}{3}), &
(0,\frac{1}{6}) \otimes (\frac{1}{2},\frac{2}{3}), &
(0,\frac{1}{6}) \otimes (\frac{1}{2},\frac{1}{6}), &
(0,\frac{1}{6}) \otimes (\frac{1}{2},\frac{5}{6}), \\
(0,\frac{1}{3}) \otimes (0,\frac{1}{2}), &
(0,\frac{1}{2}) \otimes (0,\frac{1}{3}), &
(0,\frac{1}{2}) \otimes (0,\frac{1}{6}), &
(0,\frac{1}{2}) \otimes (\frac{1}{2},\frac{1}{3}), \\
(0,\frac{1}{2}) \otimes (\frac{1}{2},\frac{1}{6}), &
(0,\frac{1}{6}) \otimes (0,\frac{1}{2}), &
(\frac{1}{2},\frac{1}{3}) \otimes (0,\frac{1}{2}), &
(\frac{1}{2},\frac{1}{6}) \otimes (0,\frac{1}{2}), \\
(0,\frac{1}{6}) \otimes (0,\frac{1}{3}), &
(0,\frac{1}{6}) \otimes (0,\frac{2}{3}), &
(0,\frac{1}{3}) \otimes (0,\frac{1}{6}), &
(0,\frac{2}{3}) \otimes (0,\frac{1}{6})
\end{array}
\right.
\end{array}
\ee
The absolute and relative size of these 30 couplings obviously depend on the
value of the deformation parameters, as reflected in
eqs. (\ref{C123}--\ref{Omega61}).

Performing an analysis similar to the $\theta\theta^2\theta^3$ case,
we find out the number of inequivalent shifts for the
$\theta^2\theta^2\theta^2$
coupling. For the non--deformed  case there are 8 different couplings, namely

\be
\vec{f_{23}}=
\left[
\begin{array}{l}
l_1=1\;{\rm or}\;l_2=1\;{\rm or}\;l_3=1\;\;
\left\{
\begin{array}{lll}
g_2^{(0)} \otimes g_2^{(0)} \otimes \hat g_2^{(0)}, &
g_2^{(0)} \otimes g_2^{(0)} \otimes \hat g_2^{(1)},&
g_2^{(0)} \otimes g_2^{(1)} \otimes \hat g_2^{(0)}, \\
g_2^{(0)} \otimes g_2^{(1)} \otimes \hat g_2^{(1)}, &
g_2^{(1)} \otimes g_2^{(1)} \otimes \hat g_2^{(0)}, &
g_2^{(1)} \otimes g_2^{(1)} \otimes \hat g_2^{(1)}
\end{array}
\right. \\
l_1=l_2=l_3=2 \;:\;\;\;
\begin{array}{lll}
g_2^{(0)} \otimes g_2^{(0)} \otimes \hat g_2^{(0)}, &
g_2^{(0)} \otimes g_2^{(0)} \otimes \hat g_2^{(1)} & .
\end{array}
\end{array}
\right.
\ee

For deformations the number is increased to 12
different couplings given by the following shifts

\be
\vec{f_{23}}=
\left[
\begin{array}{l}
l_1=1\;{\rm or}\;l_2=1\;{\rm or}\;l_3=1\;\;
\left\{
\begin{array}{lll}
g_2^{(0)} \otimes g_2^{(0)} \otimes \hat g_2^{(0)}, &
g_2^{(0)} \otimes g_2^{(0)} \otimes \hat g_2^{(1)}, &
g_2^{(0)} \otimes g_2^{(1)} \otimes \hat g_2^{(0)}, \\
g_2^{(0)} \otimes g_2^{(1)} \otimes \hat g_2^{(1)}, &
g_2^{(1)} \otimes g_2^{(1)} \otimes \hat g_2^{(0)}, &
g_2^{(1)} \otimes g_2^{(1)} \otimes \hat g_2^{(1)}, \\
g_2^{(1)} \otimes g_2^{(0)} \otimes \hat g_2^{(0)}, &
g_2^{(1)} \otimes g_2^{(0)} \otimes \hat g_2^{(1)}, &
g_2^{(1)} \otimes g_2^{(2)} \otimes \hat g_2^{(0)}, \\
g_2^{(1)} \otimes g_2^{(2)} \otimes \hat g_2^{(1)} & &
\end{array}
\right. \\
l_1=l_2=l_3=2 \;:\;\;\;
\begin{array}{lll}
g_2^{(0)} \otimes g_2^{(0)} \otimes \hat g_2^{(0)}, &
g_2^{(0)} \otimes g_2^{(0)} \otimes \hat g_2^{(1)} & .
\end{array}
\end{array}
\right.
\ee
We have performed a similar analysis for all the $Z_n$ orbifolds, the results
are in Appendix 1. In all cases we have checked by computer that the number of
different Yukawa couplings is correct.

\section{A comparative study of the $[SO(4)]^3$ and $[SU(4)]^2$ $Z_4$
orbifolds}

Although most of the aspects of orbifold Yukawa couplings have been
 adequately
illustrated in the previous section by the $Z_6$--I example, there are still
 some
interesting features that can be exhibited in the framework of a $Z_4$
orbifold.
 In
particular we will see the physical meaning of a (1,2) modulus (absent in the
 $Z_6$--I
case) and its effect in the Yukawa coupling values. Furthermore, the comparison
 of the
Yukawa couplings of a $Z_4$ orbifold, formulated in an $[SO(4)]^3$ lattice,
with
 those of
a $Z_4$ orbifold, formulated in an $[SU(4)]^2$ lattice, will show us which
 properties of the
couplings depend on the chosen lattice and which do not. Moreover, the
 $[SO(4)]^3$
case provides an example of a non--Coxeter orbifold.
The twist of a $Z_4$ orbifold in an orthogonal complex basis has the form
(see Table 1)
\begin{eqnarray}
\theta={\rm diag}(e^{i\alpha},e^{i\alpha},e^{-2i\alpha}),\;\;\;\;
\alpha=\frac{2\pi}{4} .
\label{dz41}
\end{eqnarray}
Again, the lattice $\Lambda$ can get deformations compatible with the twist
 $\theta$. These
degrees of freedom correspond to the Hermitian part of the five (1,1) moduli
surviving compactification,
 \(N_{1\bar1},\;\;N_{2\bar2},\;\;N_{3\bar3},\;\;N_{1\bar2},
\;\;N_{2\bar1}\) with  $N_{i\bar j}=|i>_R \otimes \alpha_{\bar j L}^{-1}|0>_L$,
and the (1,2) modulus \( N_{33}=|3>_R \otimes \alpha_{3L}^{-1} |0>_L\).
(Notice that no $N_{ij}$ moduli appeared in the $Z_6$--I case.)
Untwisted moduli can be easily expressed in terms of $g_{mn}$, $b_{mn}$ (
 $m,\;n$ = 1,...,6),
$i.e.$ the internal metric and torsion respectively. It is easy to check,
however, that $N_{33}$ contains only $g_{mn}$ degrees of freedom, more
precisely $(g_{55}-g_{66})$ and  $g_{56}$. Therefore both Re$(N_{33})$
and Im$(N_{33})$ correspond to deformation parameters. In order to see what
these parameters are, let us choose first an $[SO(4)]^3$ root lattice, with
 basis
$(e_1,...,e_6)$, as a lattice on which the twist $\theta$, see eq.
(\ref{dz41}),
acts crystallographically as %
\begin{eqnarray}
\begin{array} {lll}
\theta e_1 = e_2 ,& \theta e_3 = e_4 ,& \theta e_5 = -e_5, \\
\theta e_2 = -e_1, & \theta e_4 = -e_3, & \theta e_6 = -e_6 .
\end {array}
\label{t41}
\end{eqnarray}
Then, as in subsection 2.1, $P$ invariance impose the following relations
\begin{eqnarray}
\begin{array}{ll}
|e_1|=|e_2|, & |e_3|=|e_4| ,\\
\alpha_{ij}=0\;\;\; i=1,2,3,4\;\;j=5,6 & \alpha_{14}=-\alpha_{23},\\
\alpha_{12}=\alpha_{34}=0, & \alpha_{13}=\alpha_{24}
\end{array}
\label{rpd41}
\end{eqnarray}
where $\alpha_{ij}=\cos \theta_{ij}$ and $e_i e_j= |e_i| |e_j| \cos
 \theta_{ij}$. Therefore we can
take the seven deformation degrees of freedom as
\begin{eqnarray}
\begin {array}{c}
R_i=|e_i|\;\;\;\; i=1,3,5,6, \\
\alpha_{13},\;\alpha_{14},\;\alpha_{56} .
\end {array}
\label{pd41}
\end{eqnarray}

Now it is easy to see that the two deformation parameters coming from $N_{33}$
correspond to a variation of the relative size of $|e_5|$ and $|e_6|$  and to
 the
$\theta_{56}$ angle; thus allowing for a rhomboid-like lattice from  the
 original third
$SO(4)$ sublattice. It is remarkable however that, as will be seen shortly,
 the
deformation parameters coming from $N_{33}$ are not involved in the Yukawa
 couplings.

Let us briefly summarize the main results of the $[SO(4)]^3$ $Z_4$
orbifold. They have been obtained performing an analysis similar to that
 followed in the
previous section for the $Z_6$--I one.
There are 16 fixed points under $\theta$ in this orbifold, given by
\be
f_1^{(ijk)} = g_1^{(i)}\otimes g_1^{(j)}\otimes g_1^{(k)} \;;\;\;\;\; i,j=0,2
\;;\;\;k=0,1,2,3 \label{ft41}
\ee
where
\[
\begin{array}{llll}
g_1^{(0)}=(0,0)\;, & g_1^{(1)}= (\frac{1}{2},0)\;, &
g_1^{(2)}=(\frac{1}{2},\frac{1}{2}) \;, & g_1^{(3)}= (0, \frac{1}{2}) .
\end{array}
\]
Under $\theta$ each fixed point is associated with a conjugation class in a
one--to--one correspondence. Under $\theta^2$
there are 16 fixed tori that are the product of 16 fixed points in the
sublattice
$(e_1,e_2,e_3,e_4)$ by the 2--torus defined by the sublattice $(e_5,e_6)$ (with
 or without
deformations). Six of these 16 fixed points are connected to the others through
 $\theta$
rotations. The fixed points are (in the first two $SO(4)$'s)
\be
f_2^{(ij)} = g_2^{(i)} \otimes g_2^{(j)} \;;\;\;\; i,j=0,1,2,3
\label{ftt41}
\ee
with $g_2^{(i)}=g_1^{(i)}$. And we can see that
\[
\begin{array}{ccccccc}
g_2^{(0)}\otimes g_2^{(1)} & \sim & g_2^{(0)}\otimes g_2^{(3)} ,& &
 g_2^{(1)}\otimes g_2^{(0)}&\sim
& g_2^{(3)}\otimes g_2^{(0)},\\
g_2^{(2)}\otimes g_2^{(1)} & \sim & g_2^{(2)}\otimes g_2^{(3)} ,& &
 g_2^{(1)}\otimes g_2^{(2)}&\sim
& g_2^{(3)}\otimes g_2^{(2)},\\
g_2^{(1)}\otimes g_2^{(1)} & \sim & g_2^{(3)}\otimes g_2^{(3)} ,& &
 g_2^{(1)}\otimes g_2^{(3)}&\sim
& g_2^{(3)}\otimes g_2^{(1)} .
\end{array}
\]
Note that in the two first $SO(4)$'s $g_2^{(0)}$ and $g_2^{(2)}$ are fixed
 points under
$\theta$ while $\theta g_2^{(1)} \rightarrow g_2^{(3)}$. Consequently there
 are 10 $\theta^2$
conjugation classes and, as was explained in subsection 2.3, only symmetric
combinations of fixed points (for the conjugation classes with more than one
 fixed point)
take part in the Yukawa couplings.
For this orbifold all the twisted couplings are of the $\theta \theta \theta^2$
 type
and the selection rule reads
\be
f_1 + f_2 - (I + \theta) f_3 \in \Lambda ,
\label{rs41}
\ee
where $f_3$ is the $\theta^2$ fixed point. Denoting the fixed points by
\begin{eqnarray}
\left.
\begin {array}{rcl}
f_1 &=& g_1^{(i_1)}\otimes g_1^{(j_1)}\otimes g_1^{(k_1)}\\
f_2 &=& g_1^{(i_2)}\otimes g_1^{(j_2)}\otimes g_1^{(k_2)}\\
f_3 &=& g_2^{(i_3)}\otimes g_2^{(j_3)}\otimes [\alpha(e_5)+\beta(e_6)]
\end {array}
\right\}
\;\;
\begin {array} {l}
i_1,i_2,j_1,j_2=0,2,\\
k_1,k_2,i_3,j_3=0,1,2,3\\
\alpha,\beta \in R ,
\end {array}
\label{fp12341}
\end {eqnarray}
see eqs. (\ref{ft41}--\ref{ftt41}), the selection rule is simply
\begin{eqnarray}
\left.
\begin {array}{c}
i_1+i_2+2i_3=0 \\
j_1+j_2+2j_3=0 \\
k_1=k_2
\end {array}
\right\}\;\; mod.\;4 .
\label{rsm41}
\end{eqnarray}
The number of allowed couplings is 160.
It is clear now that the third $SO(4)$ lattice always enters in the couplings
as
 the fixed
torus associated with the $\theta^2$ field. Then the coupling in
this invariant plane is of the untwisted type  and, consequently, the
 deformation parameters
for the third $SO(4)$ sublattice ($i.e.\;\; R_5$, $R_6$, $\alpha_{56}$;  see
 eq. (\ref{pd41}))
do not affect the value of the coupling. Two of
these parameters are precisely those coming from $N_{33}$. Remarkably enough we
 have checked that
this is a general property for all the orbifolds: (1,2) moduli are not involved
 in the expressions
of the Yukawa couplings. It looks as though there is a selection rule
 (unknown to us)
forbidding this kind of dependences.

For the case where $f_3$ is also a $\theta$ fixed
point, the value of the coupling in the 2--3 picture is

\begin {eqnarray}
C_{\theta\theta\theta^2} &=& N \sum_{v \in (f_2-f_3+\Lambda)_\bot} \exp [-\frac
 {1}{4\pi} (
|v_1|^2 + |v_2|^2 )] \nonumber\\
& = & N \sum_{v \in (f_2-f_3+\Lambda)_\bot} \exp [-\frac {1}{4\pi}
 \vec{v}^{\top} M \vec{v}]
 \\
& = & N\;\;  \vartheta
\left[
\begin {array}{c}
\vec{f_{23}} \nonumber\\
0
\end{array}
\right]
[ 0 , \Omega ] ,
\label{ac41}
\end {eqnarray}
where $(f_2-f_3  + \Lambda)_{\bot}$ selects only $(f_2-f_3+\Lambda)$ shifts
that
 are
orthogonal to the invariant sublattice, $i.e.$ the third $SO(4)$ lattice.
Thus, $( f_2 - f_3  + \Lambda)_{\bot}$ has non--zero components in the first
two $SO(4)$'s only. Similarly $\vec{f_{23}}$ represents the four components  of
$(f_2-f_3)$ in the basis $(e_1,...,e_4)$ of the first $SO(4)$ lattices. Finally
\begin{eqnarray}
\begin{array}{c}
N = \sqrt{V_{\perp}}\; \frac{1} {2\pi}\; \frac {\Gamma ^2
 (\frac{3}{4})}{\Gamma^2 (\frac
{1}{4})}
\\ \\
M= (-4\pi^2i) \Omega =
\left(
\begin{array}{cccc}
R_1^2 & 0 & R_1R_3\alpha_{13} & R_1R_3\alpha_{14} \\
0     & R_1^2 & -R_1R_3\alpha_{14} & R_1R_3\alpha_{13} \\
R_1R_3\alpha_{13} & -R_1R_3\alpha_{14} & R_3^2 & 0 \\
R_1R_3\alpha_{14} & R_1R_3\alpha_{13} & 0 & R_3^2
\end{array}
\right)
\end{array}
\label{mac41}
\end{eqnarray}
where $V_{\perp}$ is the volume of the unit cell of the first two
$SO(4) \times SO(4)$ sublattice orthogonal to the invariant plane.

If $f_3$ is not fixed by $\theta$, see eq. (\ref{mac41}), the result is exactly
 the same but
multiplying $C_{\theta\theta\theta^2}$ by $\sqrt{2}$. Clearly the number of
 effective
deformation parameters is 4. The number of {\em different} Yukawa couplings,
 from the
160 allowed ones, is 6 (without deformations), corresponding to

\be
\vec{f_{23}}=
\left[
\begin{array}{l}
l_3=1\;:\;
\begin{array}{lll}
g_1^{(0)} \otimes g_1^{(0)}, & g_1^{(2)} \otimes g_1^{(0)}, & g_1^{(2)} \otimes
 g_1^{(2)},
\end{array}\\
l_3=2\;:\;
\begin{array}{lll}
g_2^{(0)} \otimes g_2^{(1)}, & g_2^{(2)} \otimes g_2^{(1)}, & g_2^{(1)} \otimes
 g_2^{(1)}
\end{array}
\end{array}
\right.
\ee
and 10 (when deformations are considered), namely

\be
\vec{f_{23}}=
\left[
\begin{array}{l}
l_3=1\;:\;
\begin{array}{llll}
g_1^{(0)} \otimes g_1^{(0)}, & g_1^{(2)} \otimes g_1^{(0)}, & g_1^{(0)} \otimes
 g_1^{(2)},
& g_1^{(2)} \otimes g_1^{(2)},
\end{array}\\
l_3=2\; \left\{
\begin{array}{lll}
g_2^{(0)} \otimes g_2^{(1)}, & g_2^{(2)} \otimes g_2^{(1)}, & g_2^{(1)} \otimes
 g_2^{(1)}, \\
g_2^{(1)} \otimes g_2^{(0)}, & g_2^{(1)} \otimes g_2^{(2)}, & g_2^{(1)} \otimes
 g_2^{(3)} .
\end{array}
\right.
\end{array}
\right.
\ee

We would like to compare all the previous results with those of the $Z_4$
 orbifold based on
a Coxeter twist acting on an $[SU(4)]^2$ root lattice. This will illustrate
what
 aspects of the
orbifold dynamics are independent of the chosen lattice and what aspects do
not.
Furthermore, for the $[SU(4)]^2$ $Z_4$ orbifold, the lattice cannot be
 decomposed as the
direct product of an invariant sublattice under $\theta^2$ times an orthogonal
 sublattice,
as happened in the $[SO(4)]^3$ case. This peculiarity, which is shared by
 other
orbifolds, introduces some additional complications which we would like to
show.
 The Coxeter
element in the $[SU(4)]^2$ root lattice is of the form
\begin{eqnarray}
\begin{array} {lll}
\theta e_1= e_2, & \theta e_2 = e_3, & \theta e_3 = -e_1-e_2-e_3, \\
\theta e_4= e_5, & \theta e_5 = e_6, & \theta e_6 = -e_4-e_5-e_6 .
\end{array}
\label{t42}
\end{eqnarray}
The 7 deformation parameters coming from
$(N_{1\bar1},\;\;N_{2\bar2},\;\;N_{3\bar3},\;\;N_{1\bar2},\;\;N_{2\bar1},\;\;N_{
 33})$ are
\begin{eqnarray}
\begin {array}{c}
R_i = |e_i|,\;\;i=1,4 \\
\alpha_{12},\;\; \alpha_{14},\;\;\alpha_{15},\;\;
\alpha_{16},\;\; \alpha_{45},
\end {array}
\label{pd42}
\end{eqnarray}
where $( e_1,e_2,e_3)$ is the basis of the first $SU(4)$, and $(e_4,e_5,e_6)$
the basis of
the second one. Equation (\ref{pd42}) should be compared with
eq. (\ref{pd41}), $i.e.$
 its analogue
in the $[SO(4)]^3$ case. Clearly the geometrical interpretation of the
 deformation parameters
is different for each one. Other parameters of the $SU(4)^2$ lattice are
related
to the previous ones by
\begin{eqnarray}
\begin{array}{ll}
|e_1| = |e_2| = |e_3|, & |e_4| = |e_5| = |e_6|, \\
\alpha_{23}=\alpha_{12}, & \alpha_{34} = \alpha_{16}, \\
\alpha_{13}=-1-2\alpha_{12}, & \alpha_{24} = \alpha_{35} =
-\alpha_{14}-\alpha_{15}-\alpha_{16,}\\
\alpha_{25}=\alpha_{36}=\alpha_{14},\;\;\; & \alpha_{56}=\alpha_{45}, \\
\alpha_{26}=\alpha_{15}, & \alpha_{46}=-1-2\alpha_{45} .
\end{array}
\label{opd42}
\end{eqnarray}
It is important to point out that the $[SU(4)]^2$ lattice cannot be
consistently
 deformed
into an $[SO(4)]^3$ one. To see this, note that the invariant sublattice under
 the action of
the Coxeter element (\ref{t42}) is generated by $(e_1+e_3)$ and $(e_4+e_6)$. If
 such a
deformation existed, these vectors  could be identified with the basis of the
 invariant
$SO(4)$ sublattice in the $[SO(4)]^3$ case. Now, it can be shown that we cannot
 construct a
basis of $[SU(4)]^2$ with  $(e_1+e_3)$ , $(e_4+e_6)$ and four additional
lattice
 vectors
orthogonal to these (with or without deformations). In fact, it is easy to
 check that the
$[SU(4)]^2$ Coxeter element (\ref{t42}) has the same form as the twist $\theta$
 of the
$[SO(4)]^3$ case, $i.e.$ eq. (\ref{t41}), when acting in the following set of
 lattice vectors

\begin{eqnarray}
\begin{array}{lll}
\tilde e_1 = e_1+e_2, & \tilde e_3 = e_1+e_3, & \tilde e_5 = e_5 + e_6 ,\\
\tilde e_2 = e_2+e_3, & \tilde e_4 = e_4+e_5, & \tilde e_6 = e_4 + e_6 .
\end{array}
\label{bo42}
\end{eqnarray}
Notice that when deformations are included,  $(\tilde e_1,\tilde e_2,\tilde
 e_4,\tilde e_5)$
remain orthogonal to $(\tilde e_3,\tilde e_6)$. Actually,
$(\tilde e_1,\tilde e_2,\tilde e_3,\tilde e_4,\tilde e_5,\tilde e_6)$ generate
an $[SO(4)]^3$ sublattice of the $[SU(4)]^2$ lattice but they are not a basis
of
the whole lattice. Anyway the $\tilde e_i$ will be of help below.
The number of fixed points is the same in both cases. For the case at hand,
 $[SU(4)]^2$,
there are 16 fixed points under $\theta$ which can be expressed as
\be
f_1^{(ij)}= g_1^{(i)} \otimes g_1^{(j)} \;\;;\;\;\;\;  i,j=0,1,2,3
\label{pft42}
\ee
with
\[
\begin{array}{llll}
g_1^{(0)} = (0,0,0) , & g_1^{(1)} = (\frac{1}{4}, \frac{1}{2}, \frac{3}{4}), &
g_1^{(2)} = (\frac{1}{2}, 0 ,\frac{1}{2}), &
g_1^{(3)} = (\frac{3}{4}, \frac{1}{2}, \frac{1}{4}) .
\end{array}
\]
Under $\theta^2$ there is a fixed torus generated by  $(e_1+e_3)$ and
 $(e_4+e_6)$. Then we
can form 16 fixed tori as products of this fixed torus by the following 16
 $\theta^2$ fixed
points (six of them connected to the others by $\theta$ rotations)
\begin{eqnarray}
\begin{array}{ccccccc}
g_2^{(0)} \otimes g_2^{(0)} ,& & g_2^{(2)} \otimes g_2^{(2)}, & & g_2^{(0)}
\otimes g_2^{(2)} ,& & g_2^{(2)} \otimes g_2^{(0)} ,\\
g_2^{(0)} \otimes g_2^{(1)} &\sim & g_2^{(0)} \otimes
g_2^{(3)} ,& & g_2^{(2)} \otimes g_2^{(1)} &\sim & g_2^{(2)} \otimes g_2^{(3)},
 \\
g_2^{(1)} \otimes g_2^{(0)} &\sim & g_2^{(3)} \otimes
g_2^{(0)} ,& & g_2^{(1)} \otimes g_2^{(2)} &\sim & g_2^{(3)} \otimes g_2^{(2)},
 \\
g_2^{(1)} \otimes g_2^{(1)} &\sim & g_2^{(3)} \otimes
g_2^{(3)} ,& & g_2^{(1)} \otimes g_2^{(3)} &\sim & g_2^{(3)} \otimes g_2^{(1)}
\end{array}
\label{pftt42}
\end{eqnarray}
where
\[
g_2^{(0)} = (0,0,0), \;\; g_2^{(1)} = \frac {1}{2}(1,1,0), \;\; g_2^{(2)} =
\frac{1}{2}(1,0,1), \;\; g_2^{(3)} = \frac{1}{2} (0,1,1) .
\]
Note that $g_2^{(0)}$ and $g_2^{(2)}$ are fixed under $\theta$ but $g_2^{(1)}$
 and $g_2^{(3)}$ are
connected by a $\theta$ rotation, $\theta g_2^{(1)} \rightarrow g_2^{(3)}$.
As in the $[SO(4)]^3$ case there are 10 conjugation classes. The space group
selection rule also has the same form %
\be
f_1 + f_2 - (I + \theta) f_3 \in \Lambda .
\label{rs42}
\ee
Denoting the fixed points by
\begin{eqnarray}
\begin {array}{c}
f_1 = g_1^{(i_1)}\otimes g_1^{(j_1)},\;\;
f_2 = g_1^{(i_2)}\otimes g_1^{(j_2)},\;\;
f_3 = [g_2^{(i_3)}\otimes g_2^{(j_3)}]\otimes[\alpha(e_1+e_3)+\beta(e_4+e_6)],
 \\
i_1,i_2,i_3,j_1,j_2,j_3=0,1,2,3,\;\;\alpha,\beta \in R
\end {array}
\end{eqnarray}
eq. (\ref{rs42}) can be expressed as
\begin{eqnarray}
\left.
\begin {array}{lcr}
i_1 + i_2 + 2i_3 &=& 0\\
j_1 + j_2 + 2j_3 &=& 0
\end {array}
\right\}
\;\; mod.\; 4 .
\label{rsm42}
\end{eqnarray}
(Note that in spite of eqs. (\ref{rs42}--\ref{rsm42}) being formally identical
 with
(\ref{rs41}--\ref{rsm41}) the meaning of the vectors implicitely involved is
 quite
different.) The number of allowed couplings is again the same, 160. The Yukawa
 coupling, if
$f_3$ is fixed by $\theta$, is given by

\begin {eqnarray}
C_{\theta\theta\theta^2} &=&
\bar N \sum_{v \in (f_2-f_3+\Lambda)_\bot} \exp [-\frac {1}{4\pi}
\vec{v}^{\top}
 \bar
 M \vec{v}] .
\label{ac42}
\end {eqnarray}
As usual, the arrows denote components in the lattice basis $(e_1,...e_6)$.
The subscript $\bot$ means that only $v$ shifts orthogonal to the invariant
 plane (defined
by $(e_1+e_3)$ and $(e_4+e_6)$) have to be considered. If $f_3$ is not fixed by
 $\theta$ the
previous expression has to be multiplied by a $\sqrt {2}$ factor. $\bar N$ and
 $\bar M$ are
given by
\be
\begin{array}{c}
\bar N = \sqrt{V_{\perp}} \; \frac{1} {2\pi} \; \frac {\Gamma ^2
 (\frac{3}{4})}{\Gamma^2
(\frac {1}{4})}
\\ \\
\bar M =
\left(
\begin {array}{cccccc}
a & b & -a-2b & e & f & g \\
b & a & b & -e-f-g & e & f \\
-a-2b & b & a & g & -e-f-g & e \\
e & -e-f-g & g & c & d & -c-2d \\
f & e & -e-f-g & d & c & d \\
g & f & e & -c-2d & d & c
\end {array}
\right) \\
\begin {array}{lllllll}
a = R_1^2 &
b = R_1^2 \alpha_{12} &
c = R_4^2 &
d = R_4^2 \alpha_{45} &
e = R_1 R_4 \alpha_{14} &
f = R_1 R_4 \alpha_{15} &
g = R_1 R_4 \alpha_{16} .
\end{array}
\end{array}
\label{mac42}
\ee
where $V_{\perp}$ is the volume of the sublattice orthogonal to the invariant
 plane
(see below).
By addition of lattice vectors we can always choose $f_2$ and $f_3$ in
 (\ref{ac42}) such that
$f_2-f_3$ is orthogonal to the invariant plane. Then $f_2-f_3$ can be expressed
 in the
"basis" (\ref{bo42}) as
\be
f_2-f_3=x_1 \tilde e_1 +x_2 \tilde e_2 + x_4 \tilde e_4 + x_5 \tilde e_5\;.
\label{23bo}
\ee
We can check that $x_i=0, \frac{1}{2}$ (up to lattice vectors) for all the
 choices of
$f_2$, $f_3$. However it is amusing to see that many of the possibilities are
in
 fact
equivalent. Consider, for definiteness, the case $f_2-f_3 = 0$, $i.e.\; x_i=0$
 in
(\ref{23bo}). Now we can add to $f_2-f_3$ any shift contained in the invariant
plane,
\be
f_2-f_3 = \alpha (e_1+e_3) + \beta (e_4 + e_6)\;,\;\;\; \alpha, \beta \in R .
\label{23planoinv}
\ee
Demanding $v=f_2-f_3+\Lambda$ to be orthogonal to the invariant plane we find
 constraints
for $\alpha$ and $\beta$. A shift \( \sum_{i} a_i e_i \) is orthogonal to the
 invariant
plane if it satisfies the condition \( a_1-a_2+a_3=0\) and \( a_4 -a_5+a_6=0 \)
 (with or
without deformations). Then
\be
v=\alpha(e_1+e_3)+\beta(e_4+e_6)+\sum_{i=1}^{6} n_i e_i \;,\;\;\;\; n_i\in Z
\label{exv42}
\ee
is orthogonal if
\be
(\alpha,\beta) =
 (0,0),\;\;(0,\frac{1}{2}),\;\;(\frac{1}{2},0),\;\;(\frac{1}{2},\frac{1}{2}),
\label{vv42}
\ee
up to lattice vectors. Then $v$ can be expressed as
\begin {eqnarray}
v &=& (n_1+\alpha) \tilde e_1 + (n_3+\alpha) \tilde e_2 +
 (n_4+\beta) \tilde e_4 + (n_6+ \beta) \tilde e_5 .
\label{difv42}
\end{eqnarray}
Therefore we have to sum up four possibilities for $(x_1,x_2,x_4,x_5)$, namely
\be
(0,0,0,0),\;\;(\frac{1}{2},\frac{1}{2},0,0),\;\;(0,0,\frac{1}{2},\frac{1}{2}),\;
 \;
(\frac{1}{2},\frac{1}{2},\frac{1}{2},\frac{1}{2}) .
\label{sv42}
\ee
This is characteristic of the lattices that cannot be decomposed as the direct
 product of an
invariant sublattice times an orthogonal sublattice. In particular it did not
 happen in the
$[SO(4)]^3$ lattice.
In order to write the coupling we have to add to each case in (\ref{sv42})
lattice vectors orthogonal to the invariant plane, $i.e.$ of the form
\be
u_{\bot}= n_1\tilde e_1+n_2 \tilde e_2 + n_4 \tilde e_4 +n_5 \tilde e_5,
\label{vo42}
\ee
as is reflected in (\ref{difv42}). Now we can express the coupling
 (\ref{ac42}), which
contained a $6 \times 6\;\bar M$ matrix, as a sum of four $\vartheta$ functions
 defined in the
four-dimensional lattice $(\tilde e_1, \tilde e_2, \tilde e_4, \tilde e_5)$
\begin {eqnarray}
C_{\theta\theta\theta^2} &=&
\bar N \sum_{\tilde f_{23}}\;\; \sum_{\tilde v \in (\tilde
f_{23}+\Lambda_\bot)}
 \exp
 [-\frac {1}{4\pi}
\vec{\tilde{v}}^{\top}  \bar M' \vec{\tilde{v}}] \nonumber\\
& = &\bar N \sum_{\tilde f_{23}} \vartheta
\left[
\begin {array}{c}
\vec{\tilde f_{23}} \\
0
\end{array}
\right]
[ 0 , \Omega' ]
\label{acjab42}
\end {eqnarray}
where $\vec{\tilde{v}}$ and $\vec{\tilde f_{23}}$ are the components in
$(\tilde e_1, \tilde e_2, \tilde e_4, \tilde e_5)$ of $v$ and $(f_2-f_3)$
respectively. $\vec{\tilde f_{23}}$ runs over the possibilities displayed in
(\ref{sv42}), and
$\Omega'$ is given by
\be
\Omega' = i \frac {1}{4 \pi^2}\bar M' = i \frac {1}{4 \pi^2}
\left(
\begin {array}{cccc}
\bar a & 0 &\bar b & \bar c \\
0 &\bar a & -\bar c & \bar b \\
\bar b & -\bar c & \bar d & 0 \\
\bar c & \bar b & 0 & \bar d
\end {array}
\right) \;\;\;
\begin{array}{rcl}
\bar a&=& 2 R_1^2 (1+\alpha_{12})\\
\bar b&=& R_1 R_4 (\alpha_{14} -\alpha_{16})\\
\bar c&=& R_1 R_4 (\alpha_{14}+2\alpha_{15}+\alpha_{16})\\
\bar d&=& 2 R_4^2 (1+\alpha_{45}) .
\end{array}
\label{omega42}
\ee
Note that there are 4 effective deformation parameters, as in the $[SO(4)]^3$
 case.
Besides (\ref{sv42}), there are three other inequivalent possibilities for
$f_3-f_2$, namely

\be
\begin{array} {llll}
\{\;(0,0,0,\frac {1}{2}), & (0,0,\frac{1}{2},0), &
 (\frac{1}{2},\frac{1}{2},0,\frac{1}{2}),&
(\frac{1}{2},\frac{1}{2},\frac{1}{2},0)\;\},\\
\{\;(0,\frac {1}{2},0,0), & (0,\frac {1}{2},\frac{1}{2},\frac {1}{2}), &
(\frac{1}{2},0,0,0),& (\frac{1}{2},0,\frac{1}{2},\frac {1}{2})\;\},\\
\{\;(0,\frac {1}{2},0,\frac {1}{2}), & (0,\frac {1}{2},\frac{1}{2},0), &
(\frac{1}{2},0,0,\frac{1}{2}),& (\frac{1}{2},0,\frac{1}{2},0)\;\} .
\end {array}
\label{osv42}
\ee
Taking into account that the coupling gets a factor $\sqrt{2}$ if $f_3$ is not
 fixed by
$\theta$, this gives 8 different Yukawa couplings when deformations are
 considered, and 6
without deformations (the two first possibilities in (\ref{osv42}) are equal).
 This differs from
the $[SO(4)]^3$ case, where there were 10 and 6 respectively. Note that the
 matrix $\Omega'$,
eq. (\ref{omega42}), appearing in the coupling is formally identical to that
 of $[SO(4)]^3$,
eq. (\ref{mac41}). However, as we have seen, the structure of possible shifts
is
 very different. In
any case the number of effective deformation parameters is the same for both
cases.

\section{Conclusions}

We have calculated the complete twisted Yukawa couplings for all
the $Z_n$ orbifold constructions in the most general case, i.e.
when deformations of the compactified space are considered. This
includes a certain number of tasks. Namely, determination of the allowed
couplings, calculation of the explicit dependence of the Yukawa
couplings values on the moduli expectation values (i.e. the
parameters determining the size and shape of the compactified
space), etc. Some progress in this direction has recently been
made but without arriving at such explicit expressions as those given
in this paper. This is an essential ingredient in order to
relate theory and observation. In particular it
allows a counting of the {\em different} Yukawa couplings
for each orbifold (with and without deformations), which
is crucial to determine the phenomenological
viability of the different schemes, since it is directly related
to the fermion mass hierarchy. In this sense some orbifolds
(e.g. $Z_3$, $Z_4$, $Z_6$--I, $Z_8$--I, $Z_{12}$--I)
have much better phenomenological prospects than
others (e.g. $Z_7$, $Z_6$--II, $Z_8$-II, $Z_{12}$--II).
The results for the whole set of Coxeter
orbifolds are summarized in Table 1. Other facts concerning
the phenomenological profile of $Z_n$ orbifolds are also
discussed, e.g. the existence of non--diagonal entries in the
fermion mass matrices, which is related to a non--trivial
structure of the Kobayashi--Maskawa matrix. In this sense
non--prime orbifolds are favoured over prime
ones which do not have off--diagonal entries in
the mass matrices at this fundamental level.

The results of this paper give the precise form in which moduli
fields are coupled to twisted matter. This is essential in order
to study in detail other important issues. Namely, the
supersymmetry breaking mechanism by gaugino condensation (in
which the moduli develop an additional non--perturbative
superpotential), and cosmological implications (note that the
moduli are also coupled to gravity in a Jordan-Brans-Dicke--like
way). The level of explicitness given in the paper is also necessary
for more theoretical matters
(e.g. the study of the transformation properties of the Yukawa
couplings under target--space modular transformations like
$R\rightarrow 1/R$ ). Concerning the last aspect we have found some
appealing results, such as the fact that (1,2) moduli never appear
in the expressions of the Yukawa couplings. Likewise, (1,1) moduli associated
with fixed tori which are involved in the Yukawa coupling, do not affect
the value of the coupling. It is worth noticing that the above mentioned
moduli are precisely the only ones which contribute to the string loop
corrections to gauge coupling constants \cite{pipa}.

\vspace{.2 cm}

\noindent{\bf ACKNOWLEDGEMENTS}

\noindent The work of J.A.C. was supported in part by the C.I.C.Y.T., Spain.
The work of F.G. was supported by an F.P.I. grant, Spain. C.M. is grateful
to the members of the Departamento de F\'{\i}sica de Part\'{\i}culas,
 Universidad
de Santiago de Compostela, Spain, for their kind hospitality. F.G. thanks
J. Mas for very useful discussions.
\newpage

\noindent\underline{\bf{APPENDIX 1}}

\begin{quotation}\noindent{
We follow a notation as compact as possible. The precise meaning of all the
 concepts
appearing here is explained in detail in the text for the $Z_6$--I and $Z_4$
 examples.
}\end{quotation}

\vspace{0.8 cm}
\noindent{\underline{\bf{ORBIFOLD $Z_3$}}}
\vspace{.5 cm}

\noindent{\underline {Twist}
$\;\; \theta={\rm
diag}(e^{i\alpha},e^{i\alpha},e^{-2i\alpha}) ,\;\;\;\;
\alpha=\frac{2\pi}{3} $}

\noindent{\underline {Lattice}
$\;\; [SU(3)]^3  $}

\noindent{\underline {Coxeter element}}
\[
\begin{array}{lll}
\theta e_i=e_{i+1}, & \theta e_{i+1}= -e_i-e_{i+1}, & i=1,3,5
\end {array}
\]

\noindent{\underline {Deformation parameters}}

\indent{Relations}
\[
\begin{array}{c}
|e_i|^2=|e_{i+1}|^2, \;\;\; \alpha_{i,i+1}=-\frac{1}{2}, \;\;\;
\alpha_{i,j}=\alpha_{i+1,j+1},\\
\alpha_{i,j}+\alpha_{i,j+1}+\alpha_{i+1,j}=0,\;\;\;\;
i,j=1,3,5\;\;\;i<j
\end{array}
\]
\[
\alpha_{ij}\equiv\cos(\theta_{ij})
\]

\indent{Degrees of freedom (9)}
\[
R_i = |e_i|, \;\; \alpha_{i,j},\;\; \alpha_{i,j+1}, \;\; i,j=1,3,5 \;\; i<j
\]
\noindent{\underline{Lattice basis ($e_i$) in terms of  orthogonal basis
 ($\tilde e_i$)}}

\indent{Not necessary in this case.}

\vspace{.5 cm}

\noindent{\underline {Fixed points of $\theta$} (27)}
\[
\begin {array}{c}
  f_1^{(ijk)}= g_1^{(i)} \otimes g_1^{(j)} \otimes g_1^{(k)}\;,\;\; i,j,k=
0,1,2
 ,\\
  g_1^{(0)} = (0,0)\;,\;\; g_1^{(1)} =(\frac{1}{3},\frac{2}{3})\;,\;\;
  g_1^{(2)} = (\frac{2}{3},\frac{1}{3})
\end {array}
\]

\vspace{.5 cm}
\noindent {\underline {Coupling $\theta\theta\theta$}}

\indent {Selection rule}
\[
f_1+f_2+f_3 \in \Lambda
\]

\indent {Denoting}
\[
\begin {array}{l}
f_1 = g_1^{(i_1)}\otimes g_1^{(j_1)}\otimes g_1^{(k_1)}\\
f_2 = g_1^{(i_2)}\otimes g_1^{(j_2)}\otimes g_1^{(k_2)}\\
f_3 = g_1^{(i_3)}\otimes g_1^{(j_3)}\otimes g_1^{(k_3)}
\end {array}
\;\;\;\;
\begin {array}{c}
i_1,i_2,i_3 = 0,1,2\\
j_1,j_2,j_3 = 0,1,2\\
k_1,k_2,k_3 = 0,1,2
\end {array}
\]

\indent{the selection rule reads}
\[
\left.
\begin {array}{lcr}
i_1 + i_2 + i_3 &=& 0\\
j_1 + j_2 + j_3 &=& 0\\
k_1 + k_2 + k_3 &=& 0
\end {array}
\right\}
\;\; mod.\; 3
\]

\indent{Number of allowed couplings: 729}

\indent{Expression of the coupling}
\begin {eqnarray*}
C_{\theta\theta\theta} &=&
 N \sum_{v \in (f_3-f_2+\Lambda)} \exp [-\frac {1}{4\pi} \; \sin (\frac {2
\pi} {3})\; |v|^2]  \\
 &=&
 N \sum_{\vec{u} \in Z^6} \exp [-\frac {\sqrt{3}}{8\pi} \;
 (\vec{f_{23}}+\vec{u})^{\top}  M
(\vec{f_{23}}+\vec{u})] \\
 &=&
 N \; \vartheta
\left[
\begin {array}{c}
\vec {f_{23}} \\
\ 0 \\
\end {array}
\right]
[0,\Omega]
\end {eqnarray*}

\indent {with}
\begin {eqnarray*}
\Omega &=& i \frac {\sqrt{3}}{8 \pi^2} M ,\;\;\;\;
N= \sqrt{V_{\Lambda}}\; \frac{3^{3/4}}{8 \pi^3} \;
\frac{\Gamma^6(\frac{2}{3})}{\Gamma^3 (\frac{1}{3})}\\
\Omega &=& i \frac {\sqrt{3}}{8 \pi^2}
\left(
\begin {array}{cccccc}
R_1^2 & -\frac {R_1^2} {2} & R_1 R_3 \alpha_{13} & R_1R_3 \alpha_{14} &
R_1 R_5 \alpha_{15} & R_1 R_5 \alpha_{16} \\
- \frac{R_1^2}{2} & R_1^2 & R_1R_3 \alpha_{23} &
R_1R_3 \alpha_{13}  & R_1R_5 \alpha_{25} & R_1R_5 \alpha_{15}\\
R_1R_3 \alpha_{13} & R_1R_3 \alpha_{23} & R_3^2 & -\frac
{R_3^2}{2} & R_3R_5 \alpha_{35} & R_3R_5 \alpha_{36}\\
R_1R_3 \alpha_{14} & R_1R_3 \alpha_{13} & -\frac{R_3^2}{2} & R_3^2 &
R_3R_5\alpha_{45} & R_3R_5 \alpha_{35}\\
R_1R_5 \alpha_{15} & R_1R_5 \alpha_{25} & R_3R_5 \alpha_{35} &
R_3R_5 \alpha_{45} & R_5^2 & -\frac{R_5^2}{2} \\
R_1 R_5 \alpha_{16} & R_1 R_5 \alpha_{15} & R_3R_5 \alpha_{36} &
R_3R_5 \alpha_{35} & -\frac{R_5^2}{2} & R_5^2
\end {array}
\right)
\end {eqnarray*}
\[
\alpha_{23}=-(\alpha_{13}+\alpha_{14})\;,\;\;
\alpha_{25}=-(\alpha_{15}+\alpha_{16})\;,\;\;
\alpha_{45}=-(\alpha_{35}+\alpha_{36})
\]

\indent{Number of effective parameters: 9}

\indent{Number of different couplings without deformations: 4}

\indent{corresponding to the following $\vec f_{23} $ shifts}
\[
\vec{f_{23}}=
g_1^{(0)}\otimes g_1^{(0)} \otimes g_1^{(0)} ,\;\;
g_1^{(1)}\otimes g_1^{(0)} \otimes g_1^{(0)} ,\;\;
g_1^{(1)}\otimes g_1^{(1)} \otimes g_1^{(0)} ,\;\;
g_1^{(1)}\otimes g_1^{(1)} \otimes g_1^{(1)}
\]

\indent{Number of different couplings with deformations: 14}

\indent{corresponding to the following $\vec f_{23} $ shifts}
\[
\vec{f_{23}}= \left\{
\begin{array}{l}
g_1^{(0)}\otimes g_1^{(0)} \otimes g_1^{(0)} ,\;\;
g_1^{(1)}\otimes g_1^{(0)} \otimes g_1^{(0)} ,\;\;
g_1^{(0)}\otimes g_1^{(1)} \otimes g_1^{(0)} ,\;\;
g_1^{(0)}\otimes g_1^{(0)} \otimes g_1^{(1)} ,\\
g_1^{(1)}\otimes g_1^{(1)} \otimes g_1^{(0)} ,\;\;
g_1^{(1)}\otimes g_1^{(0)} \otimes g_1^{(1)} ,\;\;
g_1^{(0)}\otimes g_1^{(1)} \otimes g_1^{(1)} ,\;\;
g_1^{(1)}\otimes g_1^{(2)} \otimes g_1^{(0)} ,\\
g_1^{(1)}\otimes g_1^{(0)} \otimes g_1^{(2)} ,\;\;
g_1^{(0)}\otimes g_1^{(1)} \otimes g_1^{(2)} ,\;\;
g_1^{(1)}\otimes g_1^{(1)} \otimes g_1^{(1)} ,\;\;
g_1^{(1)}\otimes g_1^{(1)} \otimes g_1^{(2)} ,\\
g_1^{(1)}\otimes g_1^{(2)} \otimes g_1^{(2)} ,\;\;
g_1^{(1)}\otimes g_1^{(2)} \otimes g_1^{(1)}
\end{array}
\right.
\]

\vspace {1.0 cm}
\noindent{\underline{\bf{ORBIFOLD $Z_4$}}}

\indent{See Section 3}

\vspace {1.0 cm}
\noindent{\underline{\bf{ORBIFOLD $Z_6$--I}}}

\indent{See Section 2}

\vspace {1.0 cm}
\noindent{\underline{\bf{ORBIFOLD $Z_6$--II}}}

\vspace{.5 cm}

\noindent{\underline {Twist}
$ \theta={\rm
diag}(e^{i\alpha},e^{2i\alpha},e^{-3i\alpha}) ,\;\;\;\;
\alpha=\frac{2\pi}{6} $}

\noindent{\underline {Lattice}
$ SU(6) \otimes SU(2) $}

\noindent{\underline {Coxeter element}}
\[
\begin{array}{llll}
\theta e_i=e_{i+1}, & i=1,...,4, &
\theta e_5=-e_1-e_2-e_3-e_4-e_5, & \theta e_6=-e_6
\end{array}
\]

\noindent{\underline {Deformation parameters}}

\indent{Relations}
\[
\begin{array}{ll}
|e_1|=|e_2|=|e_3|=|e_4|=|e_5|,  &
\alpha_{12}=\alpha_{23}=\alpha_{34}=\alpha_{45}=
 -\frac{1}{2}(1+\alpha_{14}+2\alpha_{15}),\\
\alpha_{15}=\alpha_{13}=\alpha_{24}=\alpha_{35},&
\alpha_{16}=-\alpha_{26}=\alpha_{36}=-\alpha_{46}=\alpha_{56},\\
\alpha_{14}=\alpha_{25} &
\end{array}
\]
\[
\alpha_{ij}\equiv\cos(\theta_{ij})
\]

\indent{Degrees of freedom (5)}
\[
\begin {array}{lllll}
R_1= |e_1|, & R_6= |e_6|,  &
\alpha_{14},& \alpha_{15}, &
\alpha_{16}
\end {array}
\]

\vspace{.5 cm}

\noindent{\underline{Lattice basis ($e_i$) in terms of  orthogonal basis
 ($\tilde e_i$)}}

\begin{eqnarray*}
e_i&=& \sum_{j=1,3,5} A_j[\cos(\varphi_j+(i-1) b_j \alpha) \tilde e_j
+ \sin(\varphi_j+(i-1) b_j \alpha) \tilde e_{j+1}] \;\;\;\; i=1,...,5 \\
e_6&=& R_6 [\cos(\varphi_3 + \Delta) \tilde e_5 +  \sin(\varphi_3 + \Delta)
 \tilde e_6 ]
\end{eqnarray*}

\indent{with $\alpha=\frac{\pi}{3}$  and $b_1=1$, $b_2=2$, $ b_3=3$}
\[
\begin{array}{llll}
\cos (\Delta) = \frac{ \sqrt{3} \alpha_{16}}{\sqrt{1+2\alpha_{15}}}, &
A_1= \frac {R_1}{\sqrt{6}} \sqrt{1-3\alpha_{14}-4\alpha_{15}},&
A_3= \frac {R_1}{\sqrt{2}} \sqrt{1+\alpha_{14}},&
A_5= \frac {R_1}{\sqrt{3}} \sqrt{1+2 \alpha_{15}}
\end{array}
\]

\indent{$\varphi_1,\;\varphi_2,\;\varphi_3$ are free parameters.}

\vspace{.5 cm}
\noindent{\underline {Fixed points of $\theta$} (12)}
\[
\begin {array}{c}
f_1^{(ij)}= g_1^{(i)}\otimes \hat g_1^{(j)}\;,\;\;\; i=0,1,...,5,\;\;\;j = 0,1
\end {array}
\]
\[
\begin{array}{lll}
g_1^{(0)} = (0,0,0,0,0)\;,& g_1^{(1)} =\frac{1}{6} (5,4,3,2,1)\;,&
g_1^{(2)} =\frac{1}{6} (4,8,6,4,2)\;,\\
g_1^{(3)} =\frac{1}{6} (3,6,9,6,3)\;,&
g_1^{(4)} =\frac{1}{6} (2,4,6,8,4)\;,&
g_1^{(5)} =\frac{1}{6} (1,2,3,4,5)\;,\\
\hat g_1^{(0)} =(0)\;,& \hat g_1^{(1)} =(\frac {1}{2}) &
\end {array}
\]

\vspace{.5 cm}
\noindent{\underline {Fixed points of $\theta^2$} (9)}

\indent{Fixed torus: $\alpha(e_1+e_3+e_5)+\beta(e_6)\;,\;\;\;\; \alpha,\;\beta
\in R$}

\[
\begin{array}{c}
f_2^{(i)}=  g_2^{(i)}\otimes[\alpha(e_1+e_3+e_5)+\beta(e_6)] \;,\;\;\;
 i=0,1,...,8,\;\;
\alpha,\beta \in R \\
\end {array}
\]
\[
\begin {array}{lll}
g_2^{(0)} =(0,0,0,0,0), & \tilde g_2^{(1)} =\frac{1}{3}(0,1,1,2,2), &
g_2^{(2)} =\frac{1}{3}(0,2,2,1,1), \\
g_2^{(3)}=\frac{1}{3}(1,0,-2,0,1) ,&
g_2^{(4)} =\frac{1}{3}(1,1,2,2,0) ,& g_2^{(5)}=\frac{1}{3}(2,2,1,1,0), \\
g_2^{(6)} =\frac{1}{3}(-1,0,2,0,-1), & g_2^{(7)}=\frac{1}{3}(2,-2,0,2,-2), &
g_2^{(8)} =\frac{1}{3}(1,-1,0,1,-1)
\end{array}
\]

\indent {Note that $\theta : g_2^{(1)}\rightarrow g_2^{(4)},
g_2^{(2)}\rightarrow g_2^{(5)}, \;\; \theta: g_2^{(3)}\rightarrow g_2^{(6)}$}

\indent {Number of conjugation classes: 6}

\vspace{.5 cm}
\noindent{\underline {Fixed points of $\theta^3$} (14)}

\indent{Fixed torus: $\alpha(e_1+e_4)+\beta(e_2+e_5)\;,\;\; \alpha,\;\beta
\in R$}

\[
\begin{array}{c}
f_3^{(ij)}= g_3^{(i)} \otimes \hat g_3^{(j)} \otimes
 [\alpha(e_1+e_4)+\beta(e_2+e_5)]
\;,\;\;\; i=0,1,...,7,\;\;\;j=0,1,\;\;\; \alpha,\beta \in R \\
\end {array}
\]
\[
\begin {array}{lll}
g_3^{(0)} =(0,0,0,0,0)\;, & g_3^{(1)}=\frac{1}{2} (1,1,1,0,0)\;,&
g_3^{(2)} =\frac{1}{2} (1,1,0,-1,-1)\;, \\
g_3^{(3)} =\frac{1}{2} (0,1,1,1,0)\;,&
g_3^{(4)} =\frac{1}{2} (1,0,0,1,0)\;, & g_3^{(5)} =\frac{1}{2} (0,0,1,1,1)\;,\\
g_3^{(6)} =\frac{1}{2} (0,1,0,0,1)\;, & g_3^{(7)} =\frac{1}{2} (1,0,1,0,1)\;,&
 \\
\hat g_3^{(0)} = (0)\;, &
\hat g_3^{(1)} = (\frac{1}{2}) &
\end{array}
\]

\indent {Note that in the $SU(6)$ lattice $\theta : g_3^{(1)}
\rightarrow g_3^{(3)} \rightarrow g_3^{(5)}$ and $\theta : g_3^{(2)}
\rightarrow g_3^{(4)} \rightarrow g_3^{(6)}$}

\indent {Number of conjugation classes: 8}

\vspace{.5 cm}
\noindent {\underline {Coupling $\theta\theta^2\theta^3$}}

\indent {Selection rule}
\[
f_1+(I+\theta)f_2-(I+\theta+\theta^2)f_3 \in \Lambda
\]

\indent {Denoting}
\[
\left.
\begin {array}{rcl}
f_1 &=& g_1^{(i_1)}\otimes \hat g_1^{(j_1)} \\
f_2 &=& g_2^{(i_2)}\otimes [\alpha (e_1+e_3+e_5)+\beta (e_6)]\\
f_3 &=& g_3^{(i_3)}\otimes \hat g_3^{(j_3)} \otimes [\gamma (e_1+e_4)+\delta
 (e_2+e_5)]
\end {array}
\right\}\;\;\;
\begin {array} {l}
i_1=0,1,...,5\;,\\
j_1,j_3=0,1\;,\\
i_2=0,1,...,8\;,\\
i_3=0,1,...,7\;,\\
\alpha,\beta,\gamma,\delta \in R
\end {array}
\]

\indent{the selection rule reads}
\[
\left.
\begin {array}{l}
i_1+2i_2+3i_3=0 \\
j_1=j_3
\end {array}
\right\}
\;\;
mod.\;6
\]

\indent{Number of allowed couplings: 48}

\indent{Expression of the coupling}

\begin {eqnarray*}
C_{\theta\theta^2\theta^3} &=& \sqrt{l_2 \; l_3}\;\;
 N \; \sum_{v \in (f_3-f_2+\Lambda)_{\perp}} \exp [-\frac
{\sqrt{3}}{4\pi}  \;|v_1|^2 ]
\end {eqnarray*}
\begin{quotation}
\noindent{where $l_i$ is the number of elements in the $f_i$ conjugation class
 and
$(f_3-f_2+\Lambda)_{\perp}$  denotes elements orthogonal to the two
invariant planes}
\end{quotation}

\[
(f_3-f_2+\Lambda)_{\perp} = \sum_{i=1}^{6} (h_1^i+n_1^i)(\frac{1}{2}
 e_1+e_2+e_3+ \frac {1}{2}
e_4)+ (h_2^i+n_2^i)(- e_1-e_2+e_4+ e_5)
\]

\indent{where denoting $\vec{\bar{f_{23}^i}}\equiv(h_1^i,h_2^i)$,
 $\vec{\bar{f_{23}^i}}$ is always}

\[
\begin {array}{lll}
\vec{\bar{f_{23}^1}}=(0,0) & \vec{\bar{f_{23}^2}}=(0,\frac{1}{2}) &
\vec{\bar{f_{23}^3}}=(\frac{1}{3},\frac{1}{3}) \\
\vec{\bar{f_{23}^4}}=(\frac{1}{3},\frac{5}{6}) &
 \vec{\bar{f_{23}^5}}=(\frac{2}{3},\frac{2}{3}) &
\vec{\bar{f_{23}^6}}=(\frac{2}{3},\frac{5}{6})
\end{array}
\]

\indent{with $n_1^i,n_2^i \in Z$. The coupling takes the final form}
\begin {eqnarray*}
C_{\theta\theta^2\theta^3} &=& \sqrt{l_2 \; l_3}\;\;
 N \;\sum_{i} \;\; \sum_{u \in Z^2} \exp [-\frac
{\sqrt{3}}{4\pi}  \;(\vec{\bar{f_{23}^i}}+ \vec{u})^{\top} M
 (\vec{\bar{f_{23}^i}}+ \vec{u})] \\
&=& \sqrt{l_2 \; l_3}\;\;
 N \;\sum_{i}  \vartheta
\left[
\begin{array}{c}
\vec{\bar{f_{23}^i}} \\
0
\end {array}
\right]
[0, \Omega]
\end {eqnarray*}

\indent{with}

\[
\Omega = \;i \frac{\sqrt{3}}{4\pi^2} M =\;i \frac{\sqrt{3}}{2\pi^2}
 R_1^2 (1-3\alpha_{14}-4\alpha_{15})
\left(
\begin {array}{rr}
\frac{1}{4} & -\frac{1}{4} \\
-\frac{1}{4} & 1
\end {array}
\right) ,\;\;\;\;\;
N= \sqrt{V_{\perp}}\; \frac{1}{2 \pi} \;
\sqrt{ \frac{\Gamma(\frac{5}{6}) \Gamma(\frac{2}{3})}
{\Gamma(\frac{1}{3}) \Gamma(\frac{1}{6})} }
\]

\indent{with $V_{\perp}$ the volume of the unit cell generated by
$\{\frac{1}{2} e_1+e_2+e_3+ \frac {1}{2}e_4, e_1+e_2-e_4- e_5\}$}

\indent{Number of effective parameters: 1}

\indent{Number of different couplings without deformations: 4}

\indent{Number of different couplings with deformations: 4}

\vspace{.5 cm}
\noindent {\underline {Coupling $\theta\theta\theta^4$}}

\indent {Selection rule}
\[
f_1+f_2-(I+\theta)f_3 \in \Lambda
\]

\indent {Denoting}
\[
\left.
\begin {array}{rcl}
f_1 &=& g_1^{(i_1)}\otimes \hat g_1^{(j_1)}\\
f_2 &=& g_1^{(i_2)}\otimes \hat g_1^{(j_2)}\\
f_3 &=& g_2^{(i_3)}\otimes [\alpha (e_1+e_3+e_5)+\beta (e_6)]
\end {array}
\right\}\;\;\;
\begin {array} {l}
i_1,i_2=0,1,...,5\;,\\
j_1,j_2=0,1\;,\\
i_3=0,1,...,8\;,\\
\alpha,\beta \in R
\end {array}
\]

\indent{the selection rule reads}
\[
\left.
\begin {array}{l}
i_1+i_2+4i_3=0 \\
j_1=j_2
\end {array}
\right\}
\;\;
mod.\;6
\]

\indent{Number of allowed couplings: 72}

\indent{Expression of the coupling}

\begin {eqnarray*}
C_{\theta\theta\theta^4} &=& \sqrt{l_3}\;\;
 N \; \sum_{v \in (f_3-f_2+\Lambda)_{\perp}} \exp [-\frac
{\sqrt{3}}{8\pi}  \; (|v_1|^2+|v_2|^2)]
\end {eqnarray*}
\begin{quotation}
\noindent{where $(f_3-f_2+\Lambda)_{\perp}$ denotes that the coset
elements must be orthogonal to the
($e_1+e_3+e_5,e_6$) plane}
\end{quotation}
\[
(f_3-f_2+\Lambda)_{\perp} = \sum_{i=1}^{3} [
 (h_1^i+n_1^i)(e_1-e_3)+(h_2^i+n_2^i)(e_2+e_3)+
(h_3^i+n_3^i)(e_3+e_4)+(h_4^i+n_4^i)(e_5-e_3) ]
\]
\begin{quotation}
\noindent{where denoting $\vec{\bar{f_{23}^i}}=(h_1^i,...,h_4^i)$
there are two possible tri--plets
of values for $\vec{\bar{f_{23}^i}}$ depending on the values of $f_2,f_3$}
\end{quotation}

\[
\begin{array}{l}
\begin{array}{lll}
\vec{\bar{f_{23}^1}}=(0,0,0,0)
 &\vec{\bar{f_{23}^2}}=(\frac{1}{3},0,0,\frac{1}{3}) &
\vec{\bar{f_{23}^3}}=(\frac{2}{3},0,0,\frac{2}{3})\\
\end{array}\\
{\rm and}\\
\begin{array}{lll}
\vec{\bar{f_{23}^1}}=(\frac{1}{3},\frac{2}{3},\frac{1}{3},\frac{2}{3}) &
\vec{\bar{f_{23}^2}}=(\frac{2}{3},\frac{2}{3},\frac{1}{3},0) &
\vec{\bar{f_{23}^3}}=(0,\frac{2}{3},\frac{1}{3},\frac{1}{3}) \\
\end{array}
\end{array}
\]

\indent{with $n_1^i,n_2^i,n_3^i,n_4^i\;\in\;Z$. Finally the coupling takes the
 form}
\begin{eqnarray*}
C_{\theta\theta\theta^4} &=& \sqrt{l_3}\;\;
 N \sum_i \;\sum_{\vec{u} \in Z^4} \exp [-\frac
{\sqrt{3}}{8\pi}  \;(\vec{\bar{f_{23}^i}}+\vec{u})^{\top} M
 (\vec{\bar{f_{23}^i}}+\vec{u})] \\
 &=& \sqrt{l_3}\;\;
 N \;\sum_{i}  \vartheta
\left[
\begin{array}{c}
\vec{\bar{f_{23}^i}} \\
0
\end {array}
\right]
[0, \Omega]
\end {eqnarray*}

\indent{with}
\[
N= \sqrt{V_{\perp}}\; \frac{1}{2 \pi} \;
\sqrt{ \frac{\Gamma(\frac{5}{6}) \Gamma(\frac{2}{3})}
{\Gamma(\frac{1}{3}) \Gamma(\frac{1}{6})} }\;,\;\;\;\;
\Omega = \; i \frac{\sqrt{3}}{8\pi^2} M
\]

\begin{quotation}
\noindent{$V_{\perp}$ is the unit cell volume
of the sublattice orthogonal to the invariant plane}
\end{quotation}

\[
\begin{array}{l}
\Omega = \; i \frac{\sqrt{3}}{8\pi^2}
\left(
\begin {array}{cccc}
2a & -a & \frac{a+c-2b}{2} & -a \\
-a & b & c & \frac{a+c-2b}{2}\\
\frac{a+c-2b}{2} & c & b & -a \\
-a & \frac{a+c-2b}{2} & -a & 2a
\end {array}
\right)
\begin{array}{l}
a=R_1^2(1-\alpha_{15})\\
b=R_1^2(1-\alpha_{14}-2\alpha_{15})\\
c=R_1^2(\alpha_{14}+\alpha_{15})
\end{array}
\end{array}
\]

\indent{Number of effective parameters: 3}

\indent{Number of different couplings without deformations: 4}

\indent{Number of different couplings with deformations: 4}

\vspace {1.0 cm}
\noindent{\underline{\bf{ORBIFOLD $Z_7$}}}
\vspace{.5 cm}

\noindent{\underline {Twist}
$ \theta={\rm
diag}(e^{i\alpha},e^{2i\alpha},e^{-3i\alpha}) ,\;\;\;\;
\alpha=\frac{2\pi}{7} $}

\noindent{\underline {Lattice}
$ SU(7) $}

\noindent{\underline {Coxeter element}}
\[
\begin{array}{lll}
\theta e_i = e_{i+1}, & i=1,...,5, &
\theta e_6 = -e_1-e_2-e_3-e_4-e_5-e_6
\end{array}
\]

\noindent{\underline {Deformation parameters}}

\indent{Relations}
\[
\begin{array}{ll}
|e_1|=|e_2|=|e_3|=|e_4|=|e_5|=|e_6|, &
\alpha_{12}=\alpha_{23}=\alpha_{34}=\alpha_{45}=\alpha_{56},\\
\alpha_{13}=\alpha_{24}=\alpha_{35}=\alpha_{46}=\alpha_{16},&
\alpha_{14}=\alpha_{25}=\alpha_{36}=\alpha_{15}=\alpha_{26}
= -\frac{1}{2}-\alpha_{12}-\alpha_{13}
\end{array}
\]
\[
\alpha_{ij}\equiv\cos(\theta_{ij})
\]

\indent{Degrees of freedom (3)}
\[
\begin {array}{lll}
R= |e_1|, & \alpha_{12}, & \alpha_{13}
\end {array}
\]

\vspace{.5 cm}
\noindent{\underline{Lattice basis ($e_i$) in terms of orthogonal basis
($\tilde
 e_i$)}}

\[
e_i = \sum_{j=1,3,5} R_j[\cos((i-1)b_j\alpha + \varphi_j) \tilde e_j +
\sin((i-1)b_j\alpha + \varphi_j) \tilde e_{j+1} ] \;\;\;\;\; i=1,...,6
\]

\indent{with $\alpha=\frac{2\pi}{7}$ and $b_1=1,\; b_3=2,\; b_5=4$}
\[
\begin{array}{c}
R_1^2= R^2 [\alpha_{12} (\alpha_5^2-\alpha_1^2) +
 \alpha_{13}(\alpha_5^2-\alpha_3^2)+
\frac{1}{2}\alpha_5^2] \\
R_3^2= R^2 [\alpha_{12} (\alpha_1^2-\alpha_3^2) +
 \alpha_{13}(\alpha_1^2-\alpha_5^2)+
\frac{1}{2}\alpha_1^2] \\
R_5^2= R^2 [\alpha_{12} (\alpha_3^2-\alpha_5^2) +
 \alpha_{13}(\alpha_3^2-\alpha_1^2)+
\frac{1}{2}\alpha_3^2] \\
\alpha_i^2= \frac{4}{7}[1-\cos(b_i \alpha)]\;,\;\;\;i=1,3,5
\end{array}
\]

\indent{$\varphi_1,\;\varphi_2,\;\varphi_3$ are free parameters.}

\vspace{.5 cm}
\noindent{\underline {Fixed points of $\theta$} (7)}
\[
\begin {array}{lll}
f_1^{(0)}= (0,0,0,0,0,0), &
f_1^{(1)}= \frac{1}{7}(6,5,4,3,2,1), &
f_1^{(2)}= \frac{1}{7}(5,3,1,6,4,2), \\
f_1^{(3)}= \frac{1}{7}(4,1,5,2,6,3), &
f_1^{(4)}= \frac{1}{7}(3,6,2,5,1,4), &
f_1^{(5)}= \frac{1}{7}(2,4,6,1,3,5), \\
f_1^{(6)}= \frac{1}{7}(1,2,3,4,5,6)& &
\end {array}
\]

\vspace{.5 cm}
\noindent {\underline {Coupling $\theta\theta^2\theta^4$}}

\indent {Selection rule}
\[
f_1+2f_2-3f_3 \in \Lambda
\]

\indent {Denoting}
\[
\left.
\begin {array}{l}
f_1 = f_1^{(i_1)}\\
f_2 = f_1^{(i_2)}\\
f_3 = f_1^{(i_3)}\\
\end {array}
\right\}\;\;\;
\begin {array} {l}
i_1,i_2,i_3 =0,1,...,6\;,\\
\end {array}
\]

\indent{the selection rule reads}
\[
\begin {array}{l}
i_1+2i_2-3i_3=0
\end {array}
\;\;
mod.\;7
\]

\indent{Number of allowed couplings: 49}

\indent{Expression of the coupling}

\begin {eqnarray*}
C_{\theta\theta^2\theta^4} &=&
 N \; \sum_{v \in (f_3-f_2+\Lambda)} \exp \left[ -\frac
{1}{4\pi}\;\sin(\alpha)\sin(2 \alpha)\sin(3 \alpha) \;
\left( \frac{|v_1|^2}{\sin^2(3\alpha)} +\frac{|v_2|^2}{\sin^2(\alpha)} +
\frac{|v_3|^2}{\sin^2(2\alpha )} \right) \right] \\
 &=& N \; \sum_{\vec{u} \in Z^6} \exp [-\frac
{1}{4\pi}\; (\vec{f_{23}}+\vec{u})^{\top} M (\vec{f_{23}}+\vec{u})]  \\
&=& N \; \vartheta
\left[
\begin{array}{c}
\vec{f_{23}} \\
0
\end {array}
\right]
[0, \Omega]
\end {eqnarray*}

\[
\Omega= i \frac{1}{4\pi^2} M \;\;\;\;
N= \sqrt{V_{\Lambda}}\; \left[ \frac{1}{2 \pi} \right]^{3/2} \;
\left[ \frac{\Gamma(\frac{3}{7}) \Gamma(\frac{5}{7}) \Gamma(\frac{6}{7}) }
{\Gamma(\frac{1}{7}) \Gamma(\frac{2}{7}) \Gamma(\frac{4}{7})}  \right]^{3/2}
\]

\[
\begin{array}{l}
\Omega= i \frac{1}{4\pi^2} \;\sin(\alpha) \sin(2 \alpha) \sin(3 \alpha)\;
\left(
\begin {array}{rrrrrr}
a & b & c & d & d & c \\
b & a & b & c & d & d \\
c & b & a & b & c & d \\
d & c & b & a & b & c \\
d & d & c & b & a & b \\
c & d & d & c & b & a
\end {array}
\right)
\;\;\;
\begin{array}{l}
a = \frac{R_1^2}{\sin^2(3\alpha)}+ \frac{R_2^2}{\sin^2(\alpha)}+
 \frac{R_3^2}{\sin^2(2\alpha)} \\
\\
b = \frac{R_1^2 \cos(\alpha) }{\sin^2(3\alpha)}+ \frac{R_2^2
 \cos(2\alpha)}{\sin^2(\alpha)}+
\frac{R_3^2 \cos(3\alpha)}{\sin^2(2\alpha)} \\
\\
c = \frac{R_1^2 \cos(2\alpha)}{\sin^2(3\alpha)}+
\frac{R_2^2 \cos(3\alpha)}{\sin^2(\alpha)}+ \frac{R_3^2
 \cos(\alpha)}{\sin^2(2\alpha)} \\
\\
d =\frac{R_1^2 \cos(3\alpha)}{\sin^2(3\alpha)}+ \frac{R_2^2
 \cos(\alpha)}{\sin^2(\alpha)}+
\frac{R_3^2 \cos(2\alpha)}{\sin^2(2\alpha)}
\end{array}
\end{array}
\]

\indent{Number of effective parameters: 3}

\indent{Number of different couplings without deformations: 2}

\indent{corresponding to the following $\vec f_{23} $ shifts}
\[
\vec{f_{23}}=\{f_1^{(0)},f_1^{(1)}\}
\]

\indent{Number of different couplings with deformations: 4}

\indent{corresponding to the following $\vec f_{23} $ shifts}
\[
\vec{f_{23}}=\{f_1^{(0)},f_1^{(1)},f_1^{(2)},f_1^{(3)}\}
\]

\vspace {1.0 cm}
\noindent{\underline{\bf{ORBIFOLD $Z_8$--I}}}
\vspace{.5 cm}

\noindent{\underline {Twist}
$ \theta={\rm
diag}(e^{i\alpha},e^{2i\alpha},e^{-3i\alpha}) ,\;\;\;\;
\alpha=\frac{2\pi}{8} $}

\noindent{\underline {Lattice}
$ SO(5) \otimes SO(9) $}

\noindent{\underline {Coxeter element}}
\[
\begin{array}{lll}
\theta e_1 = e_1+2 e_2, & \theta e_2= -e_1-e_2, &
\theta e_3= e_4 ,\\
\theta e_4=e_5 ,&
\theta e_5 = e_3+e_4+e_5+2 e_6, & \theta e_6 = -e_3-e_4-e_5-e_6
\end{array}
\]

\noindent{\underline {Deformation parameters}}

\indent{Relations}
\[
\begin{array}{lll}
|e_1|=\sqrt{2}|e_2|, & |e_3|=|e_4|=|e_5| , &
-2\alpha_{56} |e_6|=|e_3|, \\
\alpha_{12}= -\frac {1}{\sqrt{2}}, &
\alpha_{35}=0, & \alpha_{34}=\alpha_{45} ,\\
\alpha_{36}=\alpha_{46} ,&
\alpha_{36}= \frac {1}{2 \alpha_{56}} - \alpha_{56}, &
\alpha_{34}=\frac {1}{4 \alpha_{56}^2} - 1, \\
\alpha_{ij}=0\;\;\;\;i=1,2\;\;\;\;j=3,4,5,6 & &
\end{array}
\]
\[
\alpha_{ij}\equiv\cos(\theta_{ij})
\]

\indent{Degrees of freedom (3)}
\[
\begin {array}{lll}
R_1= |e_1|, & R_3= |e_3|, & \alpha_{56}
\end {array}
\]

\vspace{.5 cm}
\noindent{\underline{Lattice basis ($e_i$) in terms of orthogonal basis
($\tilde
 e_i$)}}
\[
\begin{array}{rcl}
e_1&=& \frac {R_1}{2} \{ [(2+\sqrt{2})^{1/2} \cos (\varphi_1)
 +(2-\sqrt{2})^{1/2} \sin
(\varphi_1)] \tilde e_1 +\\
& &+[-(2+\sqrt{2})^{1/2} \sin (\varphi_1) +(2-\sqrt{2})^{1/2} \cos
(\varphi_1)] \tilde e_2 \} \\
e_2&=& \frac {R_1}{2} \{ -[(2+\sqrt{2})^{1/2} \cos (\varphi_1)
 +(2-\sqrt{2})^{1/2} \sin
(\varphi_1)] \tilde e_1 +\\
& &+[(2+\sqrt{2})^{1/2} \sin (\varphi_1) +(2-\sqrt{2})^{1/2} \cos
(\varphi_1)] \tilde e_2 \} \\
e_3&=&A[ \cos(\varphi_2) \tilde e_3 + \sin(\varphi_2) \tilde e_4 ]+
B[\cos( \varphi_3) \tilde e_5 +  \sin(\varphi_3) \tilde e_6]\\
e_4&=&A [\cos(\alpha +\varphi_2) \tilde e_3 +  \sin(\alpha +\varphi_2) \tilde
 e_4]
-B[ \cos(\alpha + \varphi_3) \tilde e_5 + \sin(\alpha + \varphi_3) \tilde
e_6]\\
e_5&=&-A [\sin(\varphi_2) \tilde e_3 -  \cos(\varphi_2) \tilde e_4 ]
-B[\sin( \varphi_3) \tilde e_5 - \cos( \varphi_3) \tilde e_6]\\
e_6&=&\frac{A}{2}[-\cos(\varphi_2)+ \sin(\varphi_2)- 2\cos(\varphi_2)
 \cos(\alpha)]  \tilde e_3 +\\
& &+\frac{A}{2}[-\cos(\varphi_2)- \sin(\varphi_2)- 2\sin(\varphi_2)
 \cos(\alpha)]  \tilde e_4 +\\
& &+\frac{B}{2}[-\cos(\varphi_3)+ \sin(\varphi_3)+ 2\cos(\varphi_3)
 \cos(\alpha)]  \tilde e_5 +\\
& &+\frac{B}{2}[-\cos(\varphi_3)- \sin(\varphi_3)+ 2\sin(\varphi_3)
 \cos(\alpha)]  \tilde e_6
\end{array}
\]

\indent{with $\alpha=\frac{2\pi}{8}$ and}
\[
\begin{array}{cc}
A= R_3 \left[ \frac{1+\sqrt{2}}{2} - \frac{1}{4 \sqrt{2} \alpha_{56}^2}
 \right]^{1/2},  &
B= R_3 \left[ \frac{1-\sqrt{2}}{2} +\frac{1}{4 \sqrt{2} \alpha_{56}^2}
 \right]^{1/2}
\end{array}
\]

\indent{$\varphi_1,\;\varphi_2,\;\varphi_3$ are free parameters.}

\vspace{.5 cm}
\noindent{\underline {Fixed points of $\theta$} (4)}
\[
\begin {array}{c}
f_1^{(ij)}= g_1^{(i)} \otimes \hat g_1^{(j)} \;,\;\;\; i=0,1,\;\;\;j = 0,1
\end {array}
\]
\[
\begin{array}{llll}
g_1^{(0)} = (0,0)\;,& g_1^{(1)} =\frac{1}{2} (1,0)\;,&
\hat g_1^{(0)} =(0,0,0,0)\;,& \hat g_1^{(1)}=\frac{1}{2} (1,0,1,0)
\end {array}
\]

\vspace{.5 cm}
\noindent{\underline {Fixed points of $\theta^2$} (16)}

\[
\begin{array}{c}
f_2^{(ij)}=  g_2^{(i)} \otimes \hat g_2^{(j)}  \;,\;\;\; i,j=0,1,2,3
\end {array}
\]
\[
\begin {array}{llll}
g_2^{(0)}=(0,0), & g_2^{(1)}=\frac{1}{2}(0,1), &
g_2^{(2)}=\frac{1}{2}(1,0), & g_2^{(3)}=\frac{1}{2}(1,1),\\
\hat g_2^{(0)}=(0,0,0,0), & \hat g_2^{(1)}=\frac{1}{2}(0,1,1,0) ,&
\hat g_2^{(2)}=\frac{1}{2}(1,0,1,0), & \hat g_2^{(3)}=\frac{1}{2}(1,1,0,0)
\end{array}
\]

\indent {Note that $\theta :g_2^{(1)}\rightarrow g_2^{(3)}$ and
$\theta : \hat g_2^{(1)} \rightarrow \hat g_2^{(3)}$.}

\indent {Number of conjugation classes: 10}

\vspace{.5 cm}
\noindent{\underline {Fixed points of $\theta^3$} (4)}

\indent{The same as for $\theta$.}
\[
\begin {array}{c}
f_3^{(ij)}= g_3^{(i)} \otimes \hat g_3^{(j)} \;,\;\;\; i=0,1,\;\;\;j = 0,1
\end {array}
\]
\[
\begin{array}{llll}
g_3^{(0)} = (0,0)\;,& g_3^{(1)} =\frac{1}{2} (1,0)\;,&
\hat g_3^{(0)} =(0,0,0,0)\;,& \hat g_3^{(1)}=\frac{1}{2} (1,0,1,0)
\end {array}
\]

\vspace{.5 cm}
\noindent{\underline {Fixed points of $\theta^4$} (16)}

\indent{Fixed torus: $\alpha(e_1)+\beta(e_2)\;,\;\; \alpha,\;\beta
\in R$}

\[
\begin{array}{c}
f_4^{(i)}= [\alpha(e_1)+\beta(e_2)] \otimes \hat g_4^{(i)} \;,\;\;\;
i=0,1,...,15,\;\;\;\alpha,\;\beta \in R
\end {array}
\]
\[
\begin {array}{llll}
\hat g_4^{(0)}=(0,0,0,0) ,&
\hat g_4^{(1)}=\frac{1}{2} (1,0,1,1), &
\hat g_4^{(2)}=\frac{1}{2} (1,0,0,0), &
\hat g_4^{(3)}=\frac{1}{2} (1,0,0,1), \\
\hat g_4^{(4)}=\frac{1}{2} (1,1,0,0), &
\hat g_4^{(5)}=\frac{1}{2} (0,0,0,1), &
\hat g_4^{(6)}=\frac{1}{2} (0,1,0,0), &
\hat g_4^{(7)}=\frac{1}{2} (0,0,1,1), \\
\hat g_4^{(8)}=\frac{1}{2} (1,0,1,0), &
\hat g_4^{(9)}=\frac{1}{2} (1,1,0,1), &
\hat g_4^{(10)}=\frac{1}{2} (0,0,1,0), &
\hat g_4^{(11)}=\frac{1}{2} (0,1,0,1), \\
\hat g_4^{(12)}=\frac{1}{2} (0,1,1,0), &
\hat g_4^{(13)}=\frac{1}{2} (0,1,1,1) &
\hat g_4^{(14)}=\frac{1}{2} (1,1,1,0), &
\hat g_4^{(15)}=\frac{1}{2} (1,1,1,1)
\end{array}
\]

\indent {Note that in the $SO(9)$ lattice}
\[
\begin{array}{ll}
\theta:\hat g_4^{(4)} \rightarrow \hat g_4^{(12)}\;, &
\theta:\hat g_4^{(1)} \rightarrow \hat g_4^{(3)} \rightarrow \hat g_4^{(9)}
\rightarrow \hat g_4^{(11)}\;,\\
\theta:\hat g_4^{(2)} \rightarrow \hat g_4^{(6)} \rightarrow \hat g_4^{(10)}
\rightarrow \hat g_4^{(14)}\;,&
\theta:\hat g_4^{(5)} \rightarrow \hat g_4^{(7)} \rightarrow \hat g_4^{(13)}
\rightarrow \hat g_4^{(15)}
\end {array}
\]

\indent {Number of conjugation classes: 6}

\vspace{.5 cm}
\noindent {\underline {Coupling $\theta^2\theta^2\theta^4$}}

\indent {Selection rule}
\[
f_1+f_2-(I+\theta^2)f_3 \in \Lambda
\]

\indent {Denoting}
\[
\left.
\begin {array}{rcl}
f_1 &=& g_2^{(i_1)} \otimes \hat g_2^{(j_1)} \\
f_2 &=& g_2^{(i_2)} \otimes \hat g_2^{(j_2)} \\
f_3 &=& [\alpha(e_1)+\beta(e_2)] \otimes \hat g_4^{(j_3)}
\end {array}
\right\}\;\;\;
\begin {array} {l}
i_1,i_2,j_1,j_2=0,1,2,3,\\
j_3=0,1,...,15\;,\\
\alpha,\beta \in R
\end {array}
\]

\indent{the selection rule  reads}
\[
\left.
\begin {array}{l}
i_1=i_2 \\
j_1+(-1)^{(j_3+1)} j_2 = j_3
\end {array}
\right\} \;\;\;\; mod.\;4
\]

\indent{Number of allowed couplings: 84}

\indent{Expression of the coupling}
\begin {eqnarray*}
C_{\theta^2\theta^2\theta^4} &=&\frac{F(l_1,l_2,l_3)}{2} \;\;
 N \; \{ \sum_{v \in (f_3-f_2+\Lambda)_{\perp}} \exp [-\frac
{1}{4\pi}  \; |v|^2] + \sum_{v \in (\theta f_3-f_2+\Lambda)_{\perp}} \exp
 [-\frac
{1}{4\pi}  \; |v|^2] \} \\
&=& \frac{F(l_1,l_2,l_3)}{2} \;\; N  \{ \vartheta
\left[
\begin{array}{c}
\vec{f_{23}} \\
0
\end {array}
\right]
[0, \Omega]
+
\vartheta
\left[
\begin{array}{c}
\vec{f'_{23}} \\
0
\end {array}
\right]
[0, \Omega]
\}
\end {eqnarray*}
\begin{quotation}
\noindent{where $(f_3-f_2+\Lambda)_{\perp}$ denotes that only coset
elements belonging to $SO(9)$ lattice are considered; $f_{23}=f_2-f_3$,
$f'_{23}=\theta f_2 -f_3$, the arrows denote components in the $SO(9)$ lattice.
 $l_i$
is the number of elements in the $f_i$ conjugation class (in all the cases,
 except
$l_1,l_2,l_3=2$, $f_{23}=f'_{23}$). Finally the values of $F(l_1,l_2,l_3)$ are}
\end{quotation}

\[
\begin{array}{rlrl}
l_1=l_2=l_3=1\;: & F=1 & l_1=l_2=1\;l_3=2 \;:& F=\sqrt{2} \\
l_1=l_2=1 \; l_3=4 \;: & F=2 & l_1=l_2=2 \; l_3=1\;: & F=1  \\
l_1=l_2=l_3=2\;: & F=\sqrt{2} & l_1=l_2=2 \; l_3=4 \;:& F=1 \\
l_1=1 (2) \; l_2=2 (1) \; l_3=4 \;:& F=\sqrt{2} & &
\end{array}
\]
\[
N= \sqrt{V_{\perp}}\; \frac{1}{2 \pi} \;
\frac{\Gamma^2(\frac{3}{4})}
{\Gamma^2(\frac{1}{4})} \;\;\;\;\;
\Omega = \; i \frac {1}{4 \pi^2}
\left(
\begin {array}{cccc}
a & b & 0 & c \\
b & a & b & c \\
0 & b & a & d \\
c & c & d & e
\end {array}
\right)
\begin{array}{l}
a=R_3^2 \\
b=R_3^2 [\frac{1}{4\alpha_{56}^2}-1]\\
c=R_3^2 [\frac{1}{2\alpha_{56}}-\alpha_{56}]\\
d=-\frac{R_3^2}{2}\\
e=\frac{R_3^2}{4\alpha_{56}^2}
\end{array}
\]

\indent{where $V_{\perp}$ is the volume of the $SO(9)$ lattice}

\indent{Number of effective parameters: 2}

\indent{Number of different couplings without deformations: 8}

\indent{corresponding to the following $\vec{f_{23}}$ shifts}
\[
\vec{f_{23}}= \;\; \left[
\begin {array}{l}
F=1 \; \left\{
\begin{array}{lll}
(0,0,0,0), & (\frac{1}{2},0,\frac{1}{2},0), & (\frac{1}{2},\frac{1}{2},0,0),
\end {array}
\right. \\
F=\sqrt{2} \; \left\{
\begin{array}{lc}
(\frac{1}{2},\frac{1}{2},0,0), & (\frac{1}{2},0,0,\frac{1}{2}), \\
(0,0,\frac{1}{2},\frac{1}{2}), & (0,0,0,0) \cup (\frac{1}{2},0,\frac{1}{2},0),
\end{array}
\right. \\
F=2\;\; (\frac{1}{2},0,0,0)
\end{array}
\right.
\]

\indent{Number of different couplings with deformations: 9}

\indent{corresponding to the following $\vec{f_{23}}$ shifts}
\[
\vec{f_{23}}= \;\; \left[
\begin {array}{l}
F=1 \; \left\{
\begin{array}{ll}
(0,0,0,0), & (\frac{1}{2},0,\frac{1}{2},0), \\
(\frac{1}{2},\frac{1}{2},0,0), & (0,\frac{1}{2},0,0),
\end {array}
\right. \\
F=\sqrt{2} \; \left\{
\begin{array}{cc}
(\frac{1}{2},\frac{1}{2},0,0), & (\frac{1}{2},0,0,\frac{1}{2}), \\
(0,0,\frac{1}{2},\frac{1}{2}), & (0,0,0,0) \cup (\frac{1}{2},0,\frac{1}{2},0),
\end{array}
\right. \\
F=2\;\; (\frac{1}{2},0,0,0)
\end{array}
\right.
\]

\vspace{.5 cm}
\noindent {\underline {Coupling $\theta\theta^2\theta^5$}}

\indent {Selection rule}
\[
f_1+(I+\theta)f_2-(I+\theta+\theta^2)f_3 \in \Lambda
\]

\indent {Denoting}
\[
\left.
\begin {array}{rcl}
f_1 &=& g_1^{(i_1)}\otimes \hat g_1^{(j_1)}\\
f_2 &=& g_2^{(i_2)}\otimes \hat g_2^{(j_2)}\\
f_3 &=& g_1^{(i_3)}\otimes \hat g_1^{(j_3)}
\end {array}
\right\}\;\;\;
\begin {array} {l}
i_1,i_3,j_1,j_3=0,1\;,\\
i_2,j_2=0,1,2,3\;,
\end {array}
\]

\indent{the selection rule reads}
\[
\left.
\begin {array}{lcr}
i_1+i_2+i_3&=&0 \\
j_1+j_2+j_3&=&0
\end {array}
\right\}
\;\;
mod.\;2
\]

\indent{Number of allowed couplings: 40}

\indent{Expression of the coupling}

\begin {eqnarray*}
C_{\theta\theta^2\theta^5} &=& \sqrt{l_2}\;\;
 N \; \sum_{v \in (f_3-f_2+\Lambda)_{\perp}} \exp [-\frac
{1}{4\pi}  \; (\frac {\sqrt{2}+1}{\sqrt{2}}|v_1|^2+|v_2|^2+\frac
 {\sqrt{2}-1}{\sqrt{2}}|v_3|^2)] \\
 &=& \sqrt{l_2}\;\;
 N \;\; \sum_{\vec{u} \in Z^6} \exp [-\frac
{1}{4\pi}  \; (\vec{f_{23}}+\vec{u})^{\top} M (\vec{f_{23}}+\vec{u})] \\
&=& \sqrt{l_2}\;\; N \; \vartheta
\left[
\begin{array}{c}
\vec{f_{23}} \\
0
\end {array}
\right]
[0, \Omega]
\end {eqnarray*}

\indent{with $l_2$ the number of elements in the $f_2$ conjugation class}

\[
\Omega = \; i \frac{1}{4 \pi^2} M \;\;\;\;
N= \sqrt{V_{\Lambda}}\; \left[ \frac{1}{2 \pi} \right]^{3/2}\;
\frac{\Gamma(\frac{7}{8}) \Gamma(\frac{3}{8})}
{\Gamma(\frac{1}{8}) \Gamma(\frac{5}{8})}
\frac{\Gamma^2(\frac{3}{4})}
{\Gamma^2(\frac{1}{4})}
\]

\[
\Omega = \; i \frac{1}{4 \pi^2}
\left(
\begin {array}{cccccc}
a & -a & 0 & 0 & 0 & 0 \\
-a & 2a & 0 & 0 & 0 & 0 \\
0 & 0 & b & c & 0 & e \\
0 & 0 & c & b & c & d \\
0 & 0 & 0 & c & b & e \\
0 & 0 & e & d & e & f
\end {array}
\right)
\begin{array}{l}
a=R_1^2\\
b=\frac{1}{\sqrt{2}}[(\sqrt{2}+1)  A^2 + (\sqrt{2}-1) B^2]\\
c=\frac{1}{2}[(\sqrt{2}+1)  A^2 - (\sqrt{2}-1) B^2]\\
d=-\frac{1}{2}[(\sqrt{2}+1)^2  A^2 + (\sqrt{2}-1)^2 B^2]\\
e=-\frac{1}{2\sqrt{2}}[(\sqrt{2}+1)^2  A^2 - (\sqrt{2}-1)^2 B^2] \\
f=\frac{1}{2\sqrt{2}}[(\sqrt{2}+1)^3 A^2 + (\sqrt{2}-1)^3 B^2]
\end{array}
\]

\indent{Number of effective parameters: 3}

\indent{Number of different couplings without deformations: 8}

\indent{corresponding to the following $\vec f_{23} $ shifts}
\[
\vec{f_{23}}= \; \left[
\begin{array}{l}
l_2=1 \;\left\{
\begin{array}{llll}
g_2^{(0)} \otimes \hat g_2^{(0)}, & g_2^{(0)} \otimes \hat g_2^{(2)}, &
g_2^{(2)} \otimes \hat g_2^{(0)}, & g_2^{(2)} \otimes \hat g_2^{(2)},
\end {array}
\right. \\
l_2=2 \; \left\{
\begin{array}{ll}
g_2^{(0)} \otimes \hat g_2^{(1)}, & g_2^{(1)} \otimes \hat g_2^{(0)},\\
g_2^{(2)} \otimes \hat g_2^{(1)}, & g_2^{(1)} \otimes \hat g_2^{(2)}
\end{array}
\right.
\end{array}
\right.
\]

\indent{Number of different couplings with deformations: 9}

\indent{corresponding to the following $\vec f_{23} $ shifts}
\[
\vec{f_{23}}= \; \left[
\begin{array}{l}
l_2=1 \;\left\{
\begin{array}{llll}
g_2^{(0)} \otimes \hat g_2^{(0)}, & g_2^{(0)} \otimes \hat g_2^{(2)}, &
g_2^{(2)} \otimes \hat g_2^{(0)}, & g_2^{(2)} \otimes \hat g_2^{(2)},
\end {array}
\right. \\
l_2=2 \; \left\{
\begin{array}{lll}
g_2^{(0)} \otimes \hat g_2^{(1)}, & g_2^{(1)} \otimes \hat g_2^{(0)}, &
g_2^{(1)} \otimes \hat g_2^{(1)}, \\
g_2^{(2)} \otimes \hat g_2^{(1)}, & g_2^{(1)} \otimes \hat g_2^{(2)} &
\end{array}
\right.
\end{array}
\right.
\]

\vspace {1.0 cm}
\noindent{\underline{\bf{ORBIFOLD $Z_8$--II}}}
\vspace{.5 cm}

\noindent{\underline {Twist}
$ \theta={\rm
diag}(e^{i\alpha},e^{3i\alpha},e^{-4i\alpha}) ,\;\;\;\;
\alpha=\frac{2\pi}{8} $}

\noindent{\underline {Lattice}
$ SO(4) \otimes SO(8) $}

\noindent{\underline {Twist in the lattice basis}}
\[
\begin{array}{lll}
\theta e_1=-e_1, & \theta e_2=-e_2, &
\theta e_3= e_4+e_5 ,\\
\theta e_4 = e_3+e_4+e_6, &
\theta e_5=-e_3-e_4-e_5-e_6, & \theta e_6 = -e_3-e_4
\end{array}
\]

\noindent{\underline {Deformation parameters}}

\indent{Relations}
\[
\begin{array}{lll}
|e_3|=|e_5|,  & |e_4|= \frac{1}{\sqrt{2}} [|e_3|^2+|e_6|^2]^{1/2} ,&
\alpha_{35}=0 ,\\
\alpha_{56}=0, &
\alpha_{34}= \frac {1}{\sqrt{2}} \frac{[\frac {1}{2} |e_6|^2 - \frac {3}{2}
 |e_3|^2]}
{|e_3|[|e_3|^2+|e_6|^2]^{1/2}} ,&
\alpha_{36}= \frac {[|e_3|^2-|e_6|^2]}{2|e_3||e_6|} ,\\
\alpha_{45}= \frac {1}{\sqrt{2}} \frac{[\frac {1}{2} |e_3|^2 - \frac {3}{2}
 |e_6|^2]}
{|e_3|[|e_3|^2+|e_6|^2]^{1/2}}, &
\alpha_{46}= -\frac {1}{\sqrt{2}} \frac {[|e_3|^2 + |e_6|^2]^{1/2}}{|e_6|}, &
\alpha_{ij}=0\;\;\;i=1,2\;\;j=3,4,5,6
\end{array}
\]
\[
\alpha_{ij}\equiv\cos(\theta_{ij})
\]

\indent{Degrees of freedom (5)}
\[
\begin {array}{lllll}
R_1= |e_1|, &
R_2= |e_2|, & R_3= |e_3|,  &
R_6= |e_6|, & \alpha_{12}
\end {array}
\]

\vspace{.5 cm}
\noindent{\underline{Lattice basis ($e_i$) in terms of orthogonal basis
($\tilde
 e_i$)}}
\[
\begin{array}{rcl}
e_1&=& R_1 [\sin (\varphi_1+\theta_{12}) \tilde e_1 + \cos
 (\varphi_1+\theta_{12}) \tilde e_2] \\
e_2&=& R_2 [\sin (\varphi_1) \tilde e_1 + \cos (\varphi_1) \tilde e_2] \\
e_3&=& A [\cos (\varphi_2) \tilde e_3 + \sin (\varphi_2) \tilde e_4] +
\rho_2 [\cos (\varphi_3) \tilde e_5 + \sin (\varphi_3) \tilde e_6] \\
e_4&=& \frac{A}{\sqrt{2}} [(\cos (\varphi_2)+(1+\sqrt{2})\sin (\varphi_2))
 \tilde e_3 +
(-(1+\sqrt{2}) \cos (\varphi_2)+ \sin (\varphi_2) \tilde e_4]-\\
& &-\frac{\rho_2}{\sqrt{2}} [(\cos (\varphi_3)-(1-\sqrt{2})\sin (\varphi_3))
 \tilde e_5 +
((1-\sqrt{2}) \cos (\varphi_3)+ \sin (\varphi_3) \tilde e_6] \\
e_5&=& -A [\sin (\varphi_2) \bar e_3 - \cos (\varphi_2) \tilde e_4] +
\rho_2 [\sin (\varphi_3) \tilde e_5 - \cos (\varphi_3) \tilde e_6] \\
e_6&=& -(1+\sqrt{2})A [\cos (\varphi_2) \tilde e_3 + \sin (\varphi_2) \tilde
 e_4] +
(\sqrt{2}-1)\rho_2 [\cos (\varphi_3) \tilde e_5 + \sin (\varphi_3) \tilde e_6]
\end{array}
\]

\[
\begin{array}{ll}
A= \frac {R_3}{2^{5/4}} \left[ \left(\frac{R_6}{R_3} \right) ^2
-(1-\sqrt{2})^2 \right]^{1/2},  &
\rho_2= \frac {R_6}{2^{5/4}} \left[ \left(\frac{R_3}{R_6} \right) ^2
 (1+\sqrt{2})^2
- 1 \right]^{1/2}
\end{array}
\]

\indent{$\varphi_1,\;\varphi_2,\;\varphi_3$ are free parameters.}

\vspace{.5 cm}
\noindent{\underline {Fixed points of $\theta$} (8)}
\[
\begin {array}{c}
f_1^{(ij)}= g_1^{(i)}\otimes \hat g_1^{(j)}\;,\;\;\; i=0,1,2,3,\;\;\;j = 0,1
\end {array}
\]
\[
\begin{array}{lll}
g_1^{(0)} = (0,0)\;,& g_1^{(1)} =\frac{1}{2} (1,0)\;,&
g_1^{(2)} =\frac{1}{2} (1,1)\;, \\
g_1^{(3)} =\frac{1}{2} (0,1)\;,&
\hat g_1^{(0)} =(0,0,0,0)\;,& \hat g_1^{(1)}=\frac{1}{2} (0,0,1,1)
\end {array}
\]

\vspace{.5 cm}
\noindent{\underline {Fixed points of $\theta^2$} (4)}

\indent{Fixed torus: $\alpha(e_1)+\beta(e_2)\;,\;\; \alpha,\;\beta
\in R$}

\[
\begin{array}{c}
f_2^{(i)}= [\alpha(e_1)+\beta(e_2)] \otimes \hat g_2^{(i)} \;,\;\;\;
 i=0,1,2,3,\;\;
\alpha,\beta \in R\\
\end {array}
\]
\[
\begin {array}{llll}
\hat g_2^{(0)}=(0,0,0,0), & \hat g_2^{(1)}=\frac{1}{2}(1,0,1,0), &
\hat g_2^{(2)}=\frac{1}{2}(0,0,1,1), & \hat g_2^{(3)}=\frac{1}{2}(1,0,0,1)
\end{array}
\]

\indent {Note that in the $SO(8)$ lattice
$\theta : \hat g_2^{(1)} \rightarrow \hat g_2^{(3)}$.}

\indent {Number of conjugation classes: 3}

\vspace{.5 cm}
\noindent{\underline {Fixed points of $\theta^3$} (8)}

\indent{The same as for $\theta$.}
\[
\begin {array}{c}
f_3^{(ij)}= g_3^{(i)} \otimes \hat g_3^{(j)}\;,\;\;\; i=0,1,2,3,\;\;\;j = 0,1
\end {array}
\]
\[
\begin{array}{lll}
g_3^{(0)} = (0,0)\;,& g_3^{(1)} =\frac{1}{2} (1,0)\;,&
g_3^{(2)} =\frac{1}{2} (1,1)\;, \\
g_3^{(3)} =\frac{1}{2} (0,1)\;, &
\hat g_3^{(0)} =(0,0,0,0)\;,&
\hat g_3^{(1)} =\frac{1}{2} (0,0,1,1)
\end {array}
\]

\vspace{.5 cm}
\noindent{\underline {Fixed points of $\theta^4$} (16)}

\indent{Fixed torus: $\alpha(e_1)+\beta(e_2)\;,\;\; \alpha,\;\beta
\in R$}

\[
\begin{array}{c}
f_4^{(i)}=  [\alpha(e_1)+\beta(e_2)] \otimes \hat g_4^{(i)} \;,\;\;\;
j=0,1,...,15,\;\; \alpha,\beta \in R
\end {array}
\]
\[
\begin {array}{llll}
\hat g_4^{(0)}=(0,0,0,0) ,&
\hat g_4^{(1)}=\frac{1}{2} (1,0,0,0), &
\hat g_4^{(2)}=\frac{1}{2} (0,1,0,0) ,&
\hat g_4^{(3)}=\frac{1}{2} (0,0,0,1), \\
\hat g_4^{(4)}=\frac{1}{2} (1,0,1,0), &
\hat g_4^{(5)}=\frac{1}{2} (0,1,1,0), &
\hat g_4^{(6)}=\frac{1}{2} (1,1,0,1), &
\hat g_4^{(7)}=\frac{1}{2} (1,1,0,0), \\
\hat g_4^{(8)}=\frac{1}{2} (0,0,1,1), &
\hat g_4^{(9)}=\frac{1}{2} (0,0,1,0), &
\hat g_4^{(10)}=\frac{1}{2} (0,1,1,1), &
\hat g_4^{(11)}=\frac{1}{2} (1,0,1,1), \\
\hat g_4^{(12)}=\frac{1}{2} (1,0,0,1), &
\hat g_4^{(13)}=\frac{1}{2} (1,1,1,1), &
\hat g_4^{(14)}=\frac{1}{2} (1,1,1,0), &
\hat g_4^{(15)}=\frac{1}{2} (0,1,0,0)
\end{array}
\]

\indent {Note that in the $SO(8)$ lattice}
\[
\begin{array}{ll}
\theta:\hat g_4^{(4)} \rightarrow \hat g_4^{(12)}, &
\theta:\hat g_4^{(1)} \rightarrow \hat g_4^{(5)} \rightarrow
\hat g_4^{(9)} \rightarrow \hat g_4^{(13)},\\
\theta:\hat g_4^{(2)} \rightarrow \hat g_4^{(6)} \rightarrow
\hat g_4^{(10)} \rightarrow \hat g_4^{(14)},&
\theta:\hat g_4^{(3)} \rightarrow \hat g_4^{(7)} \rightarrow
\hat g_4^{(11)} \rightarrow \hat g_4^{(15)}
\end{array}
\]

\indent {Number of conjugation classes: 6}

\vspace{.5 cm}
\noindent {\underline {Coupling $\theta\theta\theta^6$}}

\indent {Selection rule}
\[
f_1+f_2-(I+\theta)f_3 \in \Lambda
\]

\indent {Denoting}
\[
\left.
\begin {array}{rcl}
f_1 &=& g_1^{(i_1)}\otimes \hat g_1^{(j_1)}\\
f_2 &=& g_1^{(i_2)}\otimes \hat g_1^{(j_2)}\\
f_3 &=& [\alpha(e_1)+\beta(e_2)] \otimes \hat g_2^{(j_3)}
\end {array}
\right\}\;\;\;
\begin {array} {l}
i_1,i_2,j_3=0,1,2,3\;,\\
j_1,j_2=0,1\;,\\
\alpha,\beta \in R
\end {array}
\]

\indent{the selection rule reads}
\[
\left.
\begin {array}{l}
i_1=i_2 \\
j_1+j_2+j_3=0
\end {array}
\right\}
\;\;
mod.\;2
\]

\indent{Number of allowed couplings: 24}

\indent{Expression of the coupling}

\begin {eqnarray*}
C_{\theta\theta\theta^6} &=& \sqrt{l_3}\;\;
N \; \sum_{v \in (f_3-f_2+\Lambda)_{\perp}} \exp [-\frac
{\sqrt{2}}{8\pi}  \; (|v_2|^2+|v_3|^2)] \\
&=& \sqrt{l_3}\;\;
N \;\; \sum_{\vec{u} \in Z^4} \exp [-\frac
{\sqrt{2}}{8\pi} \; (\vec{\bar{f_{23}}}+\vec{u})^{\top} M
 (\vec{\bar{f_{23}}}+\vec{u})] \\
&=& \sqrt{l_3}\;\; N \; \vartheta
\left[
\begin{array}{c}
\vec{\bar{f_{23}}} \\
0
\end {array}
\right]
[0, \Omega]
\end {eqnarray*}
\begin{quotation}
\noindent{where the expression $(f_3-f_2+\Lambda)_{\perp}$ indicates that the
 coset
elements must belong to $SO(8)$ and $\bar{f_{23}}$ is the restriction of
 $f_{23}$ to $SO(8)$,
$l_3$ denotes the number of elements in the $f_3$ conjugation class, and the
 arrows
denote the components in the $SO(8)$ lattice}
\end{quotation}

\[
N= \sqrt{V_{\perp}}\;  \frac{1}{2 \pi} \;
\frac{\Gamma(\frac{7}{8}) \Gamma(\frac{5}{8})}
{\Gamma(\frac{1}{8}) \Gamma(\frac{3}{8})}
\;,\;\;
\Omega = \; i \frac {\sqrt{2}}{8\pi^2} M
\]
\[
\Omega= \; i \frac {\sqrt{2}}{8\pi^2}
\left(
\begin {array}{cccccc}
a & b & 0 & c \\
b & d & e & -\frac{d}{2} \\
0 & e & a & 0 \\
c & -\frac{d}{2} & 0 & f
\end {array}
\right)
\begin{array}{l}
a=R_3^2\\
b=-\frac{3}{4} R_3^2 + \frac{1}{4} R_6^2\\
c= \frac {1}{2} R_3^2 - \frac {1}{2} R_6^2\\
d= \frac {1}{2} R_3^2 + \frac {1}{2} R_6^2\\
e=-\frac{3}{4} R_6^2 + \frac{1}{4} R_3^2\\
f= R_6^2
\end{array}
\]

\indent{where $V_{\perp}$ is the volume of the $SO(8)$ lattice}

\indent{Number of effective parameters: 2}

\indent{Number of different couplings without deformations: 3}

\indent{Number of different couplings with deformations: 3}

\indent{corresponding to the following $\vec{\bar{f_{23}}}$ shifts}
\[
\vec{\bar{f_{23}}}= \; \left[
\begin {array}{l}
l_3=1 \; \left \{
\begin{array}{l}
g_2^{(0)},\\
g_2^{(2)},
\end{array}
\right. \\
l_3=2 \;\;\;\; g_2^{(1)}
\end{array}
\right.
\]

\vspace{.5 cm}
\noindent {\underline {Coupling $\theta\theta^3\theta^4$}}

\indent {Selection rule}
\[
f_1+f_2-(I+\theta+\theta^2+\theta^3)f_3 \in \Lambda
\]

\indent {Denoting}
\[
\left.
\begin {array}{rcl}
f_1 &=& g_1^{(i_1)}\otimes \hat g_1^{(j_1)}\\
f_2 &=& g_1^{(i_2)}\otimes \hat g_1^{(j_2)}\\
f_3 &=& [\alpha(e_1)+\beta(e_2)] \otimes \hat g_4^{(j_3)}
\end {array}
\right\}\;\;\;
\begin {array} {l}
i_1,i_2=0,1,2,3\;,\\
j_1,j_2=0,1\;, \\
j_3=0,1,...,15\;, \\
\alpha,\beta \in R
\end {array}
\]

\indent{the selection rule reads}
\[
\left.
\begin {array}{l}
i_1=i_2 \\
j_1+j_2+j_3=0
\end {array}
\right\}
\;\;
mod.\;2
\]

\indent{Number of allowed couplings: 48}

\indent{Expression of the coupling}

\begin {eqnarray*}
C_{\theta\theta^3\theta^4} &=& \sqrt{l_3}\;\;
 N \; \sum_{v \in (f_3-f_2+\Lambda)_{\perp}} \exp [-\frac
{1}{4\pi}  \; ((\sqrt{2}+1)|v_2|^2+(\sqrt{2}-1)|v_3|^2)] \\
 &=& \sqrt{l_3}\;\;
 N \;\; \sum_{\vec{u} \in Z^4} \exp [-\frac
{1}{4\pi}  \; (\vec{\bar{f_{23}}}+\vec{u})^{\top} M
 (\vec{\bar{f_{23}}}+\vec{u})]  \\
&=& \sqrt{l_3}\;\; N \; \vartheta
\left[
\begin{array}{c}
\vec{\bar{f_{23}}} \\
0
\end {array}
\right]
[0, \Omega]
\end {eqnarray*}
\begin{quotation}
\noindent{where $(f_3-f_2+\Lambda)_{\perp}$ indicates that the coset
elements must belong to $SO(8)$,
and $V_{\perp}$ is the volume of the $SO(8)$ lattice}
\end{quotation}

\[
N= \sqrt{V_{\perp}}\;  \frac{1}{2 \pi} \;
\frac{\Gamma(\frac{7}{8}) \Gamma(\frac{5}{8})}
{\Gamma(\frac{1}{8}) \Gamma(\frac{3}{8})} \;\;\;\;
\Omega = \; i \frac{1}{4\pi^2} M
\]

\[
\Omega=\; i \frac{1}{4\pi^2}
\left(
\begin {array}{cccccc}
a & b & 0 & e \\
b & c & d & d\\
0 & d & a & 0 \\
e & d & 0 & f
\end {array}
\right)
\begin{array}{l}
a=[(\sqrt{2}+1) A^2 + (\sqrt{2}-1) B^2]\\
b=\frac{1}{\sqrt{2}}[(\sqrt{2}+1) A^2 - (\sqrt{2}-1) B^2 ]\\
c=[(\sqrt{2}+2)(\sqrt{2}+1) A^2 - (\sqrt{2}-1)(\sqrt{2}-2) B^2]\\
d=-\frac{1}{\sqrt{2}}[(\sqrt{2}+1)^2 A^2 + (\sqrt{2}-1)^2 B^2 ]\\
e=-[(\sqrt{2}+1)^2 A^2 - (\sqrt{2}-1)^2 B^2]\\
f=[(\sqrt{2}+1)^3 A^2 + (\sqrt{2}-1)^3 B^2]
\end{array}
\]

\indent{Number of effective parameters: 2}

\indent{Number of different couplings without deformations: 6}

\indent{Number of different couplings with deformations: 6}

\indent{corresponding to the following $\vec{\bar{f_{23}}}$ shifts}
\[
\vec{\bar{f_{23}}}= \; \left[
\begin{array}{l}
l_3 =1 \; \left\{
\begin{array}{l}
g_4^{(0)}, \\
g_4^{(8)},
\end {array}
\right.\\
l_3=2 \;\;\;\;\; g_4^{(4)}, \\
l_3=4 \; \left\{
\begin {array}{l}
g_4^{(1)}, \\
g_4^{(2)}, \\
g_4^{(3)}
\end{array}
\right.
\end {array}
\right.
\]

\vspace {1.0 cm}
\noindent{\underline{\bf{ORBIFOLD $Z_{12}$--I}}}
\vspace{.5 cm}

\noindent{\underline {Twist}
$ \theta={\rm
diag}(e^{i\alpha},e^{4i\alpha},e^{-5i\alpha}) ,\;\;\;\;
\alpha=\frac{2\pi}{12} $}

\noindent{\underline {Lattice}
$ SU(3) \otimes F_4 $}

\noindent{\underline {Coxeter element}}
\[
\begin{array}{lll}
\theta e_1= e_2, & \theta e_2 = -e_1-e_2, &
\theta e_3= e_4 ,\\
\theta e_4= e_3+e_4+2 e_5,  & \theta e_5=e_6, &
\theta e_6 = -e_3-e_4-e_5-e_6
\end{array}
\]

\noindent{\underline {Deformation parameters}}

\indent{Relations}
\[
\begin{array}{llll}
|e_1|=|e_2| , & |e_3|=|e_4|= \sqrt{2} |e_5| = \sqrt{2} |e_6|,&
\alpha_{12}=-\frac {1}{2} ,& \alpha_{45}=-\frac{1}{\sqrt{2}} ,\\
\alpha_{34}= \alpha_{56} , &
\alpha_{35}= \alpha_{46} = \alpha_{36} ,&
\alpha_{35}= -\frac{1}{2\sqrt{2}}[1+2\alpha_{34}], &
 \alpha_{ij}=0\;\;\;i=1,2,\;\;j=1,2,3,4
\end{array}
\]
\[
\alpha_{ij}\equiv\cos(\theta_{ij})
\]

\indent{Degrees of freedom (3)}
\[
\begin {array}{lll}
R_1= |e_1|, & R_3= |e_3|, & \alpha_{34}
\end {array}
\]

\vspace{.5 cm}
\noindent{\underline{Lattice basis ($e_i$) in terms of orthogonal basis
($\tilde
 e_i$)}}
\[
\begin{array}{l}
e_1= R_1 \cos(\phi_1) \tilde e_1 + R_1 \sin(\phi_1) \tilde e_2 \\
e_2= R_1 \cos(\phi_1+\alpha) \tilde e_1 + R_1 \sin (\phi_1 +\alpha ) \tilde e_2
 \\
e_3= A \cos(\phi_2) \tilde e_3 + A \sin(\phi_2) \tilde e_4 + B \cos (\phi_3)
 \tilde e_5 +
B \cos (\phi_3) \tilde e_6 \\
e_4= A \cos(\phi_2+\beta) \tilde e_3 + A \sin(\phi_2+\beta) \tilde e_4 + B \cos
 (\phi_3+ 7\beta)
\tilde e_5 + B \cos (\phi_3+ 7 \beta) \tilde e_6 \\
e_5= \frac{A}{\sqrt{2}}[-\sin(\phi_2+\frac{5}{2}\beta) \tilde e_3 +\cos(\phi_2+
\frac{5}{2}\beta)\tilde e_4] +
\frac{B}{\sqrt{2}}[\sin(\phi_3+ \frac{11}{2}\beta) \tilde e_5
-\cos (\phi_3+\frac{11}{2}\beta) \tilde e_6 ]\\
e_6= \frac{A}{\sqrt{2}}[-\sin(\phi_2+\frac{7}{2}\beta) \tilde e_3 +\cos(\phi_2+
\frac{7}{2}\beta)\tilde e_4] +
\frac{B}{\sqrt{2}}[\sin(\phi_3+ \frac{1}{2}\beta) \tilde e_5
-\cos (\phi_3+\frac{1}{2}\beta) \tilde e_6 ]\\
\end{array}
\]

\[
\begin{array}{llll}
\alpha = \frac{2\pi}{3}, & \beta= \frac {\pi}{6}, &
A= R_3 \left[ \frac {1}{2} + \frac{\alpha_{34}}{\sqrt{3}} \right]^{1/2}, &
B= R_3 \left[ \frac {1}{2} - \frac{\alpha_{34}}{\sqrt{3}} \right]^{1/2}
\end{array}
\]

\indent{$\phi_1,\;\phi_2,\;\phi_3$ are free parameters}

\vspace{.5 cm}
\noindent{\underline {Fixed points of $\theta$} (3)}
\[
\begin {array}{c}
f_1^{(i)}= g_1^{(i)}\otimes \hat g_1^{(0)}\;,\;\;\; i=0,1,2
\end {array}
\]
\[
\begin{array}{llll}
g_1^{(0)} = (0,0)\;,& g_1^{(1)} =\frac{1}{3} (1,2)\;,&
g_1^{(2)} =\frac{1}{3} (2,1)\;, & \hat g_1^{(0)}=(0,0,0,0)\;
\end {array}
\]

\vspace{.5 cm}
\noindent{\underline {Fixed points of $\theta^2$} (3)}

\indent{The same as for $\theta$}

\[
\begin {array}{c}
f_2^{(i)}= g_2^{(i)} \otimes \hat g_2^{(0)}\;,\;\;\; i=0,1,2
\end {array}
\]
\[
\begin{array}{llll}
g_2^{(0)} = (0,0)\;,& g_2^{(1)} =\frac{1}{3} (1,2)\;,&
g_2^{(2)} =\frac{1}{3} (2,1)\;, & \hat g_2^{(0)}=(0,0,0,0)\;
\end {array}
\]

\vspace{.5 cm}
\noindent{\underline {Fixed points of $\theta^3$} (4)}

\indent{Fixed torus: $\alpha(e_1)+\beta(e_2)\;,\;\; \alpha,\;\beta
\in R$}

\[
\begin {array}{c}
f_3^{(i)}= [\alpha(e_1)+\beta(e_2)]\otimes \hat g_3^{(i)}\;,\;\;\;
 i=0,1,2,3,\;\;
\alpha,\beta \in R
\end {array}
\]
\[
\begin{array}{llll}
\hat g_3^{(0)} =(0,0,0,0)\;,& \hat g_3^{(1)} =\frac{1}{2} (1,0,0,0)\;, &
\hat g_3^{(2)} =\frac{1}{2} (0,1,0,0)\;, & \hat g_3^{(3)} =\frac{1}{2}
 (1,1,0,0)\;
\end {array}
\]

\indent{Note that in the $F_4$ lattice $\theta:\hat g_3^{(1)}  \rightarrow \hat
 g_3^{(2)}
\rightarrow \hat g_3^{(3)}$}

\indent {Number of conjugation classes: 2}

\vspace{.5 cm}
\noindent{\underline {Fixed points of $\theta^4$} (27)}

\[
\begin{array}{c}
f_4^{(ij)}=  g_4^{(i)} \otimes \hat  g_4^{(j)} \;,\;\;\;
i=0,1,2, \;\;\;\; j=0,1,...,8
\end {array}
\]
\[
\begin {array}{llll}
g_4^{(0)} = (0,0),&
g_4^{(1)} =\frac{1}{3} (1,2),&
g_4^{(2)} =\frac{1}{3} (2,1), &
\hat g_4^{(0)}=(0,0,0,0), \\
\hat g_4^{(1)}=\frac{1}{3} (2,1,2,0), &
\hat g_4^{(2)}=\frac{1}{3} (2,2,0,2), &
\hat g_4^{(3)}=\frac{1}{3} (1,0,2,2), &
\hat g_4^{(4)}=\frac{1}{3} (0,2,2,1), \\
\hat g_4^{(5)}=\frac{1}{3} (1,2,1,0), &
\hat g_4^{(6)}=\frac{1}{3} (1,1,0,1), &
\hat g_4^{(7)}=\frac{1}{3} (2,0,1,1), &
\hat g_4^{(8)}=\frac{1}{3} (0,1,1,2)
\end{array}
\]
\begin{quotation}
\noindent{Note that in the $F_4$ lattice $\theta :\hat g_4^{(1)} \rightarrow
 \hat g_4^{(3)}
\rightarrow  \hat g_4^{(5)} \rightarrow \hat g_4^{(7)}$
and $\theta: \hat g_4^{(2)} \rightarrow \hat g_4^{(4)} \rightarrow \hat
 g_4^{(6)}
\rightarrow \hat g_4^{(8)}$}
\end{quotation}

\indent {Number of conjugation classes: 9}

\vspace{.5 cm}
\noindent{\underline {Fixed points of $\theta^5$} (3)}

\indent{The same as for $\theta$}
\[
\begin {array}{c}
f_5^{(i)}= g_5^{(i)}\otimes \hat g_5^{(0)}\;,\;\;\; i=0,1,2
\end {array}
\]
\[
\begin{array}{llll}
g_5^{(0)} = (0,0)\;,& g_5^{(1)} =\frac{1}{3} (1,2)\;,&
g_5^{(2)} =\frac{1}{3} (2,1)\;, & \hat g_5^{(0)}=(0,0,0,0)\;
\end {array}
\]

\vspace{.5 cm}
\noindent{\underline {Fixed points of $\theta^6$} (16)}

\indent{Fixed torus: $\alpha(e_1)+\beta(e_2)\;,\;\; \alpha,\;\beta
\in R$}

\[
\begin {array}{c}
f_6^{(i)}= [\alpha(e_1)+\beta(e_2)] \otimes \hat g_6^{(i)}\;,\;\;\;
 i=0,1,...,15,\;\;
\alpha, \beta \in R
\end {array}
\]

\[
\begin{array}{llll}
\hat g_6^{(0)}=(0,0,0,0)\;,&
\hat g_6^{(1)} =\frac{1}{2} (1,1,1,1)\;,&
\hat g_6^{(2)} =\frac{1}{2} (0,0,0,1)\;, &
\hat g_6^{(3)} =\frac{1}{2} (0,0,1,0)\;,\\
\hat g_6^{(4)} =\frac{1}{2} (1,0,0,0)\;, &
\hat g_6^{(5)} =\frac{1}{2} (0,0,1,1)\;, &
\hat g_6^{(6)} =\frac{1}{2} (0,1,0,1)\;, &
\hat g_6^{(7)} =\frac{1}{2} (1,0,1,0)\;,\\
\hat g_6^{(8)} =\frac{1}{2} (0,1,0,0)\;,&
\hat g_6^{(9)} =\frac{1}{2} (0,1,1,1)\;,&
\hat g_6^{(10)} =\frac{1}{2} (1,1,0,1)\;,&
\hat g_6^{(11)}=\frac{1}{2} (0,1,1,0)\;,\\
\hat g_6^{(12)} =\frac{1}{2} (1,1,0,0)\;, &
\hat g_6^{(13)} =\frac{1}{2} (1,0,1,1)\;,&
\hat g_6^{(14)} =\frac{1}{2} (1,0,0,1)\;, &
\hat g_6^{(15)} =\frac{1}{2} (1,1,1,0)\;
\end {array}
\]
\begin{quotation}
\noindent{Note that in the $F_4$ lattice $\theta:\hat g_6^{(3)} \rightarrow
\hat
 g_6^{(2)}
\rightarrow \hat g_6^{(1)} \rightarrow \hat g_6^{(11)} \rightarrow \hat
 g_6^{(10)}
\rightarrow \hat g_6^{(9)}$ ,  $\theta: \hat g_6^{(7)}
\rightarrow \hat g_6^{(6)} \rightarrow \hat g_6^{(5)} \rightarrow \hat
 g_6^{(15)}
\rightarrow \hat g_6^{(14)}
\rightarrow \hat g_6^{(13)}$ and $\theta: \hat g_6^{(4)}
\rightarrow \hat g_6^{(8)}
\rightarrow \hat g_6^{(12)}$}
\end{quotation}

\indent {Number of conjugation classes: 4}

\vspace{.5 cm}
\noindent {\underline {Coupling $\theta\theta^2\theta^9$}}

\indent {Selection rule}
\[
f_1+(I+\theta)f_2-(I+\theta+\theta^2)f_3 \in \Lambda
\]

\indent {Denoting}
\[
\left.
\begin {array}{rcl}
f_1 &=& g_1^{(i_1)}\otimes \hat g_1^{(0)}\\
f_2 &=& g_2^{(i_2)}\otimes \hat g_2^{(0)}\\
f_3 &=& [\alpha(e_1)+\beta(e_2)] \otimes \hat g_3^{(j_3)}
\end {array}
\right\}\;\;\;
\begin {array} {l}
i_1,i_2,j_3=0,1,2,3\;,\\
\alpha,\beta \in R
\end {array}
\]

\indent{the selection rule reads}
\[
\begin {array}{lcr}
i_1&=&i_2
\end {array}
\]

\indent{Number of allowed couplings: 6}

\indent{Expression of the coupling}

\begin {eqnarray*}
C_{\theta\theta^2\theta^9} &=& \sqrt{l_3}\;\;
 N \; \sum_{v \in (f_3-f_2+\Lambda)_{\perp}} \exp [-\frac
{1}{4\pi} \sin(\frac{\pi}{6} \sin (\frac{\pi}{4}) \;
 (\frac{|v_2|^2}{\sin(\frac{\pi}{12})}
+\frac{|v_3|^2}{\cos(\frac{\pi}{12})})] \\
 &=& \sqrt{l_3}\;\;
 N \;\; \sum_{u \in Z^4} \exp [-\frac
{\sqrt{2}}{4\pi}  \; (\vec{\bar{f_{23}}}+\vec{u})^{\top} M
 (\vec{\bar{f_{23}}}+\vec{u})]  \\
&=& \sqrt{l_3}\;\; N \; \vartheta
\left[
\begin{array}{c}
\vec{\bar{f_{23}}} \\
0
\end {array}
\right]
[0, \Omega]
\end {eqnarray*}
\begin{quotation}
\noindent{where $(f_3-f_2+\Lambda)_{\perp}$ indicates that the coset
elements must belong to $F_4$, $l_3$ is the number of elements in the $f_3$
 conjugation class,
the arrows denote components in the $F_4$ lattice, and $V_{\perp}$ is the
volume
 of the $F_4$
lattice unit cell}
\end{quotation}

\[
N= \sqrt{V_{\perp}}\;  \frac{1}{2 \pi} \;
\frac{\Gamma(\frac{5}{6})}{\Gamma(\frac{1}{6})} \;
\left[ \frac{\Gamma(\frac{11}{12}) \Gamma(\frac{5}{12})}
{\Gamma(\frac{1}{12}) \Gamma(\frac{7}{12})} \right] ^{1/2}\;,\;\;\;
\Omega = \; i \frac{\sqrt{2}}{4\pi^2} M
\]

\[
\Omega = \; i \frac{\sqrt{2}}{4\pi^2}
\left(
\begin {array}{cccc}
a & \frac{b\sqrt{3}}{2} & -\frac{b}{\sqrt{2}} &  -\frac{b}{\sqrt{2}} \\
\frac{b\sqrt{3}}{2} & a & -\frac{a}{2} & \frac{b\sqrt{3}}{2} \\
-\frac{b}{\sqrt{2}} & -\frac{a}{2} & \frac{a}{2} & \frac{b\sqrt{3}}{4} \\
-\frac{b}{\sqrt{2}} & \frac{b\sqrt{3}}{2} & \frac{b\sqrt{3}}{4} & \frac{a}{2}
\end {array}
\right)
\begin{array}{l}
a=[A^2 \cos (\frac{\pi}{12})+ B^2 \sin (\frac{\pi}{12})]\\
b=[A^2 \cos (\frac{\pi}{12})- B^2 \sin (\frac{\pi}{12})]
\end{array}
\]

\indent{Number of effective parameters: 2}

\indent{Number of different couplings without deformations: 2}

\indent{Number of different couplings with deformations: 2}

\indent{corresponding to the following $\vec{\bar{ f_{23}}} $ shifts}
\[
\vec{\bar{f_{23}}} = \hat g_3^{(0)}, \; \hat g_3^{(1)}
\]

\indent{Note that this coupling is the same as $\theta^2\theta^3\theta^7$}

\vspace{.5 cm}
\noindent {\underline {Coupling $\theta\theta^4\theta^7$}}

\indent {Selection rule}
\[
f_1+(I+\theta+\theta^2+\theta^3)f_2-(I+\theta+\theta^2+\theta^3+\theta^4)f_3
\in
 \Lambda
\]

\indent {Denoting}
\[
\left.
\begin {array}{rcl}
f_1 &=& g_1^{(i_1)}\otimes \hat g_1^{(0)}\\
f_2 &=& g_4^{(i_2)}\otimes \hat g_4^{(j_2)}\\
f_3 &=& g_5^{(i_3)}\otimes \hat g_5^{(0)}
\end {array}
\right\}\;\;\;
\begin {array} {l}
i_1,i_2,i_3 = 0,1,2,\\
j_2= 0,1,...,8
\end {array}
\]

\indent{the selection rule reads}
\[
\begin {array}{lcr}
i_1+i_2+i_3=0\;,\;\;\; mod.\;3
\end {array}
\]

\indent{Number of allowed couplings: 6}

\indent{Expression of the coupling}

\begin {eqnarray*}
C_{\theta\theta^4\theta^7} &=& \sqrt{l_2}\;\;
 N \; \sum_{v \in (f_3-f_2+\Lambda)} \exp [-\frac
{1}{4\pi} \; (\frac{\sqrt{3}}{2} |v_1|^2 +
 2\sqrt{3}(\frac{|v_2|^2}{\cos^2(\frac{\pi}{12})}
+\frac{|v_3|^2}{\sin^2(\frac{\pi}{12})}))] \\
 &=& \sqrt{l_2}\;\;
 N \;\; \sum_{\vec{u} \in Z^6} \exp [-\frac
{\sqrt{3}}{8\pi}  \; (\vec{f_{23}}+\vec{u})^{\top} M (\vec{f_{23}}+\vec{u})]
\\
&=& \sqrt{l_2}\;\; N \; \vartheta
\left[
\begin{array}{c}
\vec{{f_{23}}} \\
0
\end {array}
\right]
[0, \Omega]
\end {eqnarray*}
\begin{quotation}
\noindent{$l_2$ is the number of elements in the $f_2$ conjugation class, and
 the arrows denote
components in the $SU(3) \otimes F_4$ lattice}
\end {quotation}

\[
N= \sqrt{V_{\Lambda}}\;  \frac{3^{1/4}}{4 \pi^2} \;
\frac{\Gamma^3(\frac{2}{3})}{\Gamma^2(\frac{1}{3})} \;
\left[ \frac{\Gamma(\frac{11}{12}) \Gamma(\frac{5}{12})}
{\Gamma(\frac{1}{12}) \Gamma(\frac{7}{12})} \right] \;,\;\;\;\;
\Omega = \; i \frac{\sqrt{2}}{4\pi^2} M
\]

\[
\begin{array}{l}
\Omega= \;i \frac{\sqrt{3}}{8\pi^2}
\left(
\begin {array}{cccccc}
R_1^2 & -\frac{R_1^2}{2} & 0 & 0 & 0 & 0 \\
-\frac{R_1^2}{2} & R_1^2 & 0 & 0 & 0 & 0 \\
0 & 0 & a & \frac{b\sqrt{3}}{2} & -\frac{b}{\sqrt{2}} &  -\frac{b}{\sqrt{2}} \\
0 & 0 & \frac{b\sqrt{3}}{2} & a & -\frac{a}{2} & \frac{b\sqrt{3}}{2} \\
0 & 0 & -\frac{b}{\sqrt{2}} & -\frac{a}{2} & \frac{a}{2} & \frac{b\sqrt{3}}{4}
 \\
0 & 0 & -\frac{b}{\sqrt{2}} & \frac{b\sqrt{3}}{2} & \frac{b\sqrt{3}}{4} &
 \frac{a}{2}
\end {array}
\right)
\begin{array}{l}
a=4[A^2 \cos (\frac{\pi}{12})+ B^2 \sin (\frac{\pi}{12})]\\
b=4[A^2 \cos (\frac{\pi}{12})- B^2 \sin (\frac{\pi}{12})]
\end{array}
\end{array}
\]

\indent{Number of effective parameters: 3}

\indent{Number of different couplings without deformations: 4}

\indent{Number of different couplings with deformations: 4}

\indent{corresponding to the following $\vec{f_{23}} $ shifts}

\[
\vec{f_{23}}= \;
\left[
\begin{array}{l}
l_2=1 \;
\left\{
\begin{array}{l}
g_4^{(0)} \otimes \hat g_4^{(0)}, \\
g_4^{(1)} \otimes \hat g_4^{(0)},
\end{array}
\right.\\
l_2=2 \;
\left\{
\begin{array}{l}
g_4^{(0)} \otimes \hat g_4^{(1)}, \\
g_4^{(1)} \otimes \hat g_4^{(1)}
\end{array}
\right.
\end{array}
\right.
\]

\vspace{.5 cm}
\noindent {\underline {Coupling $\theta^2\theta^4\theta^6$}}

\indent {Selection rule}
\[
f_1+(I+\theta^2)f_2-(I+\theta^2+\theta^4)f_3 \in \Lambda
\]

\indent {Denoting}
\[
\left.
\begin {array}{rcl}
f_1 &=& g_2^{(i_1)}\otimes \hat g_2^{(0)}\\
f_2 &=& g_4^{(i_2)}\otimes \hat g_4^{(j_2)}\\
f_3 &=& [\alpha(e_1)+\beta(e_2)] \otimes \hat g_6^{(j_3)}
\end {array}
\right\}\;\;\;
\begin {array} {l}
i_1,i_2= 0,1,2,\\
j_2= 0,1,...,8,\\
j_3= 0,1,...,15,\\
\alpha,\beta \in R
\end {array}
\]

\indent{the selection rule reads}
\[
\begin {array}{lcr}
i_1=i_2 \;,
\end {array}
\]

\indent{Number of allowed couplings: 36}

\indent{Expression of the coupling (in all the cases except $l_3=6,\; l_2=4$)}

\begin {eqnarray*}
C_{\theta^2\theta^4\theta^6} &=&
N \sqrt{l_2\;l_3} \; \sum_{v \in (f_{23}+\Lambda)_{\perp}} \exp [-\frac
{1}{4\pi} \; \frac {\sin(\frac{\pi}{3})}{\sin(\frac{\pi}{6})} \; |v|^2] \\
&=&
N \sqrt{l_2\;l_3} \; \sum_{\vec{u} \in Z^4} \exp [-\frac
{\sqrt{3}}{4\pi} \; (\vec{\bar{f_{23}}}+\vec{u})^{\top} M
 (\vec{\bar{f_{23}}}+\vec{u})] \\
&=& N \sqrt{l_2\;l_3} \;\vartheta
\left[
\begin{array}{c}
\vec{\bar{f_{23}}} \\
0
\end {array}
\right]
[0, \Omega]
\end {eqnarray*}
\begin{quotation}
\noindent{where $f_{23}=f_2-f_3$, $\bar f_{23}$ is the restriction of $f_{23}$
 to the $F_4$
lattice; $(f_{23}+\Lambda)_{\perp}$ indicates that the coset must belong to
 $F_4$
and $l_i$ is the number of elements in the $f_i$ conjugation class. $V_{\perp}$
 is
the volume of the $F_4$ unit cell}
\end{quotation}

\indent{In the $l_3=6,\;l_2=4$ case}

\begin {eqnarray*}
C_{\theta^2\theta^4\theta^6} &=&
N \sqrt{6} \; \sum_{v \in (f_2-f_3+\Lambda)_{\perp}\cup (\theta
 f_2-f_3+\Lambda)_{\perp}}
\exp [-\frac{1}{4\pi} \; \frac {\sin(\frac{\pi}{3})}{\sin(\frac{\pi}{6})} \;
 |v|^2] \\
&=& N \sqrt{6} \; \{ \vartheta
\left[
\begin{array}{c}
\vec{\bar{f_{23}}} \\
0
\end {array}
\right]
[0, \Omega] +
\vartheta
\left[
\begin{array}{c}
\vec{\bar{f'_{23}}} \\
0
\end {array}
\right]
[0, \Omega] \}
\end {eqnarray*}

\[
N= \sqrt{V_{\perp}}\;  \frac{1}{2\pi} \;
\left[ \frac{\Gamma(\frac{2}{3}) \Gamma(\frac{5}{6})}
{\Gamma(\frac{1}{6}) \Gamma(\frac{1}{3})} \right]\;,\;\;\;
\Omega = \;i \frac {\sqrt{3}}{4 \pi^2} M
\]

\[
\Omega = \;i \frac {\sqrt{3}}{4 \pi^2} R_3^2
\left(
\begin {array}{cccc}
1 & \alpha_{34} & -\frac{1}{4}[1+2\alpha_{34}] & -\frac{1}{4}[1+2\alpha_{34}]
\\
\alpha_{34} & 1 & -\frac{1}{2} & -\frac{1}{4}[1+2\alpha_{34}] \\
-\frac{1}{4}[1+2\alpha_{34}] & -\frac{1}{2} & \frac {1}{2} & \frac
 {\alpha_{34}}{2} \\
-\frac{1}{4}[1+2\alpha_{34}] & -\frac{1}{4}[1+2\alpha_{34}] & \frac
 {\alpha_{34}}{2} & \frac{1}{2}
\end {array}
\right)
\]

\indent{Number of effective parameters: 2}

\indent{Number of different couplings without deformations: 7 }

\indent{Number of different couplings with deformations: 7}

\indent{corresponding to the following $\vec{\bar{ f_{23}}} $ shifts}
\[
\vec{\bar{f_{23}}}=\; \left[
\begin{array}{l}
l_2=l_3=1\;\;(0,0,0,0) ,\\
l_2=1\;l_3=3\;\;(\frac{1}{2},0,0,0),\\
l_2=1\;l_3=6\;\;(0,0,\frac{1}{2},0),\\
l_2=4\;l_3=1\;\;(\frac{2}{3},\frac{1}{3},\frac{2}{3},0),\\
l_2=4\;l_3=3\;\;(\frac{1}{6},\frac{1}{3},\frac{2}{3},0),\\
l_2=4\;l_3=6\;\;\left\{
\begin{array}{l}
(\frac{2}{3},\frac{1}{3},\frac{1}{6},0)\cup
 (\frac{2}{3},\frac{1}{3},\frac{2}{3},\frac{1}{2}),\\
(\frac{1}{6},\frac{1}{3},\frac{1}{6},0)\cup
 (\frac{2}{3},\frac{5}{6},\frac{2}{3},\frac{1}{2})
\end{array}
\right.
\end{array}
\right.
\]

\vspace{.5 cm}
\noindent {\underline {Coupling $\theta^4\theta^4\theta^4$}}

\indent {Selection rule}
\[
f_1+f_2+f_3 \in \Lambda
\]

\indent {Denoting}
\[
\left.
\begin {array}{rcl}
f_1 &=& g_4^{(i_1)}\otimes \hat g_4^{(j_1)}\\
f_2 &=& g_4^{(i_2)}\otimes \hat g_4^{(j_2)}\\
f_3 &=& g_4^{(i_3)}\otimes \hat g_4^{(j_3)}
\end {array}
\right\}\;\;\;
\begin {array} {l}
i_1,i_2,i_3= 0,1,2,\\
j_1,j_2,j_3= 0,1,...,8,
\end {array}
\]

\indent{The selection rule reads}

\[
\begin {array}{l}
i_1+i_2+i_3=0\;\;\; mod. \; 3 \\
\left.
\begin{array}{cl}
j_{\sigma(1)}=0\;j_{\sigma(2)},j_{\sigma(3)}\neq 0 &
 j_{\sigma(2)}-j_{\sigma(3)}=4 \\
j_{\sigma(1)}\;{\rm even}\;j_{\sigma(2)},j_{\sigma(3)}\; {\rm odd} &
j_{\sigma(3)}-5=j_{\sigma(2)}+1=j_{\sigma(1)}\\
j_{\sigma(1)}\;{\rm odd}\;j_{\sigma(2)},j_{\sigma(3)}\; {\rm even} &
j_{\sigma(3)}-7=j_{\sigma(2)}-5=j_{\sigma(1)}\\
j_{\sigma(1)},j_{\sigma(2)},j_{\sigma(3)}\; {\rm odd\; or \; even} &
j_{\sigma(3)}=j_{\sigma(2)}=j_{\sigma(1)}
\end {array}
\right\}\;\;\;
\begin {array}{l}
mod.\;8 \\
\sigma\;\equiv{\rm permutation\;of\; \{1,2,3\}}
\end {array}
\end {array}
\]

\indent{Number of allowed couplings: 189}

\indent{Expression of the coupling}

\begin {eqnarray*}
C_{\theta^4\theta^4\theta^4} &=&
N \;F(l_1,l_2,l_3)\; \sum_{v \in
(f_3-f_2+\Lambda)} \exp [-\frac
{\sqrt{3}}{8\pi} |v|^2] \\
&=& N \;F(l_1,l_2,l_3)\; \sum_{\vec{u}\in Z^6} \exp [-\frac
{\sqrt{3}}{8\pi} (\vec{f_{23}}+\vec{u})^{top} M (\vec{f_{23}}+\vec{u})]\\
&=& N \; F(l_1,l_2,l_3) \; [\vartheta
\left[
\begin{array}{c}
\vec{\bar{f_{23}}} \\
0
\end {array}
\right]
[0, \Omega]
\end {eqnarray*}

\indent{where $F=1$ except for the case $f_1=f_2=f_3$ and $l_1=l_2=l_3=4$
in which $F=\frac{1}{2}$}
\[
\Omega = \; i \frac{\sqrt{3}}{8\pi}\; M \;\;\;
N= \sqrt{V_{\Lambda}}\;  \frac{3^{1/4}}{4\pi^2} \;
\left[ \frac{\Gamma^5(\frac{2}{3})}
{\Gamma^4(\frac{1}{6})} \right]
\]

\[
\begin{array}{l}
\Omega =\; i \frac{\sqrt{3}}{8\pi}\;
\left(
\begin {array}{cccccc}
R_1^2 & -\frac{R_1^2}{2} & 0 & 0 & 0 & 0 \\
- \frac {R_1^2}{2} & R_1^2 & 0 & 0 & 0 & 0 \\
0 & 0 & R_3^2 &R_3^2 \alpha_{34} & -R_3^2 \frac{1+2\alpha_{34}}{4} &
-R_3^2 \frac{1+2\alpha_{34}}{4} \\
0 & 0 & R_3^2 \alpha_{34} & R_3^2 & -R_3^2 \frac{1}{2} & -R_3^2
 \frac{1+2\alpha_{34}}{4}\\
0 & 0 & -R_3^2 \frac{1+2\alpha_{34}}{4}& -R_3^2 \frac{1}{2} & R_3^2\frac {1}{2}
& R_3^2 \frac {\alpha_{34}}{2} \\
0 & 0 & -R_3^2 \frac{1+2\alpha_{34}}{4} & -R_3^2 \frac{1+2\alpha_{34}}{4}
& R_3^2 \frac {\alpha_{34}}{2} & R_3^2 \frac{1}{2}
\end {array}
\right)
\end{array}
\]

\indent{Number of effective parameters: 3}

\indent{Number of different couplings without deformations: 6}

\indent{Number of different couplings with deformations: 6}

\indent{corresponding to the following $\vec f_{23} $ shifts}
\[
\vec{f_{23}}=\;
\left[
\begin{array}{l}
F=\frac{1}{2} \left\{
\begin{array}{l}
(0,0,\frac{2}{3},\frac{1}{3},\frac{2}{3},0),\\
(\frac{1}{3},\frac{2}{3},\frac{2}{3},\frac{1}{3},\frac{2}{3},0),
\end{array}
\right. \\
F=1 \left\{
\begin{array}{l}
(0,0,0,0,0,0),\\
(\frac{1}{3},\frac{2}{3},0,0,0,0),\\
(0,0,\frac{2}{3},\frac{1}{3},\frac{2}{3},0),\\
(\frac{1}{3},\frac{2}{3},\frac{2}{3},\frac{1}{3},\frac{2}{3},0)
\end{array}
\right.
\end{array}
\right.
\]

\vspace {1.0 cm}
\noindent{\underline{\bf{ORBIFOLD $Z_{12}$--II}}}
\vspace{.5 cm}

\noindent{\underline {Twist}
$ \theta={\rm
diag}(e^{i\alpha},e^{5i\alpha},e^{-6i\alpha}) ,\;\;\;\;
\alpha=\frac{2\pi}{12} $}

\noindent{\underline {Lattice}
$ SO(4) \otimes F_4 $}

\noindent{\underline {Coxeter element}}
\[
\begin{array}{lll}
\theta e_1=-e_1, & \theta e_2=-e_2, & \theta e_3=e_4, \\
\theta e_4=e_3+e_4+2 e_5 ,& \theta e_5= e_6, &
\theta e_6 = -e_3-e_4-e_5-e_6
\end{array}
\]

\noindent{\underline {Deformation parameters}}

\indent{Relations}
\[
\begin{array}{lll}
|e_3|=|e_4|= \sqrt{2} |e_5| = \sqrt{2} |e_6|,&
\alpha_{45}=-\frac{1}{\sqrt{2}}, &
\alpha_{34}= \alpha_{56},  \\
\alpha_{35}= \alpha_{46} = \alpha_{36}, &
\alpha_{35}= -\frac{1}{2\sqrt{2}}[1+2\alpha_{34}] &
 \alpha_{ij}=0\;\;\;i=1,2,\;\;j=3,4,5,6
\end{array}
\]
\[
\alpha_{ij}\equiv\cos(\theta_{ij})
\]

\indent{Degrees of freedom (5)}
\[
\begin {array}{lllll}
R_1= |e_1|, & R_2= |e_2|, & R_3= |e_3|, &
\alpha_{12}, &
\alpha_{34}
\end {array}
\]

\noindent{\underline{Lattice basis ($e_i$) in terms of  orthogonal basis
 ($\tilde e_i$)}}

\indent{Not necessary in this case.}

\vspace{.5 cm}
\noindent{\underline {Fixed points of $\theta$} (4)}
\[
\begin {array}{c}
f_1^{(i)}= g_1^{(i)} \otimes \hat g_1^{(0)} \;,\;\;\; i=0,1,2,3
\end {array}
\]
\[
\begin{array}{lll}
g_1^{(0)} = (0,0)\;,& g_1^{(1)} =\frac{1}{2} (0,1)\;,&
g_1^{(2)} =\frac{1}{2} (1,1)\;, \\
g_1^{(3)} =\frac{1}{2} (1,0)\;, &
\hat g_1^{(0)}=(0,0,0,0)\;&
\end {array}
\]

\vspace{.5 cm}
\noindent{\underline {Fixed points of $\theta^2$} (1)}

\indent{Fixed torus: $\alpha(e_1)+\beta(e_2)\;,\;\; \alpha,\;\beta
\in R$}

\[
\begin {array}{c}
f_{i}^{(2)}=[\alpha(e_1)+\beta(e_2)] \otimes \hat g_2^{(0)}\;,\;\;\;
\alpha,\beta \in R
\end {array}
\]
\[
\begin{array}{l}
\hat g_2^{(0)}=(0,0,0,0)
\end {array}
\]

\vspace{.5 cm}
\noindent{\underline {Fixed points of $\theta^3$} (16)}

\[
\begin {array}{c}
f_3^{(ij)}= g_3^{(i)} \otimes \hat g_3^{(j)}\;,\;\;\;i,j=0,1,2,3
\end {array}
\]
\[
\begin{array}{llll}
g_3^{(0)} = (0,0)\;,&
g_3^{(1)} =\frac{1}{2} (0,1)\;,&
g_3^{(2)} =\frac{1}{2} (1,1)\;, &
g_3^{(3)} =\frac{1}{2} (1,0)\;, \\
\hat g_3^{(0)} =(0,0,0,0)\;,&
\hat g_3^{(1)} =\frac{1}{2} (1,0,0,0)\;,&
\hat g_3^{(2)} =\frac{1}{2} (1,1,0,0)\;, &
\hat g_3^{(3)} =\frac{1}{2} (0,1,0,0)
\end {array}
\]

\indent{Note that, in $F_4$ $, \theta : \hat g_3^{(1)} \rightarrow
\hat g_3^{(3)} \rightarrow \hat g_3^{(2)}$}

\indent {Number of conjugation classes: 8}

\vspace{.5 cm}
\noindent{\underline {Fixed points of $\theta^4$} (9)}

\indent{Fixed torus: $\alpha(e_1)+\beta(e_2)\;,\;\; \alpha,\;\beta
\in R$}
\[
\begin{array}{c}
f_4^{(i)}= [\alpha(e_1)+\beta(e_2)] \otimes \hat g_4^{(i)} \;,\;\;\;
i=0,1,...,8
\end {array}
\]
\[
\begin {array}{llll}
\hat g_4^{(0)}=(0,0,0,0), &
\hat g_4^{(1)}=\frac{1}{3} (2,1,2,0), &
\hat g_4^{(2)}=\frac{1}{3} (2,2,0,2), &
\hat g_4^{(3)}=\frac{1}{3} (1,0,2,2), \\
\hat g_4^{(4)}=\frac{1}{3} (0,2,2,1), &
\hat g_4^{(5)}=\frac{1}{3} (1,2,1,0), &
\hat g_4^{(6)}=\frac{1}{3} (1,1,0,1), &
\hat g_4^{(7)}=\frac{1}{3} (2,0,1,1), \\
\hat g_4^{(8)}=\frac{1}{3} (0,1,1,2) & & &
\end{array}
\]

\indent{Note that, in $F_4$, $\theta : \hat g_4^{(1)} \rightarrow \hat
g_4^{(3)}
\rightarrow \hat g_4^{(5)} \rightarrow \hat g_4^{(7)}$
and $\theta : \hat g_4^{(2)} \rightarrow \hat g_4^{(4)} \rightarrow \hat
 g_4^{(6)}
\rightarrow \hat g_4^{(8)}$}

\indent {Number of conjugation classes: 3}

\vspace{.5 cm}
\noindent{\underline {Fixed points of $\theta^5$} (4)}

\indent{The same  as for $\theta$.}
\[
\begin {array}{c}
f_5^{(i)}= g_5^{(i)} \otimes \hat g_5^{(0)}\;,\;\;\; i=0,1,2,3
\end {array}
\]
\[
\begin{array}{lll}
g_5^{(0)} = (0,0)\;,&
g_5^{(1)} =\frac{1}{2} (0,1)\;,&
g_5^{(2)} =\frac{1}{2} (1,1)\;, \\
g_5^{(3)} =\frac{1}{2} (1,0)\;, &
\hat g_5^{(0)}=(0,0,0,0) &
\end {array}
\]

\vspace{.5 cm}
\noindent{\underline {Fixed points of $\theta^6$} (16)}

\indent{Fixed torus: $\alpha(e_1)+\beta(e_2)\;,\;\; \alpha,\;\beta
\in R$}

\[
\begin {array}{c}
f_6^{(i)}= [\alpha(e_1)+\beta(e_2)]\otimes \hat g_6^{(i)}\;,\;\;\;
 i=0,1,...,15,\;\;
\alpha,\beta \in R
\end {array}
\]
\[
\begin{array}{llll}
\hat g_6^{(0)}=(0,0,0,0)\;,&
\hat g_6^{(1)} =\frac{1}{2} (1,1,1,1)\;,&
\hat g_6^{(2)} =\frac{1}{2} (0,0,0,1)\;, &
\hat g_6^{(3)} =\frac{1}{2} (0,0,1,0)\;, \\
\hat g_6^{(4)} =\frac{1}{2} (1,0,0,0)\;, &
\hat g_6^{(5)} =\frac{1}{2} (0,0,1,1)\;, &
\hat g_6^{(6)} =\frac{1}{2} (0,1,0,1)\;, &
\hat g_6^{(7)} =\frac{1}{2} (1,0,1,0)\;, \\
\hat g_6^{(8)} =\frac{1}{2} (0,1,0,0)\;, &
\hat g_6^{(9)} =\frac{1}{2} (0,1,1,1)\;, &
\hat g_6^{(10)} =\frac{1}{2} (1,1,0,1)\;,&
\hat g_6^{(11)}=\frac{1}{2} (0,1,1,0)\;, \\
\hat g_6^{(12)} =\frac{1}{2} (1,1,0,0)\;, &
\hat g_6^{(13)} =\frac{1}{2} (1,0,1,1)\;, &
\hat g_6^{(14)} =\frac{1}{2} (1,0,0,1)\;, &
\hat g_6^{(15)} =\frac{1}{2} (1,1,1,0)
\end {array}
\]
\begin{quotation}
\noindent{Note that in $F_4$ $\theta:\hat g_6^{(3)} \rightarrow \hat g_6^{(2)}
\rightarrow \hat g_6^{(1)} \rightarrow \hat g_6^{(11)} \rightarrow \hat
 g_6^{(10)}
\rightarrow \hat g_6^{(9)}$ ,  $\theta: \hat g_6^{(7)}
\rightarrow \hat g_6^{(6)} \rightarrow \hat g_6^{(5)} \rightarrow \hat
 g_6^{(15)}
\rightarrow \hat g_6^{(14)}
\rightarrow \hat g_6^{(13)}$ and $\theta: \hat g_6^{(4)}
\rightarrow \hat g_6^{(8)}
\rightarrow \hat g_6^{(12)}$}
\end{quotation}

\indent {Number of conjugation classes: 4}

\vspace{.5 cm}
\noindent {\underline {Coupling $\theta\theta\theta^{10}$}}

\indent {Selection rule}
\[
f_1+f_2-(I+\theta)f_3 \in \Lambda
\]

\indent {Denoting}
\[
\left.
\begin {array}{rcl}
f_1 &=& g_1^{(i_1)}\otimes \hat g_1^{(0)}\\
f_2 &=& g_1^{(i_2)}\otimes \hat g_1^{(0)}\\
f_3 &=& [\alpha(e_1)+\beta(e_2)] \otimes \hat g_2^{(0)}
\end {array}
\right\}\;\;\;
\begin {array} {l}
i_1,i_2=0,1,2,3\;,
\alpha,\beta \in R
\end {array}
\]

\indent{The selection rule reads}
\[
\begin {array}{l}
i_1=i_2 \;,
\end {array}
\]

\indent{Number of allowed couplings: 4}

\indent{Expression of the coupling}

\begin {eqnarray*}
C_{\theta\theta\theta^{10}} &=&
N \; \sum_{v \in (f_3-f_2+\Lambda)_{\perp}} \exp [-\frac
{1}{4\pi} \sin(\frac{\pi}{6} |v|^2] \\
&=&
N \;\; \sum_{\vec{u} \in Z^4} \exp [ -\frac
{1}{4\pi} \sin(\frac{\pi}{6}\; (\vec{\bar{f_{23}}}+\vec{u})^{\top}M
(\vec{\bar{f_{23}}}+\vec{u})] \\
&=&
N\; \vartheta
\left[
\begin{array}{c}
\vec{\bar{f_{23}}} \\
0
\end {array}
\right]
[0, \Omega]
\end {eqnarray*}
\begin{quotation}
\noindent{where $\bar f_{23}$ is the restriction of $f_{23}$ to the $F_4$
 lattice,
$(f_3-f_2+\Lambda)_{\perp}$ indicates that the coset
elements must belong to $F_4$ and $V_{\perp}$ is the volume
of the $F_4$ unit cell. In all cases $\bar{f_{23}}=0$. Finally}
\end{quotation}

\[
\Omega = i\frac{1}{4\pi^2} \sin(\frac{\pi}{6}) M \;,\;\;\;
N= \sqrt{V_{\perp}}\;  \frac{1}{2 \pi} \;
\left[ \frac{\Gamma(\frac{11}{12}) \Gamma(\frac{7}{12})}
{\Gamma(\frac{1}{12}) \Gamma(\frac{5}{12})} \right]
\]

\[
\begin{array}{l}
\Omega =i \frac{1}{4\pi^2} \sin(\frac{\pi}{6})\; R_3^2
\left(
\begin {array}{cccc}
1 & \alpha_{34} & -\frac{1}{4}[1+2\alpha_{34}] & -\frac{1}{4}[1+2\alpha_{34}]
\\
\alpha_{34} & 1 & -\frac{1}{2} & -\frac{1}{4}[1+2\alpha_{34}] \\
-\frac{1}{4}[1+2\alpha_{34}] & -\frac{1}{2} & \frac {1}{2} & \frac
 {\alpha_{34}}{2} \\
-\frac{1}{4}[1+2\alpha_{34}] & -\frac{1}{4}[1+2\alpha_{34}] & \frac
 {\alpha_{34}}{2} & \frac{1}{2}
\end {array}
\right)
\end{array}
\]

\indent{Number of effective parameters: 2}

\indent{Number of different couplings without deformations: 1}

\indent{Number of different couplings with deformations: 1}

\indent{Note that this coupling is the same as $\theta^2\theta^5\theta^5$}

\vspace{.5 cm}
\noindent {\underline {Coupling $\theta\theta^3\theta^8$}}

\indent {Selection rule}
\[
f_1+(I+\theta+\theta^2)f_2-(I+\theta+\theta^2+\theta^3)f_3 \in \Lambda
\]

\indent {Denoting}
\[
\left.
\begin {array}{rcl}
f_1 &=& g_1^{(i_1)}\otimes \hat g_1^{(0)}\\
f_2 &=& g_3^{(i_2)}\otimes \hat g_3^{(j_2)}\\
f_3 &=& [\alpha(e_1)+\beta(e_2)] \otimes \hat g_4^{(j_3)}
\end {array}
\right\}\;\;\;
\begin {array} {l}
i_1,i_2,j_2=0,1,2,3\;,\\
j_3=0,1,...,8\;,\\
\alpha,\beta \in R
\end {array}
\]

\indent{the selection rule reads}
\[
\begin {array}{l}
i_1=i_2
\end {array}
\]

\indent{Number of allowed couplings: 6}

\indent{Expression of the coupling}

\begin {eqnarray*}
C_{\theta\theta^3\theta^8} &=&
N \; \sqrt{l_2 l_3} \; \sum_{v \in (f_3-f_2+\Lambda)_{\perp}}  \exp [-\frac
{1}{4\pi} \frac{\sin(\frac{\pi}{3})\sin(\frac{\pi}{4})}{\sin(\frac{\pi}{12})}
 |v|^2] \\
 &=& N \; \sqrt{l_2 l_3}\; \sum_{\vec{u} \in Z^4} \exp [ -\frac
{1}{4\pi} \frac{\sin(\frac{\pi}{3})\sin(\frac{\pi}{4})}{\sin(\frac{\pi}{12})}
\;
(\vec{\bar{f_{23}}}+\vec{u})^{\top} M (\vec{\bar{f_{23}}}+\vec{u})]  \\
&=&
N\;\sqrt{l_2 l_3}\; \vartheta
\left[
\begin{array}{c}
\vec{\bar{f_{23}}} \\
0
\end {array}
\right]
[0, \Omega]
\end {eqnarray*}
\begin{quotation}
\noindent{with the same notation as in the previous coupling,
$l_i$ is the number of elements of the
$f_i$ conjugation class and $V_{\perp}$ is the volume of the $F_4$ unit cell.
 Finally}
\end{quotation}

\[
\Omega = i\frac{1}{4\pi^2}
 \frac{\sin(\frac{\pi}{3})\sin(\frac{\pi}{4})}{\sin(\frac{\pi}{12})}\;M
\;,\;\;\;
N= \sqrt{V_{\perp}}\;  \frac{1}{2 \pi} \; \frac
 {\Gamma(\frac{3}{4})}{\Gamma(\frac{1}{4})}
\left[ \frac{\Gamma(\frac{11}{12}) \Gamma(\frac{7}{12})}
{\Gamma(\frac{1}{12}) \Gamma(\frac{5}{12})} \right]^{1/2}
\]

\[
\begin{array}{l}
\Omega= i\frac{1}{4\pi^2} \frac{\sin(\frac{\pi}{3})\sin(\frac{\pi}{4})}
{\sin(\frac{\pi}{12})}\; R_3^2 \left(
\begin {array}{cccc}
1 & \alpha_{34} & -\frac{1}{4}[1+2\alpha_{34}] & -\frac{1}{4}[1+2\alpha_{34}]
\\
\alpha_{34} & 1 & -\frac{1}{2} & -\frac{1}{4}[1+2\alpha_{34}] \\
-\frac{1}{4}[1+2\alpha_{34}] & -\frac{1}{2} & \frac {1}{2} & \frac
 {\alpha_{34}}{2} \\
-\frac{1}{4}[1+2\alpha_{34}] & -\frac{1}{4}[1+2\alpha_{34}] & \frac
 {\alpha_{34}}{2} & \frac{1}{2}
\end {array}
\right)
\end{array}
\]

\indent{Number of effective parameters: 2}

\indent{Number of different couplings without deformations: 4}

\indent{Number of different couplings with deformations: 4}

\indent{corresponding to the following $\vec{\bar{ f_{23}}} $ shifts}

\[
\vec{\bar{f_{23}}}= \; \left[
\begin{array}{ll}
l_2=1\;l_3=1  & (0,0,0,0), \\
l_2=3\;l_3=1  & (\frac{1}{2},0,0,0), \\
l_2=1\;l_3=4  & (\frac{2}{3},\frac{2}{3},0,\frac{2}{3}), \\
l_2=3\;l_3=4  & (\frac{1}{6},\frac{1}{6},0,\frac{2}{3})
\end{array}
\right.
\]

\vspace{.5 cm}
\noindent {\underline {Coupling $\theta^3\theta^3\theta^6$}}

\indent {Selection rule}
\[
f_1+f_2-(I+\theta^3)f_3 \in \Lambda
\]

\indent {Denoting}
\[
\left.
\begin {array}{rcl}
f_1 &=& g_3^{(i_1)}\otimes \hat g_3^{(j_1)}\\
f_2 &=& g_3^{(i_2)}\otimes \hat g_3^{(j_2)}\\
f_3 &=& [\alpha(e_1)+\beta(e_2)]\otimes \hat g_6^{(j_3)}
\end {array}
\right\}\;\;\;
\begin {array} {l}
i_1,i_2,j_1,j_2=0,1,2,3\;,\\
j_3=0,1,...,15\;,\\
\alpha, \beta \in R
\end {array}
\]

\indent{the selection rule reads}
\[
\left.
\begin {array}{l}
i_1=i_2 \\
j_1+(-1)^{(j_3+1)}j_2=j_3
\end {array}
\right\} \;\;\; mod.\; 4
\]

\indent{Number of allowed couplings: 56}

\indent{Expression of the coupling}

\indent {In all the cases, except for the case $l_1=l_2=l_3=3$}

\begin {eqnarray*}
C_{\theta^3\theta^3\theta^6} &=&
N \;F(l_1,l_2,l_3) \; \sum_{v \in (f_3- f_2+\Lambda)_{\perp}} \exp [-\frac
{1}{4\pi} \; |v|^2] \\
&=& N \; F(l_1,l_2,l_3) \; \sum_{\vec{u} \in Z^4} \exp [-\frac
{1}{4\pi} \; (\vec{\bar{f_{23}}}+\vec{u})^{\top} M
(\vec{\bar{f_{23}}}+\vec{u})] \\
&=&
N\;F(l_1,l_2,l_3) \; \vartheta
\left[
\begin{array}{c}
\vec{\bar{f_{23}}} \\
0
\end {array}
\right]
[0, \Omega]
\end {eqnarray*}
\begin{quotation}
\noindent{with the same notation as in the previous coupling.
$l_i$ is the number of elements of the
$f_i$ conjugation class. $F(l_1,l_2,l_3)$ is given by}
\end{quotation}

\[
\begin{array}{rlrl}
l_1=l_2=l_3=1 \;{\rm and}\;l_1=l_2=3 \; l_3=1 \;:& F=1  &
l_1=l_2=1 \; l_3=3 \;:& F=\sqrt{3} \\
l_1=1(3)\;l_2=3(1)\;l_3=6 \;:& F=\sqrt{2}  &
l_1=l_2=3 \; l_3=6 \;:& F=2\sqrt{2}
\end{array}
\]

\indent{In the case $l_1=l_2=l_3=3$}

\begin {eqnarray*}
C_{\theta^3\theta^3\theta^6} &=&
N \;\frac{1}{\sqrt{3}}\; \sum_{v \in \cup_{p=0}^{2} (\theta^p f_3-
 f_2+\Lambda)_{\perp}} \exp [-\frac
{1}{4\pi} \; |v|^2] \\
&=&
N\; \frac{1}{\sqrt{3}} \;\{ \vartheta
\left[
\begin{array}{c}
\vec{\bar{f_{23}}} \\
0
\end {array}
\right]
[0, \Omega] +
\vartheta
\left[
\begin{array}{c}
\vec{\bar{ f'_{23}}} \\
0
\end {array}
\right]
[0, \Omega] +
\vartheta
\left[
\begin{array}{c}
\vec{\bar{ f''_{23}}} \\
0
\end {array}
\right]
[0, \Omega] \}
\end {eqnarray*}

\indent{$f'_{23}=\theta f_2  -f_3$ and $f''_{23}= \theta^2 f_2 -f_3$}

\[
\Omega = i \frac{1}{4 \pi^2} M
\;\;\;\;
N= \sqrt{V_{\perp}}\;  \frac{1}{2 \pi} \; \left[ \frac {\Gamma(\frac{3}{4})}
{\Gamma(\frac{1}{4})} \right]^2
\]

\indent{$V_{\perp}$ is the volume of the $F_4$ unit cell}

\[
\begin{array}{l}
\Omega = i \frac{1}{4 \pi^2} \; R_3^2
\left(
\begin {array}{cccc}
1 & \alpha_{34} & -\frac{1}{4}[1+2\alpha_{34}] & -\frac{1}{4}[1+2\alpha_{34}]
\\
\alpha_{34} & 1 & -\frac{1}{2} & -\frac{1}{4}[1+2\alpha_{34}] \\
-\frac{1}{4}[1+2\alpha_{34}] & -\frac{1}{2} & \frac {1}{2} & \frac
 {\alpha_{34}}{2} \\
-\frac{1}{4}[1+2\alpha_{34}] & -\frac{1}{4}[1+2\alpha_{34}] & \frac
 {\alpha_{34}}{2} & \frac{1}{2}
\end {array}
\right)
\end{array}
\]

\indent{Number of effective parameters: 2}

\indent{Number of different couplings without deformations: 6}

\indent{Number of different couplings with deformations: 6}

\indent{corresponding to the following $\vec{\bar{ f_{23}}} $ shifts}
\[
\vec{\bar{f_{23}}}=\; \left[
\begin{array}{ll}
l_1=l_2=l_3=1 & (0,0,0,0), \\
l_1=l_2=3\; l_3=1 & (\frac{1}{2},0,0,0), \\
l_1=l_2=1\; l_3=3 & (\frac{1}{2},0,0,0), \\
l_1=l_2=l_3=3 & (0,0,0,0) \cup (\frac{1}{2},0,0,0) \cup (0,\frac{1}{2},0,0), \\
l_1=1(3)\; l_2=3(1)\; l_3=6 & (0,\frac{1}{2},\frac{1}{2},0), \\
l_1=l_2=3\; l_3=6 & (\frac{1}{2},0,\frac{1}{2},0)
\end{array}
\right.
\]
\newpage

\textwidth=18cm
\textheight=22cm
\oddsidemargin=-1cm
\evensidemargin=0cm
\topmargin=0cm
\parindent=1cm

\begin{table}
\underline{\bf TABLE 1}
$\begin{array}{|c|c|c|c|c|c|c|c|c|} \hline
Orb. & Twist\;\theta  & Lattice & \#DP & Coupling  & \#AC
&\#EDP  &\#DCR  & \#DCD \\ \hline
Z_3  & (1,1,-2)/3    & SU(3)^3 & 9 & \theta\theta\theta    & 729 &
9 & 4 & 14 \\ \hline
Z_4  & (1,1,-2)/4    & SU(4)^2 & 7 & \theta\theta\theta^2    & 160 &
4 & 6 & 10 \\
  &     & SO(4)^3 & 7 & \theta\theta\theta^2    & 160 &
4 & 6 & 8 \\ \hline
Z_6-{\rm I} & (1,1,-2)/6 & G_2^2\times SU(3) & 5 & \theta\theta^2\theta^3 & 90
&
4 & 10 & 30 \\
      &  &  & & \theta^2\theta^2\theta^2 & 369 &
5 & 8 & 12 \\ \hline
Z_6-{\rm II} & (1,2,-3)/6 & SU(6)\times SU(2) & 5 & \theta\theta^2\theta^3
& 48 & 1 & 4 & 4 \\
      &  &  & & \theta\theta\theta^4 & 72 &
3 & 4 & 4 \\ \hline
Z_7 & (1,2,-3)/7 & SU(7) & 3 & \theta\theta^2\theta^4
& 49 & 3 & 2 & 4 \\ \hline
Z_8-{\rm I} & (1,2,-3)/8 & SO(5)\times SO(9) & 3 & \theta^2\theta^2\theta^4
& 84 & 2 & 8 & 9 \\
      &  &  & & \theta\theta^2\theta^5 & 40 &
3 & 8 & 9 \\ \hline
Z_8-{\rm II} & (1,3,-4)/8 & SO(4)\times SO(8) & 5 & \theta\theta\theta^6
& 24 & 2 & 3 & 3 \\
      &  &  & & \theta\theta^3\theta^4 & 48 &
2 & 6 & 6 \\ \hline
Z_{12}-{\rm I} & (1,4,-5)/12 & SU(3)\times F_4 & 3 & \theta\theta^2\theta^9
& 6 & 2 & 2 & 2 \\
      &  &  & & \theta\theta^4\theta^7 & 6 &
3 & 4 & 4 \\
      &  &  & & \theta^2\theta^4\theta^6 & 36 &
2 & 7 & 7 \\
      &  &  & & \theta^4\theta^4\theta^4 & 189 &
3 & 6 & 6 \\ \hline
Z_{12}-{\rm II} & (1,5,-6)/12 & SO(4)\times F_4 & 5 & \theta\theta\theta^{10}
& 4 & 2 & 1 & 1 \\
      &  &  & & \theta\theta^3\theta^8 & 6 &
2 & 4 & 4 \\
      &  &  & & \theta^3\theta^3\theta^6 & 56 &
2 & 6 & 6 \\ \hline
\end{array}$
\caption{Characteristics of twisted Yukawa couplings for $Z_n$
Coxeter orbifolds (the non--Coxeter $Z_4$ one with $SO(4)^3$
lattice is given for comparison). The twist $\theta$ is
specified by the three $c_i$ parameters (one for each complex
plane rotation) appearing in $\theta=\exp (\sum c_i J_i)$.
$\#DP\equiv\#$ of deformation parameters, $\#AC\equiv\#$ of
allowed couplings, $\#EDP\equiv\#$ of effective deformation
parameters, $\#DCR\equiv\#$ of different Yukawa couplings for
the non--deformed (rigid) orbifold, $\#DCD\equiv\#$ of different
Yukawa couplings when deformations are considered.}

\end{table}

\end{document}